\documentclass[fleqn,usenatbib]{mnras}

\usepackage{newtxtext,newtxmath}

\usepackage[T1]{fontenc}
\usepackage{ae,aecompl}

\usepackage{graphicx}	
\usepackage{subfigure}

\makeatletter
    \renewcommand{\p@subfigure}{}
\makeatother

\usepackage{alphalph} 

\usepackage{caption}
\usepackage{amsmath}	
\usepackage{bm}
\usepackage{gensymb}
\usepackage{enumitem}
\usepackage{xspace}
\usepackage{multirow}
\usepackage[dvipsnames,table]{xcolor}
\usepackage[utf8]{inputenc}
\usepackage{pdflscape} 
\usepackage{longtable}
\usepackage{booktabs}
\usepackage{afterpage}
\usepackage{xurl}

\newcommand\hii{\ion{H}{II}}
\newcommand\ha{$\rm{H}\alpha$}
\newcommand\hb{$\rm{H}\beta$}
\newcommand\lha{$L(\rm{H}\alpha)$}
\newcommand\lhaobs{$L(\rm{H}\alpha)^{obs}$}
\newcommand\lhaobsI{$L(\rm{H}\alpha)^{obs}_0$}
\newcommand\lhamod{$L(\rm{H}\alpha)^{mod}$}

\newcommand\ergs{\rm{~erg~s^{-1}}}
\newcommand{\ch}{$\chi^2$}
\newcommand{\msun}{M$_\odot$}
\newcommand{\tm}{$age^{\rm mod}$}
\newcommand{\avm}{$A_V^{\rm mod}$}
\newcommand{\mm}{$mass^{\rm mod}$}
\newcommand{\ts}{$age^{\rm stoch}$}
\newcommand{\avs}{$A_V^{\rm stoch}$}
\newcommand{\ms}{$mass^{\rm stoch}$}

\makeatletter
\newcommand{\aftertwo}[1]{\afterpage{\if@firstcolumn #1
  \else\afterpage{#1}\fi}}
\makeatother

\title[The PARSEC view of the centre of an early spiral]{The PARSEC view of star formation in 
galaxy centres: from 
protoclusters to star clusters in an early-type spiral  }

\author[A. Prieto et al.]{
Almudena Prieto,$^{1,2,3}$\thanks{E-mail: almudena.prieto@iac.es}
Gladis Magris C.,$^{4}$
Gustavo Bruzual,$^{5}$
\newauthor
Juan A. Fern\'andez-Ontiveros,$^{6}$
Andreas Burkert$^{3,7}$
\vspace{0.2cm}\\
$^{1}$Instituto de Astrof\'isica de Canarias (IAC),  La Laguna, Tenerife, Spain\\
$^{2}$Departamento de Astrof\'isica, Universidad de La Laguna, La Laguna, Tenerife, Spain\\
$^{3}$Universit\"ats-Sternwarte M\"unchen,  D-81679 M\"unchen, Germany\\
$^{4}$Centro de Investigaciones de Astronom\'ia (CIDA),  M\'erida, 5101, Venezuela\\
$^{5}$Instituto de Radioastronom{\'i}a y Astrof{\'i}sica (IRyA), UNAM, Campus Morelia, Michoac\'an, C.P. 58089, M{\'e}xico\\
$^{6}$Centro de Estudios de F\'isica del Cosmos de Arag\'on (CEFCA), Plaza San Juan 1, 44001, Teruel, Spain\\
$^{7}$Max-Planck-Institut f\"ur extraterrestrische Physik, D-85741 Garching, Germany}

\date{Accepted XXX. Received YYY; in original form ZZZ}

\pubyear{2021}

\begin{document}
\label{firstpage}
\pagerange{\pageref{firstpage}--\pageref{lastpage}}
\maketitle

\begin{abstract}
Understanding star formation in galaxies requires resolving the physical scale on which
star formation often occurs: the scale of star clusters. We present a multiwavelength, 
eight-parsec resolution study of star formation in the circumnuclear star cluster and molecular
gas rings of the early-type spiral NGC\,1386. The cluster ring formed simultaneously $\sim 4$ Myr ago. 
The clusters have similar properties in terms of mass and star
formation rate, resembling those of \hii\ regions in the Milky Way disc. The molecular
CO gas resolves into long filaments, which define a secondary ring detached from the cluster ring. Most clusters are in CO voids. Their
separation with respect the CO filaments is reminiscent of
that
seen in galaxy spiral arms. By analogy, we propose that a density wave through the
disc of this galaxy may have produced 
this gap
in the central kpc. The 
CO filaments fragment into strings of dense, unresolved clouds with no evidence of a stellar
counterpart. These clouds may be the sites of a future population of clusters in the ring.
The free-fall time of these clouds, $\sim$\,10 Myr, is close to the orbital time of the CO ring.
This coincidence could lead to a synchronous bursting ring, as is the case for the current
ring. The inward spiralling morphology of the CO filaments and co-spatiality with
equivalent kpc-scale dust filaments are suggestive of their role as matter carriers from
the galaxy outskirts to feed the molecular ring and a moderate active nucleus.

\end{abstract}



\begin{table*}
\begin{center}
 \caption{\label{sample}{NGC 1386 general properties.}}
 \begin{tabular}{cccccccc}
 \hline
  \multirow{2}{*}{    Name} & \multirow{2}{*}{    Type} & \multirow{2}{*}{    Class} & \multicolumn{1}{c}{    D}& \multicolumn{1}{c}{    1$''$}& \multicolumn{1}{c}{    FWHM} & \multicolumn{1}{c}{    Star-formation} & \multicolumn{1}{c}{    Star-formation} \\
  & & & \multicolumn{1}{c}{[Mpc]} & \multicolumn{1}{c}{[pc]} & \multicolumn{1}{c}{[pc]} & \multicolumn{1}{c}{    morphology} & \multicolumn{1}{c}{{    radius to centre} [pc]}\\[0.1cm]
 \hline\\[-0.3cm]
 NGC 1386 & SB(s)a          & Sy 2          & 15.3$^a$ &  73.4   & 6.7$^b$   & Ring     & 960 \\[0.1cm]
 \hline\\[-0.3cm]
 \multicolumn{8}{l}{{\it Notes.} $^a${\citet{jensen+03}}. $^b$FWHM corresponds to the size of the most compact object found in the FOV,}\\
 \multicolumn{8}{l}{usually in the VLT\,\textsf{NaCo}\,\textit{Ks}-band images.}\\
 \end{tabular}
\end{center}
\end{table*}

\section{Introduction}\label{s:intro}

Young massive star clusters are thought to be the common 
sites
of massive star formation (Lada \& Lada 2003; but see also \citep{krumholz+19}).  Clusters are gravitationally bound collections of thousands 
of stars which  result from  the fragmentation of  molecular  gas into massive clouds, which in turn further collapse 
into
the critical mass, density and gas temperature needed to trigger star formation. Young massive star clusters are 
characterised by 
ages 
<\,100\,Myr, 
masses 
in the $\sim 10^4$--$10^6\, \rm{M_\odot}$ range, and 
sizes in the range  
$\sim\,1$--$6\, \rm{pc}$. Starburst and 
interacting galaxies are the most common places to find them \citep{1995Natur.375..742M,1995ApJ...446L...1O,2010AJ....140...75W}. 
However, they are also found in different environments such as  tidal tails \citep{2011ApJ...731...93M}, circumnuclear rings 
\citep{1996AJ....111.2248M}, dwarf galaxies \citep{2011MNRAS.415.2388A} and often at the centre of  galaxies, including our 
Milky Way \mbox{\citep{2010IAUS..266...58B,2010A&A...511A..18S}}. Young massive clusters have been detected in high-$z$ galaxies 
up to the reionization era \citep{2023ApJ...945...53V}.

Young star clusters, and by extension the star-formation process, are thought to 
share
a set of common properties 
\citep{2010ARA&A..48..431P,krumholz+19}. Most of these characteristics are inherited from the natal molecular clouds 
where these clusters were formed
and remain similar in very different environments 
\citep[see][and references therein]{2010ARA&A..48..547F}; 
for example, in
spiral (Milky Way, M33 and M31), irregular 
(the Small and Large 
Magellanic Clouds, NGC\,6822), starburst (IC\,10), dwarf elliptical (NGC\,185, NGC\,205)
and  local galaxies (M51, M83, NGC\,253, NGC\,1569). 
Young star clusters are expected to be associated with large reservoirs of molecular gas, either in situ or in their 
surroundings. They are often seen in aggregations or groups; e.g.\ in circumnuclear rings in the centres of galaxies, 
in spiral arms and in interacting systems. Ideally, the study of star formation in these regions will largely benefit 
from the characterisation of these basic units
(the star clusters themselves and 
their
parental molecular clouds). For this
purpose, an
angular resolution in line with typical cluster sizes
(a few parsecs)
over a wide spectral range, typically 
from the UV to radio wavelengths, 
enables 
us to isolate clusters 
into 
groups and measure their individual 
ages, masses, sizes, star formation rates (SFR) and
possibly
their temporal relation to the surrounding molecular gas.  

The  PARSEC project\footnote{\url{https://www.iac.es/en/projects/central-parsec-galaxies-using-high-spatial-resolution-techniques}}, 
which 
is 
a  parsec-scale multiwavelength investigation  of the 
centres
 of the nearest galaxies, includes as one of its goals the study 
of some of the nearest nuclear star-forming regions in galaxies with a variety of 
galaxy morphologies
and nuclear activity, most of them 
being 
early-type galaxies. 
A major difference between PARSEC and other 
multiwavelength surveys, e.g.\ PHANGS (\citealt{leroy+23}), WISDOM (Davis et al. 2022) is 
its angular resolution  and  the galactic region under study. The PARSEC 
survey's 
resolution is restricted to a few parsecs in the 0.4--20 $\mu$m range, and, depending on the availability 
of high angular resolution data, it also covers the 
cm--mm and high energy ranges. PHANGS focuses on the 
large-scale 
view of star formation in spiral arms and bulges, which in 
general limits its resolution to tens of parsecs (\citealt{leroy+23}), whereas PARSEC instead focuses on the 
central nucleus
(kpc to a few hundred parsecs). 

PARSEC, using a typical scale of a few parsecs, from the UV to the mid-IR, by   resorting to HST in the UV--optical, and 
diffraction-limited and interferometry data from 8--10 m telescopes in the 1--20 $\mu$m range, has resolved  some of 
the nearest nuclear star-forming regions 
into 
their basic building units
(star clusters). 
These include the central 300 pc of 
the archetypal starburst galaxy NGC\,253, resolved into 37 individual star clusters bursting  from their dust cocoons. Most of 
these clusters were previously unknown due to confusion and/or extinction effects. A first determination of the size of these 
clusters, $\sim$\,1.5\,pc FWHM, became 
possible 
\citep{2009MNRAS.392L..16F}. 
In the central kpc of NGC\,1052, a massive elliptical galaxy, a nuclear population of 27 star clusters aged $\sim\,7$ Myr was 
discovered. These clusters are broadly dispersed in the region and no hint 
of 
a  driving agent producing this young population was 
found \citep{2011MNRAS.411L..21F}. The 
well-populated 
circumnuclear stellar ring in the early-type spiral NGC\,1097 was resolved into several hundred star clusters. The 
ages 
of these clusters, each individually dated, indicate that they belong to at least four different bursts of star formation,
 each lasting 10--20 Myr, over a total time span of 100 Myr \citep{prieto+19}.
Moving to older clusters, 115 
13-Gyr-old  
globular clusters were spatially resolved for the first time in the central kpc of M87. 
As for NGC\,1052, these clusters are broadly dispersed in the centre of M87, some of them 
being 
as close as a few tens of 
parsecs 
\citep{montes+14}.

\begin{figure*}
\begin{center}
    \includegraphics[scale=0.8]{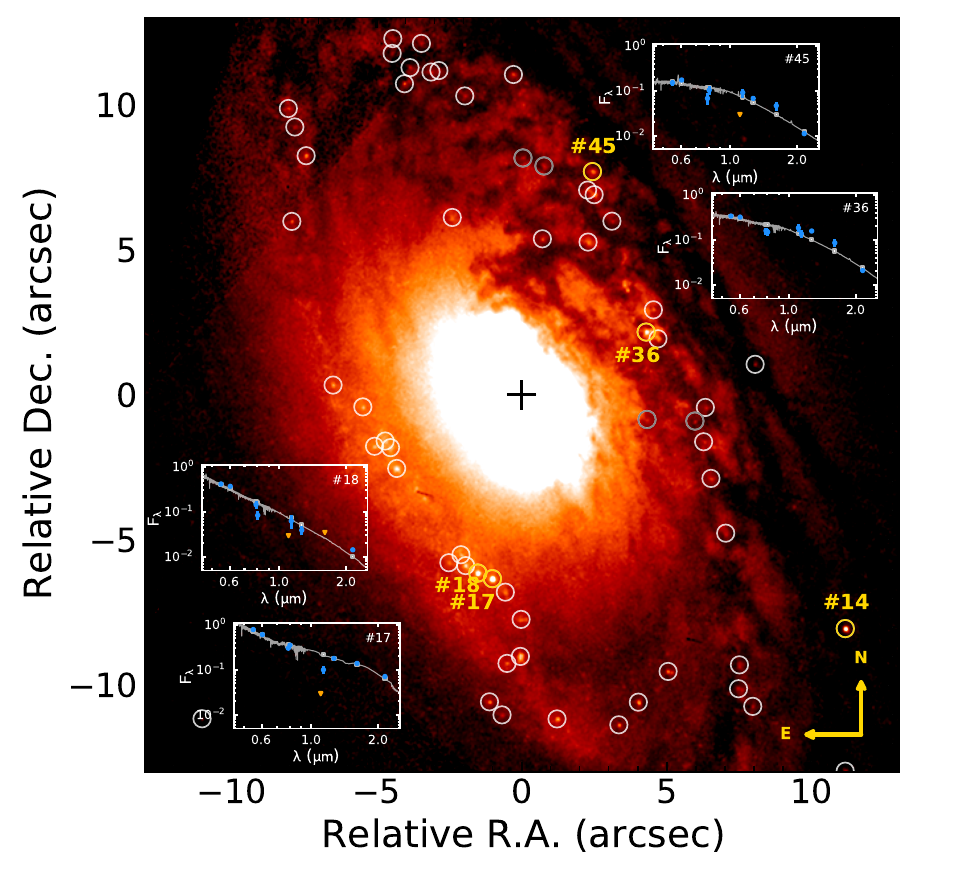}
      \includegraphics[width=1\columnwidth, scale=1]{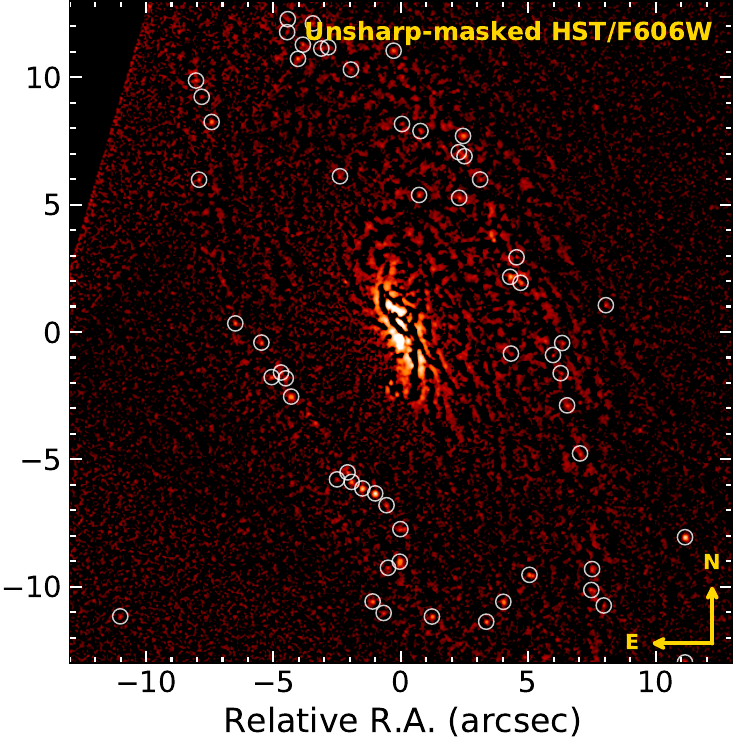}
      \includegraphics[width=1.\columnwidth, scale=1]{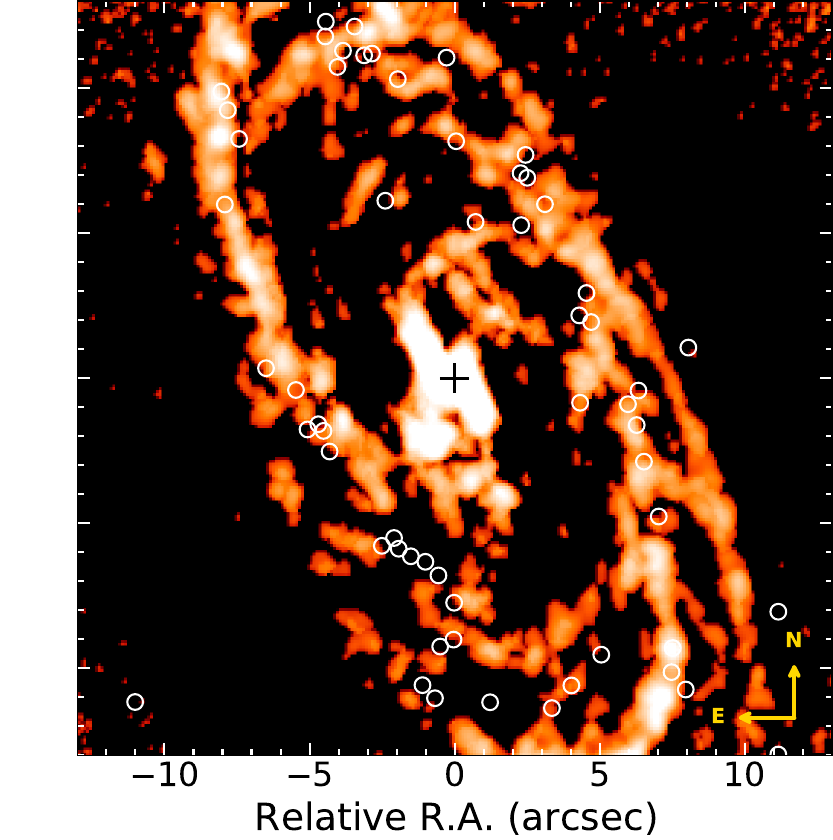}
    \caption{\label{image_V}{\it Upper panel} \textit{HST}/F606W image of the central  $26\arcsec \times26\arcsec$ 
in NGC\,1386 ($1\arcsec = $ 73.4 pc). 
Sixty-one 
clusters are detected by the \textsc{daofind} algorithm, marked with circles. 
The optical-IR SEDs (in units of $10^{-17} \ergs\,cm^{-2}\,$\AA$^{-1}$) for a representative group of clusters are shown. {\it Lower-left panel:} unsharp-masked 
\textit{HST}/F606W image of the central $26\arcsec \times26\arcsec$ showing all the clusters. {\it Lower-rigth panel:} 
ALMA/CO$(2-1)$ image of the same field of view as in the \textit{HST}/F606W image. The position of the optical continuum 
light of the clusters is marked with the same circles as in previous panels. The spatial displacement between the clusters 
continuum emission and CO is best appreciated in Fig.\,\ref{image_CO_large},
 where circle's size are the clusters upper limit size, 
FWHM $\approx 0.1\arcsec$, in the \textsf{NaCo}-K band image.
    }
\end{center}
\end{figure*}

In this paper, we focus on a moderately populated  circumnuclear young star cluster ring in the 
nearby 
 Sa galaxy NGC\,1386 
(15.3 Mpc, \citealt{jensen+03}). This central 1 kpc radius ring contains  61 clusters, each individually   studied on scales 
< 7\,pc FWHM in the 0.5--4 $\mu$m continuum emission range, and $\sim\,9$ pc FWHM in the \ha\ line.
These data are complemented with 
submillimetre 
CO(2-1) molecular gas emission,
resolved to scales of 40\,pc FWHM.  The combined 
multiwavelength 
dataset and the proximity of this galaxy allow for a clear spatial separation of the molecular 
gas into what we call a second circumnuclear ring
enclosing,
 but not overlapping,
 with the stellar one (Fig.\,\ref{image_V}).
Our data are described in Section \ref{data}. The main properties of the star clusters are discussed in Sections \ref{s:methods} and \ref{s:results}, and the 
molecular ring in Section \ref{s:CO}. An overview of the molecular--stellar system is proposed in Section \ref{s:conclusion}.

\begin{table*}
  \small
  \centering
  \caption[optical and IR dataset]{Optical (\textit{HST}) and IR (VLT) dataset for the galaxy NGC\,1386$^a$.}\label{tab_obs}
  \begin{minipage}{\textwidth}
  \renewcommand{\thefootnote}{\alph{footnote}}
  \let\footnotesize\footnotesize
  \begin{tabular*}{\textwidth}{rrcrrlclcll}
    \addlinespace[0.2cm]
Instrument & 
Filter &
\multicolumn{1}{c}{ $\lambda \ - \ \Delta \lambda$} & 
\multicolumn{1}{c}{Scale} & 
\multicolumn{1}{c}{Exposure} & 
\multicolumn{1}{c}{WFS} & 
\multicolumn{1}{r}{Seeing} & 
\multicolumn{1}{c}{ZP} & 
\multicolumn{1}{c}{$M_{\rm{nucl}} \pm Err$} & 
\multicolumn{1}{c}{FWHM} & 
Date \\

 & & \multicolumn{1}{c}{[$\rm{\mu m}$]} & \multicolumn{1}{c}{[$\,''\, \rm{px^{-1}}$]} & \multicolumn{1}{c}{[s]} & & & \multicolumn{1}{c}{[mag]} & \multicolumn{1}{c}{[mag]} & & [\textsc{dd mm yyyy}]\\

\toprule
\multicolumn{11}{c}{\cellcolor[gray]{0.9} \bf \sf NGC 1386} \\
\addlinespace[0.08cm]
\emph{HST}/WFPC2  &   F502N      & 0.501 - 0.011  & 0.0455  &  800    & \multicolumn{2}{c}{\dotfill} & 18.0    &    15.5 + 0.3    & 0.11 & 28 06 1997 \\
\emph{HST}/WFPC2  &   F547M      & 0.548 - 0.048  & 0.0455  &	360   & \multicolumn{2}{c}{\dotfill} & 21.7    &    17.8 + 0.2    & 0.09 & 1997--1999\footnotemark[1] \\
\emph{HST}/WFPC2  &   F606W      & 0.600 - 0.150  & 0.0455  &  160    & \multicolumn{2}{c}{\dotfill} & 22.90   &   17.09 + 0.08   & 0.11 & 14 06 1995 \\
\emph{HST}/WFPC2  &   F658N      & 0.659 - 0.007  & 0.0455  & 2000    & \multicolumn{2}{c}{\dotfill} & 18.2    &    14.8 + 0.2    & 0.12 & 1997--1999\footnotemark[1] \\
\emph{HST}/WFPC2  &   F791W      & 0.787 - 0.122  & 0.0455  &	80    & \multicolumn{2}{c}{\dotfill} & 21.5    &    16.5 + 0.1    & 0.11 & 28 06 1997 \\
\emph{HST}/WFPC2  &   F814W      & 0.800 - 0.152  & 0.0455  &	80    & \multicolumn{2}{c}{\dotfill} & 21.6    &    16.4 + 0.1    & 0.11 & 25 02 1999 \\
\emph{HST}/NIC1   &   F110M      & 1.103 - 0.139  & 0.043   &  512    & \multicolumn{2}{c}{\dotfill} & 21.05   &   15.60 + 0.09   & 0.14 & 01 04 1998 \\
\emph{HST}/NIC2   &   F110W      & 1.128 - 0.384  & 0.075   &	32    & \multicolumn{2}{c}{\dotfill} & 22.48   &   14.38 + 0.03   & 0.14 & 01 04 1998 \\
\emph{HST}/NIC2   &   F160W      & 1.600 - 0.280  & 0.075   &  256    & \multicolumn{2}{c}{\dotfill} & 21.84   &   13.33 + 0.03   & 0.16 & 01 04 1998 \\
VLT/NaCo          &   J	         & 1.265 - 0.250  & 0.0271  & 1200    &      VIS       & 0.8         & 24.14   &   15.78 + 0.03   & 0.21 & 28 11 2005 \\
VLT/NaCo          &   Ks	 & 2.18 - 0.35    & 0.0271  &  600    &      VIS       & 0.9         & 23.06   &   14.02 + 0.02   & 0.09 & 28 11 2005 \\
VLT/NaCo          &   L$'$	 & 3.80 - 0.62    & 0.0271  & 1050    &      VIS       & 1.1	     & 22.12   &   11.78 + 0.01   & 0.16 & 03 12 2005\\
\bottomrule
\multicolumn{11}{l}{{\it Notes.} $^a$Combined images for NGC~1386 from the F547M and F658N datasets adquired in 28 06 1997 ($80\, \rm{s}$ and $800\, \rm{s}$) and 25 02 1999 ($280\, \rm{s}$ and $1200\, \rm{s}$).}\\
\multicolumn{11}{l}{Columns correspond to: telescope/instrument, filter name, central wavelength and full-width at half-maximum of the filter ($\lambda - \Delta \lambda$), pixel scale, integration}\\
\multicolumn{11}{l}{time, Wave-Front Sensor dichroic (WFS), Seeing, zero point magnitude (ZP), Vega magnitude of the nucleus, full-width at half-maximum of the most}\\
\multicolumn{11}{l}{compact object on each image (\textsc{fwhm}) and observation date. The optical nuclear peak  was the reference for the adaptive optics system.}\\
 
  \end{tabular*}
  \end{minipage}
\end{table*}

\section{Data}\label{data}

\subsection{VLT IR and ALMA data} \label{IR_obs}
Near-IR images were obtained with the 
AO-assisted 
instrument \textsf{NaCo} at the \textsf{ESO} VLT. 
The 
field of view
(FOV) covers $27.7\arcsec \times 27.7\arcsec$. 
The  optical nuclear peak was used as reference for the AO correction. Data were collected 
in
the \textit{J}, 
\textit{H} and \textit{Ks} bands using the \textit{jitter} technique \citep{1999ASPC..172..333D}. The angular 
resolution measured in the pointlike star clusters is $\approx\,0.1\arcsec$ FWHM (Table\,\ref{sample}).
The millimetre information was obtained with ALMA in the CO$(2-1)$ line and the continuum at 230 GHz (Programme 
ID: 2016.1.01279.S). The analysis was done on the basis of the observatory-provided science calibrated and 
continuum-subtracted datacube.  Continuum emission is detected in the nucleus only. 
CO$(2-1)$ line emission is detected in the nuclear region and in a circumnuclear ring 
partially
enclosing the cluster ring. 
In this ring, the 
CO resolves into long, coherent filaments, which in turn resolve into multiple point-like clouds, 
intermixed with diffuse gas (Fig.\ \ref{image_V}). The FWHM of these clouds is $\lesssim\,0.5\,\arcsec,\,\sim 37\,\rm{pc}$, 
which we interpret as an upper limit to their size, since it is of the order of the angular resolution of the 
ALMA observations, the beam size at 230 GHz is $\sim\,0.44\arcsec\,\times\,0.48\arcsec$, with a largest 
recoverable
angular scale 
of $4.6\arcsec$. 

\subsection{IR--Optical HST data}\label{HST_obs}
Additionally, \textit{J}-, \textit{H}- and \textit{K}-band IR images from NICMOS,
\textit{V}- to \textit{I}-band optical images from WFPC2 (Table\,\ref{tab_obs}),
and narrow-band filter images 
centred 
on the H$\alpha$+[\textsc{N\,ii}] emission-line blend (F656N, F658N) 
were 
used. A continuum-subtracted H$\alpha$+[\textsc{N\,ii}] image was obtained using the \textit{V} and \textit{I} bands 
to interpolate the continuum level. This interpolation was based on the photometric calibration of the broad-band filters, 
while assuming a power law for the continuum emission.
\begin{figure}
\begin{center}
    \includegraphics[width=1.\columnwidth]{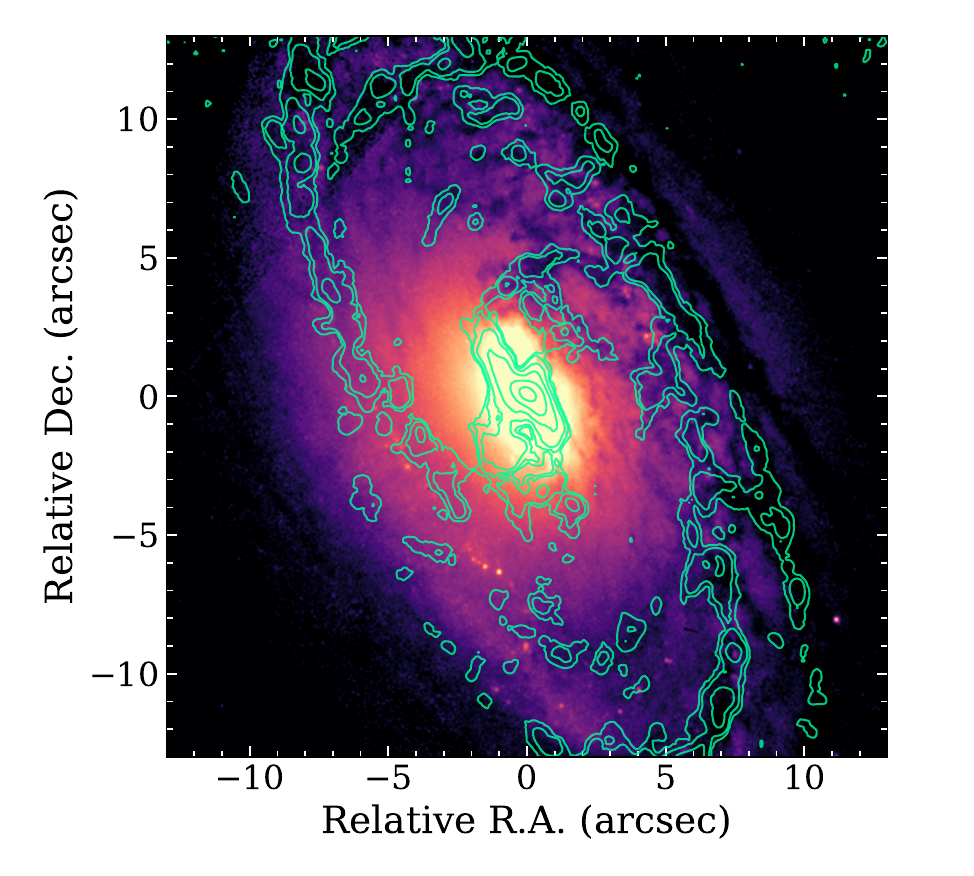} 
    \caption{\textit{HST}/F606W image of the central $13\arcsec \times 13\arcsec$ of NGC\,1386 plus ALMA/CO$(2-1)$ contours 
(in green). }
    \label{CO_contours}
\end{center}
\end{figure}

\subsection{Data reduction}\label{reduction}
Reduction of ground-based IR data includes sky subtraction, registration and combination of frames corresponding to 
each dataset using the \textsc{eclipse} and \textsc{esorex} packages provided by \textsf{ESO} (European Organisation 
for Astronomical Research in the Southern Hemisphere). Further analysis of the data was performed using 
\textsc{iraf}\footnote{\textsc{iraf} is distributed by the National Optical Astronomy Observatories, which are operated 
by the Association of Universities for Research in Astronomy, Inc., under cooperative agreement with the National Science 
Foundation.} (Image Reduction and Analysis Facility), \textsc{idl} (Interactive Data Language) and the \textit{Python 
Programming Language}.\footnote{\url{http://python.org}} The estimation of photometric zero points was based on observations
 of calibrators selected from the lists of \citet[][VIS and N20C80 dichroics]{1998AJ....116.2475P} and 
\citet[][N90C10 dichroic]{1996A&AS..119..547V} for the \textit{J} and \textit{Ks} filters.
These values are in agreement with those given by the `Quality control and data processing group' at \textsf{ESO}. 
Conversion factors from counts per second to flux units were derived using the photometric zero points and the FUV to 
IR Vega spectra from \citet{2007ASPC..364..315B}.

In the optical range, reduced and calibrated images were taken directly from the \textit{HST} archive. Prior to the 
image registration process, we corrected geometric distortions for the WFPC2 image using the \textit{MultiDrizzle} package
 in PyRAF \citep{2002hstc.conf..337K}.

\subsection{Image registration}\label{alignment}
The image registration of the complete IR--optical database is based on the location of point-like star clusters 
detected in all images. 
Clusters were selected among the brightest, more isolated and well distributed sources over the FOV.
This method has been described in detail by \citet{2014MNRAS.442.2145P}. The final alignment is found by minimising the 
cluster relative shifts among different images. The variance of these relative shifts provides an estimate of the alignment 
error, including uncorrected distortions and possibly small shifts of the cluster position with wavelength. The alignment error 
estimate is  $0.01\arcsec$. 
The nuclear optical peak emission---so far identified in the literature as the active nucleus---was avoided for alignment 
purposes. As shown by \citet{2014MNRAS.442.2145P}, this peak is a bright \ha\ blob, 
and 
the active nucleus is fully obscured in the optical by a dust filament crossing it and only becomes visible longward of 2\,$\mu$m.  
As no clusters are seen in the ALMA  data, the alignment of the CO and continuum images  with the optical--IR dataset was based 
on the assumption that the $2\,\mu$m peak emission---identified as the galaxy active nucleus---coincides with the continuum 
peak emission at 230 GHz.  We are confident of this assumption as the nuclear and ring dust filaments identified  in the 
\textit{HST} 
optical images (only HST F606W is shown in Fig.\,\ref{image_V}) do overlap with the CO filaments 
\citep[][and Fig.\,\ref{CO_contours}]{2014MNRAS.442.2145P}.

\aftertwo{
\onecolumn
     \begin{landscape}
  \footnotesize
  \centering
  \setlength{\LTcapwidth}{1.35\textheight}
  \begin{longtable}{rrrrrrrrrrrrrrrrrrrrrc@{}r@{}c@{}}
\caption{Main characteristics of star clusters identified in NGC\,1386. Relative R.A. and Dec. correspond to the shift in equatorial coordinates with respect to the active nucleus ($\star$; R.A.\,$= 03h36m46.253s$, Dec.\,$= -35d59m58.64s$). Cluster size is the half-light radius. Upper limits correspond to $3\sigma$. No aperture correction has been applied to the measured fluxes, for a discussion on this point see Section\,\ref{phot}.}
\label{phot_n1386}\\

\multirow{2}{0.2cm}[0.03cm]{\#} &
\multicolumn{2}{c}{Relative} &
\multicolumn{2}{c}{F547M} &
\multicolumn{2}{c}{F606W} &
\multicolumn{2}{c}{F791W} &
\multicolumn{2}{c}{F814W} &
\multicolumn{2}{c}{F110M} &
\multicolumn{2}{c}{F110W} &
\multicolumn{2}{c}{\textit{J}} &
\multicolumn{2}{c}{F160W} &
\multicolumn{2}{c}{\textit{Ks}} &
\multicolumn{1}{c}{Size} &
\multicolumn{1}{c}{\textsc{w(H${\alpha}$)}} &
\multicolumn{1}{c}{log \lha} \\

& \multicolumn{1}{c}{R.A.} & \multicolumn{1}{c}{Dec.} & \multicolumn{1}{c}{flux} & \multicolumn{1}{c}{err} & \multicolumn{1}{c}{flux} & \multicolumn{1}{c}{err} & \multicolumn{1}{c}{flux} & \multicolumn{1}{c}{err} & \multicolumn{1}{c}{flux} & \multicolumn{1}{c}{err} & \multicolumn{1}{c}{flux} & \multicolumn{1}{c}{err} & \multicolumn{1}{c}{flux} & \multicolumn{1}{c}{err} & \multicolumn{1}{c}{flux} & \multicolumn{1}{c}{err} & \multicolumn{1}{c}{flux} & \multicolumn{1}{c}{err} & \multicolumn{1}{c}{flux} & \multicolumn{1}{c}{err} \\
& & & \multicolumn{18}{c}{\cellcolor[gray]{0.9} [$\rm{\mu Jy}$]} & \multicolumn{1}{c}{[pc]} & \multicolumn{1}{c}{[\AA]} & \multicolumn{1}{c}{[erg s$^{-1}$]}  \\

\toprule
  0   &$ 11$\farcs42 &$ -18$\farcs84 &\multicolumn{1}{c}{\dotfill} &\multicolumn{1}{c}{\dotfill} & $1.0$ & $0.1$ &\multicolumn{1}{c}{\dotfill} &\multicolumn{1}{c}{\dotfill} & $<\,$0.9 & $0.0$ &\multicolumn{1}{c}{\dotfill} &\multicolumn{1}{c}{\dotfill} &\multicolumn{1}{c}{\dotfill} &\multicolumn{1}{c}{\dotfill} & $<\,$0.3 & $0.0$ &\multicolumn{1}{c}{\dotfill} &\multicolumn{1}{c}{\dotfill} & $<\,$0.3 & $0.0$ & $6.8$ & $-1000$ &\multicolumn{1}{c}{\dotfill}\\
  1   &$ 11$\farcs15 &$ -12$\farcs97 & $0.8$ & $0.2$ & $0.9$ & $0.1$ & $0.9$ & $0.3$ & $1.2$ & $0.4$ &\multicolumn{1}{c}{\dotfill} &\multicolumn{1}{c}{\dotfill} &\multicolumn{1}{c}{\dotfill} &\multicolumn{1}{c}{\dotfill} & $<\,$0.3 & $0.0$ &\multicolumn{1}{c}{\dotfill} &\multicolumn{1}{c}{\dotfill} & $<\,$0.3 & $0.0$ & $6.7$ & $-0.8$ &\multicolumn{1}{c}{\dotfill}\\
  2   &$ 3$\farcs35 &$ -11$\farcs37 & $1.2$ & $0.2$ & $1.3$ & $0.1$ & $1.0$ & $0.3$ & $1.0$ & $0.3$ &\multicolumn{1}{c}{\dotfill} &\multicolumn{1}{c}{\dotfill} &\multicolumn{1}{c}{\dotfill} &\multicolumn{1}{c}{\dotfill} & $0.6$ & $0.2$ &\multicolumn{1}{c}{\dotfill} &\multicolumn{1}{c}{\dotfill} & $0.3$ & $0.1$ & $6.7$ & $27.4$ & $35.85$\\
  3   &$ 1$\farcs22 &$ -11$\farcs17 & $1.5$ & $0.2$ & $1.8$ & $0.2$ & $0.6$ & $0.3$ & $1.0$ & $0.3$ &\multicolumn{1}{c}{\dotfill} &\multicolumn{1}{c}{\dotfill} &\multicolumn{1}{c}{\dotfill} &\multicolumn{1}{c}{\dotfill} & $2.1$ & $0.1$ &\multicolumn{1}{c}{\dotfill} &\multicolumn{1}{c}{\dotfill} & $0.3$ & $0.1$ & $11.1$ & $-2.5$ &\multicolumn{1}{c}{\dotfill}\\
  4   &$ -11$\farcs01 &$ -11$\farcs16 & $0.3$ & $0.1$ & $0.6$ & $0.1$ & $0.7$ & $0.3$ & $0.8$ & $0.3$ &\multicolumn{1}{c}{\dotfill} &\multicolumn{1}{c}{\dotfill} &\multicolumn{1}{c}{\dotfill} &\multicolumn{1}{c}{\dotfill} & $0.5$ & $0.1$ &\multicolumn{1}{c}{\dotfill} &\multicolumn{1}{c}{\dotfill} & $0.1$ & $0.1$ &\multicolumn{1}{c}{\dotfill} & $-29.8$ &\multicolumn{1}{c}{\dotfill}\\
  5   &$-0$\farcs68 &$ -11$\farcs03 & $0.6$ & $0.2$ & $0.7$ & $0.1$ & $<\,$0.6 & $0.0$ & $0.5$ & $0.3$ &\multicolumn{1}{c}{\dotfill} &\multicolumn{1}{c}{\dotfill} &\multicolumn{1}{c}{\dotfill} &\multicolumn{1}{c}{\dotfill} & $<\,$0.4 & $0.0$ &\multicolumn{1}{c}{\dotfill} &\multicolumn{1}{c}{\dotfill} & $<\,$0.3 & $0.0$ & $6.7$ & $35.1$ & $35.77$\\
  6   &$ 7$\farcs96 &$ -10$\farcs73 & $0.4$ & $0.1$ & $0.6$ & $0.1$ & $0.8$ & $0.3$ & $0.8$ & $0.3$ &\multicolumn{1}{c}{\dotfill} &\multicolumn{1}{c}{\dotfill} &\multicolumn{1}{c}{\dotfill} &\multicolumn{1}{c}{\dotfill} & $1.2$ & $0.2$ &\multicolumn{1}{c}{\dotfill} &\multicolumn{1}{c}{\dotfill} & $<\,$0.3 & $0.0$ & $6.7$ & $-71.1$ &\multicolumn{1}{c}{\dotfill}\\
  7   &$ 4$\farcs02 &$ -10$\farcs59 & $2.0$ & $0.2$ & $2.0$ & $0.2$ & $1.2$ & $0.4$ & $1.5$ & $0.4$ &\multicolumn{1}{c}{\dotfill} &\multicolumn{1}{c}{\dotfill} &\multicolumn{1}{c}{\dotfill} &\multicolumn{1}{c}{\dotfill} & $1.0$ & $0.1$ &\multicolumn{1}{c}{\dotfill} &\multicolumn{1}{c}{\dotfill} & $1.6$ & $0.1$ & $7.2$ & $84.3$ & $36.50$\\
  8   &$ -1$\farcs11 &$ -10$\farcs58 & $1.4$ & $0.2$ & $1.6$ & $0.2$ & $2.1$ & $0.4$ & $1.4$ & $0.4$ &\multicolumn{1}{c}{\dotfill} &\multicolumn{1}{c}{\dotfill} &\multicolumn{1}{c}{\dotfill} &\multicolumn{1}{c}{\dotfill} & $2.4$ & $0.2$ &\multicolumn{1}{c}{\dotfill} &\multicolumn{1}{c}{\dotfill} & $1.2$ & $0.1$ & $9.1$ & $12.9$ & $35.66$\\
  9   &$ 7$\farcs47 &$ -10$\farcs13 & $0.7$ & $0.1$ & $0.9$ & $0.1$ & $<\,$0.6 & $0.0$ & $0.7$ & $0.3$ &\multicolumn{1}{c}{\dotfill} &\multicolumn{1}{c}{\dotfill} &\multicolumn{1}{c}{\dotfill} &\multicolumn{1}{c}{\dotfill} & $0.7$ & $0.1$ &\multicolumn{1}{c}{\dotfill} &\multicolumn{1}{c}{\dotfill} & $0.7$ & $0.1$ & $6.7$ & $185.7$ & $36.48$\\
 10   &$ 5$\farcs05 &$ -9$\farcs53 & $1.4$ & $0.2$ & $1.5$ & $0.1$ & $1.3$ & $0.4$ & $1.5$ & $0.3$ & $<\,$1.2 & $0.0$ & $<\,$2.0 & $0.0$ & $0.4$ & $0.1$ & $<\,$0.9 & $0.0$ & $<\,$0.3 & $0.0$ & $6.8$ & $66.0$ & $36.28$\\
 11   &$ 7$\farcs51 &$ -9$\farcs30 & $0.7$ & $0.1$ & $0.9$ & $0.1$ & $0.9$ & $0.3$ & $1.4$ & $0.3$ & $<\,$1.2 & $0.0$ & $<\,$2.0 & $0.0$ & $0.9$ & $0.1$ & $2.1$ & $0.3$ & $2.4$ & $0.1$ & $9.4$ & $463.4$ & $36.91$\\
 12   &$-0$\farcs50 &$ -9$\farcs25 & $0.7$ & $0.2$ & $0.9$ & $0.2$ & $1.0$ & $0.3$ & $0.6$ & $0.3$ & $<\,$1.2 & $0.0$ & $<\,$2.0 & $0.0$ & $2.7$ & $0.2$ &\multicolumn{1}{c}{\dotfill} &\multicolumn{1}{c}{\dotfill} & $1.0$ & $0.1$ & $7.9$ & $73.6$ & $36.15$\\
 13   &$-0$\farcs04 &$ -9$\farcs01 & $3.0$ & $0.3$ & $3.0$ & $0.2$ & $1.5$ & $0.4$ & $1.8$ & $0.4$ & $<\,$1.2 & $0.0$ & $<\,$2.0 & $0.0$ & $1.7$ & $0.2$ & $<\,$0.9 & $0.0$ & $0.7$ & $0.1$ & $15.2$ & $28.2$ & $36.19$\\
 14   &$ 11$\farcs15 &$ -8$\farcs05 & $3.6$ & $0.3$ & $5.7$ & $0.3$ & $11.6$ & $1.0$ & $12.5$ & $0.9$ & $<\,$1.2 & $0.0$ & $<\,$2.0 & $0.0$ & $14.3$ & $0.2$ & $<\,$1.0 & $0.0$ & $10.6$ & $0.2$ & $6.8$ & $23.0$ & $36.51$\\
 15   &$-0$\farcs02 &$ -7$\farcs74 & $0.8$ & $0.2$ & $1.1$ & $0.1$ & $0.9$ & $0.3$ & $1.2$ & $0.3$ & $<\,$1.2 & $0.0$ & $<\,$2.0 & $0.0$ & $3.9$ & $0.2$ & $3.1$ & $0.5$ & $1.6$ & $0.1$ & $26.7$ & $20.9$ & $35.66$\\
 16   &$-0$\farcs56 &$ -6$\farcs80 & $0.7$ & $0.2$ & $1.0$ & $0.1$ & $1.2$ & $0.4$ & $0.6$ & $0.3$ & $<\,$1.2 & $0.0$ & $<\,$2.0 & $0.0$ & $4.5$ & $0.5$ & $<\,$2.4 & $0.0$ & $1.0$ & $0.2$ & $8.8$ & $31.3$ & $35.81$\\
 17   &$ -1$\farcs01 &$ -6$\farcs33 & $7.9$ & $0.4$ & $7.4$ & $0.3$ & $6.3$ & $0.7$ & $7.6$ & $0.8$ & $<\,$1.2 & $0.0$ & $4.3$ & $0.9$ & $9.6$ & $0.4$ & $12.0$ & $0.8$ & $10.9$ & $0.2$ & $8.8$ & $3.9$ & $35.75$\\
 18   &$ -1$\farcs51 &$ -6$\farcs14 & $4.2$ & $0.3$ & $4.5$ & $0.3$ & $3.0$ & $0.6$ & $1.8$ & $0.4$ & $<\,$1.2 & $0.0$ & $2.7$ & $0.9$ & $2.1$ & $0.5$ & $<\,$3.0 & $0.0$ & $2.2$ & $0.2$ & $14.2$ & $-14.6$ &\multicolumn{1}{c}{\dotfill}\\
 19   &$ -1$\farcs93 &$ -5$\farcs88 & $2.2$ & $0.3$ & $1.9$ & $0.2$ & $1.3$ & $0.4$ & $1.6$ & $0.4$ & $<\,$1.2 & $0.0$ & $<\,$2.0 & $0.0$ & $1.1$ & $0.4$ & $<\,$2.4 & $0.0$ & $<\,$0.6 & $0.0$ & $19.0$ & $-3.0$ &\multicolumn{1}{c}{\dotfill}\\
 20   &$ -2$\farcs51 &$ -5$\farcs78 & $0.6$ & $0.2$ & $0.7$ & $0.2$ & $<\,$0.9 & $0.0$ & $<\,$0.9 & $0.0$ & $<\,$1.2 & $0.0$ & $<\,$2.0 & $0.0$ & $2.2$ & $0.5$ & $2.0$ & $0.8$ & $1.3$ & $0.2$ & $8.8$ & $9.5$ & $35.19$\\
 21   &$ -2$\farcs09 &$ -5$\farcs51 & $1.1$ & $0.2$ & $1.2$ & $0.2$ & $0.7$ & $0.3$ & $1.4$ & $0.4$ & $<\,$1.2 & $0.0$ & $<\,$2.0 & $0.0$ & $<\,$1.2 & $0.0$ & $<\,$2.1 & $0.0$ & $<\,$0.6 & $0.0$ & $8.8$ & $-23.8$ &\multicolumn{1}{c}{\dotfill}\\
 22   &$ 7$\farcs03 &$ -4$\farcs77 & $0.5$ & $0.2$ & $0.9$ & $0.2$ & $<\,$0.9 & $0.0$ & $0.8$ & $0.3$ & $<\,$1.2 & $0.0$ & $<\,$2.0 & $0.0$ & $<\,$0.9 & $0.0$ & $<\,$1.8 & $0.0$ & $<\,$0.6 & $0.0$ & $83.8$ & $351.6$ & $36.73$\\
 23   &$ 6$\farcs52 &$ -2$\farcs88 & $1.2$ & $0.2$ & $1.2$ & $0.2$ & $0.7$ & $0.3$ & $1.1$ & $0.3$ & $<\,$1.2 & $0.0$ & $<\,$2.0 & $0.0$ & $4.7$ & $0.5$ & $4.3$ & $0.9$ & $1.3$ & $0.2$ & $183.7$ & $159.9$ & $36.54$\\
 24   &$ -4$\farcs31 &$ -2$\farcs53 & $2.6$ & $0.3$ & $2.7$ & $0.2$ & $2.2$ & $0.5$ & $2.4$ & $0.5$ & $<\,$1.2 & $0.0$ & $2.3$ & $0.9$ & $2.9$ & $0.5$ & $4.0$ & $0.9$ & $2.8$ & $0.3$ & $73.3$ & $30.4$ & $36.20$\\
 25   &$ -4$\farcs52 &$ -1$\farcs82 & $1.0$ & $0.2$ & $1.3$ & $0.2$ & $1.5$ & $0.4$ & $1.5$ & $0.4$ & $<\,$1.2 & $0.0$ & $3.2$ & $0.8$ & $3.3$ & $0.5$ & $2.7$ & $0.9$ & $1.4$ & $0.3$ & $99.6$ & $229.1$ & $36.80$\\
 26   &$ -5$\farcs07 &$ -1$\farcs77 & $0.8$ & $0.2$ & $1.3$ & $0.2$ & $0.9$ & $0.4$ & $1.0$ & $0.4$ & $<\,$1.2 & $0.0$ & $<\,$2.0 & $0.0$ & $3.8$ & $0.6$ & $1.9$ & $1.1$ & $<\,$0.9 & $0.0$ & $139.3$ & $9.3$ & $35.34$\\
 27   &$ 6$\farcs27 &$ -1$\farcs62 & $0.9$ & $0.2$ & $1.0$ & $0.2$ & $0.6$ & $0.3$ & $0.7$ & $0.3$ & $<\,$1.2 & $0.0$ & $<\,$2.0 & $0.0$ & $3.2$ & $0.3$ & $<\,$2.1 & $0.0$ & $0.6$ & $0.2$ & $171.4$ & $31.3$ & $35.76$\\
 28   &$ -4$\farcs70 &$ -1$\farcs59 & $1.2$ & $0.2$ & $1.4$ & $0.2$ & $1.6$ & $0.4$ & $0.9$ & $0.3$ & $2.5$ & $0.6$ & $<\,$2.0 & $0.0$ & $1.6$ & $0.5$ & $<\,$3.0 & $0.0$ & $<\,$0.9 & $0.0$ & $109.7$ & $112.7$ & $36.51$\\
 29   &$ 5$\farcs97 &$-0$\farcs90 & $0.4$ & $0.1$ & $<\,$0.3 & $0.0$ & $<\,$0.9 & $0.0$ & $<\,$0.9 & $0.0$ &\multicolumn{1}{c}{\dotfill} &\multicolumn{1}{c}{\dotfill} & $<\,$2.0 & $0.0$ & $2.1$ & $0.6$ & $<\,$3.0 & $0.0$ & $<\,$0.9 & $0.0$ & $166.1$ & $172.4$ & $35.98$\\
 30   &$ 4$\farcs32 &$-0$\farcs85 & $<\,$0.6 & $0.0$ & $<\,$0.6 & $0.0$ & $<\,$1.2 & $0.0$ & $<\,$1.2 & $0.0$ & $<\,$1.2 & $0.0$ & $<\,$2.0 & $0.0$ & $5.2$ & $1.0$ & $<\,$5.7 & $0.0$ & $1.0$ & $0.4$ & $97.9$ & $-265.8$ &\multicolumn{1}{c}{\dotfill}\\
 31   &$ 6$\farcs33 &$-0$\farcs43 & $0.5$ & $0.1$ & $0.6$ & $0.1$ & $0.7$ & $0.3$ & $<\,$0.9 & $0.0$ & $<\,$1.2 & $0.0$ & $<\,$2.0 & $0.0$ & $3.3$ & $0.4$ & $<\,$1.2 & $0.0$ & $1.4$ & $0.2$ & $159.2$ & $-67.6$ &\multicolumn{1}{c}{\dotfill}\\
 32   &$ -5$\farcs47 &$-0$\farcs41 & $1.0$ & $0.2$ & $1.1$ & $0.1$ & $1.0$ & $0.3$ & $0.9$ & $0.3$ & $<\,$1.2 & $0.0$ & $2.9$ & $1.1$ & $4.4$ & $0.6$ & $1.9$ & $1.0$ & $1.3$ & $0.3$ & $166.4$ & $-16.9$ &\multicolumn{1}{c}{\dotfill}\\
 33   &$ -6$\farcs50 &$ 0$\farcs34 & $0.9$ & $0.2$ & $1.1$ & $0.2$ & $0.8$ & $0.3$ & $0.6$ & $0.3$ & $<\,$1.2 & $0.0$ & $<\,$2.0 & $0.0$ & $<\,$1.5 & $0.0$ & $<\,$2.4 & $0.0$ & $<\,$0.9 & $0.0$ & $129.8$ & $39.8$ & $35.93$\\
 34   &$ 8$\farcs05 &$ 1$\farcs05 & $0.6$ & $0.1$ & $0.6$ & $0.1$ & $<\,$0.6 & $0.0$ & $0.4$ & $0.2$ & $<\,$1.2 & $0.0$ & $<\,$2.0 & $0.0$ & $<\,$0.6 & $0.0$ & $<\,$1.2 & $0.0$ & $<\,$0.3 & $0.0$ & $6.7$ & $-34.7$ &\multicolumn{1}{c}{\dotfill}\\
 35   &$ 4$\farcs70 &$ 1$\farcs93 & $1.9$ & $0.2$ & $2.4$ & $0.2$ & $1.5$ & $0.4$ & $1.6$ & $0.4$ & $4.4$ & $1.3$ & $3.3$ & $1.2$ & $8.1$ & $0.7$ & $5.3$ & $1.5$ & $2.5$ & $0.3$ & $121.2$ & $303.3$ & $37.12$\\
 36   &$ 4$\farcs29 &$ 2$\farcs16 & $3.4$ & $0.3$ & $3.9$ & $0.3$ & $3.3$ & $0.6$ & $3.2$ & $0.6$ & $7.5$ & $1.5$ & $5.9$ & $1.2$ & $8.4$ & $0.8$ & $7.4$ & $1.4$ & $3.2$ & $0.3$ & $86.6$ & $66.6$ & $36.71$\\
 37   &$ 4$\farcs54 &$ 2$\farcs93 & $0.8$ & $0.2$ & $1.2$ & $0.3$ & $1.0$ & $0.5$ & $0.9$ & $0.5$ & $<\,$1.2 & $0.0$ & $<\,$2.0 & $0.0$ & $1.8$ & $0.8$ & $<\,$4.5 & $0.0$ & $<\,$1.2 & $0.0$ & $103.2$ & $11.1$ & $35.42$\\
 38   &$ 2$\farcs29 &$ 5$\farcs27 & $1.8$ & $0.2$ & $2.2$ & $0.2$ & $1.8$ & $0.4$ & $1.9$ & $0.4$ & $2.1$ & $0.8$ & $4.2$ & $1.0$ & $5.8$ & $0.4$ & $6.3$ & $0.9$ & $1.9$ & $0.2$ & $86.2$ & $11.6$ & $35.69$\\
 39   &$ 0$\farcs72 &$ 5$\farcs38 & $0.8$ & $0.2$ & $1.3$ & $0.2$ & $1.5$ & $0.4$ & $1.9$ & $0.4$ & $2.1$ & $1.1$ & $3.3$ & $1.0$ & $1.7$ & $0.3$ & $4.4$ & $0.7$ & $1.5$ & $0.2$ & $113.4$ & $63.2$ & $36.24$\\
 40   &$ -7$\farcs92 &$ 5$\farcs98 & $0.4$ & $0.1$ & $0.9$ & $0.1$ & $1.7$ & $0.4$ & $1.6$ & $0.4$ & $<\,$1.4 & $0.0$ & $10.9$ & $1.1$ & $4.2$ & $0.2$ & $11.5$ & $0.7$ & $4.7$ & $0.2$ & $10.1$ & $-8.2$ &\multicolumn{1}{c}{\dotfill}\\
 41   &$ 3$\farcs11 &$ 5$\farcs99 & $0.7$ & $0.1$ & $0.9$ & $0.1$ & $0.8$ & $0.3$ & $0.9$ & $0.3$ &\multicolumn{1}{c}{\dotfill} &\multicolumn{1}{c}{\dotfill} & $<\,$2.0 & $0.0$ & $2.2$ & $0.3$ & $2.6$ & $0.6$ & $0.6$ & $0.2$ & $123.7$ & $-13.4$ &\multicolumn{1}{c}{\dotfill}\\
 42   &$ -2$\farcs39 &$ 6$\farcs11 & $0.8$ & $0.2$ & $1.2$ & $0.2$ & $1.5$ & $0.4$ & $1.0$ & $0.4$ & $<\,$1.2 & $0.0$ & $2.5$ & $1.0$ & $5.3$ & $0.5$ & $3.9$ & $0.9$ & $1.2$ & $0.2$ & $146.7$ & $25.2$ & $35.79$\\
 43   &$ 2$\farcs50 &$ 6$\farcs90 & $2.0$ & $0.2$ & $1.9$ & $0.2$ & $2.2$ & $0.5$ & $1.6$ & $0.4$ & $<\,$1.2 & $0.0$ & $2.5$ & $0.7$ & $2.8$ & $0.2$ & $1.5$ & $0.5$ & $0.6$ & $0.1$ & $119.1$ & $38.9$ & $36.19$\\
 44   &$ 2$\farcs27 &$ 7$\farcs06 & $1.3$ & $0.2$ & $1.5$ & $0.2$ & $1.1$ & $0.4$ & $1.3$ & $0.3$ & $<\,$1.2 & $0.0$ & $<\,$2.0 & $0.0$ & $3.0$ & $0.2$ & $<\,$2.1 & $0.0$ & $0.6$ & $0.2$ & $125.6$ & $36.2$ & $36.02$\\
 45   &$ 2$\farcs44 &$ 7$\farcs69 & $1.5$ & $0.2$ & $2.1$ & $0.2$ & $1.4$ & $0.4$ & $2.2$ & $0.5$ & $<\,$1.2 & $0.0$ & $3.9$ & $0.9$ & $3.6$ & $0.4$ & $4.0$ & $0.9$ & $1.7$ & $0.2$ & $128.2$ & $34.1$ & $36.12$\\
 46   &$ 0$\farcs77 &$ 7$\farcs89 & $<\,$0.6 & $0.0$ & $<\,$0.6 & $0.0$ & $<\,$0.6 & $0.0$ & $<\,$0.9 & $0.0$ & $<\,$1.2 & $0.0$ & $<\,$2.0 & $0.0$ & $<\,$0.9 & $0.0$ & $<\,$1.5 & $0.0$ & $<\,$0.3 & $0.0$ & $196.5$ & $1410.9$ & $37.04$\\
 47   &$ 0$\farcs05 &$ 8$\farcs16 & $<\,$0.3 & $0.0$ & $<\,$0.3 & $0.0$ & $<\,$0.6 & $0.0$ & $<\,$0.9 & $0.0$ & $<\,$1.2 & $0.0$ & $<\,$2.0 & $0.0$ & $1.4$ & $0.3$ & $<\,$2.1 & $0.0$ & $1.1$ & $0.2$ & $17.6$ & $-1000$ &\multicolumn{1}{c}{\dotfill}\\
 48   &$ -7$\farcs43 &$ 8$\farcs24 & $1.6$ & $0.2$ & $1.9$ & $0.2$ & $1.9$ & $0.4$ & $2.3$ & $0.4$ & $<\,$1.2 & $0.0$ & $3.3$ & $0.7$ & $5.8$ & $0.2$ & $6.7$ & $0.5$ & $3.9$ & $0.1$ & $12.8$ & $5.5$ & $35.32$\\
 49   &$ -7$\farcs82 &$ 9$\farcs23 & $<\,$0.3 & $0.0$ & $0.5$ & $0.1$ & $<\,$0.6 & $0.0$ & $0.5$ & $0.2$ & $<\,$1.2 & $0.0$ & $14.0$ & $2.9$ & $<\,$0.3 & $0.0$ & $1.1$ & $0.1$ & $<\,$0.3 & $0.0$ & $7.9$ & $958.4$ & $37.06$\\
 50   &$ -8$\farcs04 &$ 9$\farcs87 & $1.0$ & $0.2$ & $1.2$ & $0.1$ & $1.1$ & $0.3$ & $1.1$ & $0.3$ &\multicolumn{1}{c}{\dotfill} &\multicolumn{1}{c}{\dotfill} &\multicolumn{1}{c}{\dotfill} &\multicolumn{1}{c}{\dotfill} & $2.1$ & $0.1$ &\multicolumn{1}{c}{\dotfill} &\multicolumn{1}{c}{\dotfill} & $2.0$ & $0.1$ & $21.2$ & $45.5$ & $36.04$\\
 51   &$ -1$\farcs96 &$ 10$\farcs30 & $0.6$ & $0.1$ & $0.9$ & $0.1$ & $0.8$ & $0.3$ & $0.5$ & $0.2$ &\multicolumn{1}{c}{\dotfill} &\multicolumn{1}{c}{\dotfill} &\multicolumn{1}{c}{\dotfill} &\multicolumn{1}{c}{\dotfill} & $0.4$ & $0.1$ &\multicolumn{1}{c}{\dotfill} &\multicolumn{1}{c}{\dotfill} & $0.4$ & $0.1$ & $18.7$ & $136.6$ & $36.37$\\
 52   &$ -4$\farcs04 &$ 10$\farcs72 & $0.9$ & $0.2$ & $1.3$ & $0.2$ & $1.2$ & $0.4$ & $1.0$ & $0.3$ &\multicolumn{1}{c}{\dotfill} &\multicolumn{1}{c}{\dotfill} &\multicolumn{1}{c}{\dotfill} &\multicolumn{1}{c}{\dotfill} & $0.6$ & $0.1$ &\multicolumn{1}{c}{\dotfill} &\multicolumn{1}{c}{\dotfill} & $0.7$ & $0.1$ & $7.5$ & $48.7$ & $36.09$\\
 53   &$-0$\farcs28 &$ 11$\farcs04 & $1.0$ & $0.2$ & $1.5$ & $0.2$ & $1.2$ & $0.4$ & $1.0$ & $0.3$ &\multicolumn{1}{c}{\dotfill} &\multicolumn{1}{c}{\dotfill} &\multicolumn{1}{c}{\dotfill} &\multicolumn{1}{c}{\dotfill} & $3.4$ & $0.3$ &\multicolumn{1}{c}{\dotfill} &\multicolumn{1}{c}{\dotfill} & $2.0$ & $0.2$ & $17.4$ & $-22.1$ &\multicolumn{1}{c}{\dotfill}\\
 54   &$ -3$\farcs12 &$ 11$\farcs12 & $1.0$ & $0.2$ & $1.1$ & $0.1$ & $1.1$ & $0.3$ & $1.0$ & $0.3$ &\multicolumn{1}{c}{\dotfill} &\multicolumn{1}{c}{\dotfill} &\multicolumn{1}{c}{\dotfill} &\multicolumn{1}{c}{\dotfill} & $3.1$ & $0.1$ &\multicolumn{1}{c}{\dotfill} &\multicolumn{1}{c}{\dotfill} & $1.8$ & $0.1$ & $13.0$ & $16.5$ & $35.56$\\
 55   &$ -2$\farcs85 &$ 11$\farcs18 & $0.6$ & $0.1$ & $0.8$ & $0.1$ & $1.5$ & $0.4$ & $0.7$ & $0.3$ &\multicolumn{1}{c}{\dotfill} &\multicolumn{1}{c}{\dotfill} &\multicolumn{1}{c}{\dotfill} &\multicolumn{1}{c}{\dotfill} & $3.0$ & $0.1$ &\multicolumn{1}{c}{\dotfill} &\multicolumn{1}{c}{\dotfill} & $0.6$ & $0.1$ & $12.0$ & $110.8$ & $36.34$\\
 56   &$ -3$\farcs85 &$ 11$\farcs28 & $0.5$ & $0.1$ & $1.0$ & $0.1$ & $1.2$ & $0.3$ & $1.1$ & $0.3$ &\multicolumn{1}{c}{\dotfill} &\multicolumn{1}{c}{\dotfill} &\multicolumn{1}{c}{\dotfill} &\multicolumn{1}{c}{\dotfill} & $1.1$ & $0.1$ &\multicolumn{1}{c}{\dotfill} &\multicolumn{1}{c}{\dotfill} & $0.4$ & $0.1$ & $14.0$ & $110.2$ & $36.37$\\
 57   &$ -4$\farcs46 &$ 11$\farcs77 & $<\,$0.3 & $0.0$ & $0.3$ & $0.1$ & $<\,$0.6 & $0.0$ & $<\,$0.6 & $0.0$ &\multicolumn{1}{c}{\dotfill} &\multicolumn{1}{c}{\dotfill} &\multicolumn{1}{c}{\dotfill} &\multicolumn{1}{c}{\dotfill} & $0.9$ & $0.2$ &\multicolumn{1}{c}{\dotfill} &\multicolumn{1}{c}{\dotfill} & $0.5$ & $0.1$ & $14.5$ & $-26.2$ &\multicolumn{1}{c}{\dotfill}\\
 58   &$ -3$\farcs45 &$ 12$\farcs11 & $0.4$ & $0.1$ & $0.7$ & $0.1$ & $0.6$ & $0.3$ & $0.6$ & $0.3$ &\multicolumn{1}{c}{\dotfill} &\multicolumn{1}{c}{\dotfill} &\multicolumn{1}{c}{\dotfill} &\multicolumn{1}{c}{\dotfill} & $3.5$ & $0.2$ &\multicolumn{1}{c}{\dotfill} &\multicolumn{1}{c}{\dotfill} & $2.5$ & $0.2$ & $13.0$ & $96.0$ & $36.14$\\
 59   &$ -4$\farcs44 &$ 12$\farcs28 & $0.7$ & $0.2$ & $0.8$ & $0.1$ & $0.5$ & $0.3$ & $0.4$ & $0.2$ &\multicolumn{1}{c}{\dotfill} &\multicolumn{1}{c}{\dotfill} &\multicolumn{1}{c}{\dotfill} &\multicolumn{1}{c}{\dotfill} & $3.1$ & $0.3$ &\multicolumn{1}{c}{\dotfill} &\multicolumn{1}{c}{\dotfill} & $1.0$ & $0.2$ & $11.6$ & $95.3$ & $36.15$\\
 60   &$ -8$\farcs26 &$ 16$\farcs24 & $<\,$0.3 & $0.0$ & $1.3$ & $0.2$ & $0.9$ & $0.3$ & $<\,$0.3 & $0.0$ &\multicolumn{1}{c}{\dotfill} &\multicolumn{1}{c}{\dotfill} &\multicolumn{1}{c}{\dotfill} &\multicolumn{1}{c}{\dotfill} & $<\,$0.3 & $0.0$ &\multicolumn{1}{c}{\dotfill} &\multicolumn{1}{c}{\dotfill} & $<\,$0.3 & $0.0$ & $9.6$ & $31.7$ & $35.88$\\

$\star$ & $0$\farcs00 & $0$\farcs00& 10.4& & 11.7& & 40.3& & 44.0& & 476.7& & 416& & 470.4& & 550& & 483.4& & 23.1& 804& 40.0\\

\bottomrule

\end{longtable}
\end{landscape}
\twocolumn}

\subsection{Detection and photometry of compact sources}\label{phot}
The WFPC2/F606W broad-band image was used as reference for alignment and source detection since it contains the 
largest number of clusters.
Usually, the \textsc{daophot} finding algorithm \citep{1987PASP...99..191S} permits 
us 
to identify point-like sources over a 
$3\sigma$ threshold on each reference image.
In the case of NGC\,1386 \textsc{daophot} could not be directly applied since the galaxy light outshines that of the clusters. 
In order to amend this effect, we applied the unsharp-masking technique \citep[e.g.][]{1994AJ....107...35H,1994PASJ...46....1S} to 
the images prior to the source identification with \textsc{daophot}. Although this method consists 
in
dividing an image by itself 
after applying 
smooth filtering, we preferred the subtraction of the original image and the smoothed one. This was done to avoid 
the noise that the classical unsharp-masked method introduces in low signal-to-noise ($S/N$) regions. In all cases, the smoothing 
kernel is a median circular filter with $r\,\sim\,0.3\arcsec$, wide enough to preserve the cluster light in the case of worse spatial 
resolution, FWHM $<0.2\arcsec$ (Table\,\ref{tab_obs}).
We set \textsc{daophot} 
to detect sources above a $3\sigma$ threshold. There is an increase of $10$--$15\%$ in the number of clusters detected with
 respect to the case with 
unfiltered images. 
Figure \ref{image_V} 
shows the WFPC2/F606W image of the nuclear region 
of NGC\,1386 and the masked image.

Photometric measurements were performed on each image for all the clusters detected in the WFPC2/F606W image. The photometry was 
obtained with \textsc{daophot}, using a circular aperture with radius 
$r\,\approx \rm{FWHM}$, to avoid contamination from the galaxy background light. 
The local background subtraction was based on the mode value estimated in a ring surrounding each cluster 
($r \approx [2$--$4]\,\times\,\rm{FWHM}$). 
The flux measurements for all the clusters are listed in Table\,\ref{phot_n1386}. 
Upper limits for undetected clusters are provided, corresponding to the $3\sigma$ background level.

Aperture corrections cannot be applied in the case of the AO near-IR fluxes as the shape of the point
spread function (PSF) is 
image dependent 
on
AO observation. For consistency, this correction is not applied to the \textit{HST} data. However, we estimate 
as follows the losses due to the apertures used in this work. The fraction of measured energy considering the apertures used in 
this study is in the $60\%$--$80\%$ range for the WFPC2 camera, and in the $80\%$--$90\%$ range for the NIC2 camera 
\citep{2005PASP..117.1049S,1995PASP..107..156H,1999nicm.rept....7H}. Thus, aperture-corrected \textit{HST} fluxes would imply an 
incremental factor of $1.3$--$1.7$ when compared with the values in Table\,\ref{phot_n1386}. 
With
regard to  
\textsf{ESO}\,/\,\textsf{NaCo} AO images, the simulated PSF obtained with version 3.2.6 of the 
ETC\footnote{\url{http://www.eso.org/observing/etc}}
indicates that the fraction of encircled energy in the apertures used in this 
study
is in the $50\%$--$60\%$ range. Putting it all 
together, a correction factor of  $\sim 1.7$--$2.0$ is expected to apply, in a kind of systematic form, to all the fluxes in this study. 

Table\,\ref{phot_n1386} provides for each cluster
its relative position 
with respect to the galaxy nucleus measured in the $K$-band, 
an upper limit FWHM
size, and the flux at different wavelengths, including the \ha\ equivalent width  [W(H$\alpha$)] and the H$\alpha$ 
luminosity (\lhaobs). \ha+[\ion{N}{ii}] fluxes were derived after  subtracting the continuum from a 1D polynomial fit to the  
emission in the adjacent \textit{V} and \textit{I} filters. The \ha\ luminosities are further corrected for the [\textsc{N\,ii}] 
contribution in the narrow-band  WFPC2/F658N filter, assumed 
to 
be 40\% of the H$\alpha$ + [\textsc{N\,ii}] blend 
\citep{1981PASP...93....5B}. 

The FWHM measured in the most compact clusters sets their size to be FWHM $\lesssim 0.09 \arcsec \approx 6.7\ \rm{pc}$. Larger sizes 
in Table\,\ref{phot_n1386} are due to low S/N and to confusion 
caused by 
overcrowding in some regions of the ring.

\section{Methodology}\label{s:methods}

\subsection{From photometry to physical properties} \label{s:fitting}
 
We use the one-dimensional version of the 
DynBaS 
code
\citep{magris+15}, a standard spectral energy distribution (SED) fitting code, 
to derive, for 
the 
stellar population
of the
cluster  
considered here,
the age,  mass and  extinction $A_V$  along the 
line of sight to
a given cluster.
 Since the SED has been corrected for the effects of MW dust, $A_V$ is intrinsic to the host galaxy. Assuming each cluster to be 
coeval associations of stars, well described by a passively evolving simple stellar population (SSP) of known metallicity $Z$, and 
fully populating  a given initial mass function (IMF), for each cluster we look for the age $t$ and the extinction ${A_V}$ 
that minimise
the goodness-of-fit merit function
\begin{equation}
\chi^2=\sum_l \frac{[F^{\rm obs}_l-a f_l(t) \times 10^{-0.4 ({A_V} (A_{\lambda_l}/A_V))}]^2}{\sigma{_l^2}}.
\label{chi2}
\end{equation}
The sum in Eq.~(\ref{chi2}) extends over all the observed filter bands for the cluster. 
$F^{\rm obs}_l$ and $\sigma_l$ are the observed flux in the $l$ band and its error, respectively. 
$f_l(t)$ corresponds to the flux of the 
model SSP at age $t$ in the $l$ band.  $A_{\lambda_l}/A_V$ is the extinction law, assumed to follow the \citet{ccm89} 
analytic formula with $R_V = 3.1$. 
For each 
$(t,\,A_V)$ 
pair, 
the coefficient $a$ is computed 
with the requirement that
$\partial\chi^2/\partial\,a$\,=\,0. 
$a$ is the cluster luminous stellar mass if the model flux $f(t)$ corresponds to an SSP of unit mass.
For clusters with \ha\ detected in emission, we include the condition \lhamod\,$\ge$ \lhaobsI, with \lhamod\ derived from the model SED\footnote{
At each age \lhamod\ is computed from the number $Q$(H) of H ionizing photons transformed to \ha\,luminosity using the calibration for Case B recombination at $T$\,=\,10000 K, \lha $= 1.365\times 10^{-12}~  Q(\rm{H})\ergs$ \citep{osterbrock89}.}.
This condition changes only slightly the parameters of a couple of clusters, it is included for completeness.
For this comparison, 
we correct the observed \ha\,luminosity, by the [NII] contribution to the \ha +[NII] blend (value in column 24
of
Table~\ref{phot_n1386})
and further  correct it 
for 
extinction
using  $A_V$ from the model
(also  as
 derived from the model SED).  This fully corrected observed \ha,  \lhaobsI\, is reported in Table\,\ref{tab:n1386_fitted_par}, 
together with
the luminosity, 
\lhamod. 
Table\,\ref{tab:n1386_fitted_par} also includes the values of $A_V$  and age derived from the fit.

To account for the uncertainty in the derived mass, age and $A_V$,  we sample the likelihood
\begin{equation}
\rm{ln} ~ \mathcal{L} = const - \frac{1}{2} ~ \chi^2,
\label{lhood}
\end{equation}
around the minimum-$\chi^2$ solution
using a fine grid in model parameter space. We then compute the marginalised posterior probability distribution 
functions (PDF) for the age, mass and $A_V$ assigned to each cluster. 
When the minimum $\chi^2$ solution is outside the 16th or 84th percentiles of
the PDF, we adopt the mode of the PDF as our solution, otherwise we take the parameters given by the minimum $\chi^2$.
The age, mass, and $A_V$ thus derived for each cluster are denoted as \tm, \mm  and \avm, respectively. 

In this 
study 
we use the Charlot \& Bruzual (hereafter C\&B) population synthesis models \citep{plat+19}. The stellar 
ingredients used in the C\&B models are explained in detail in \citet[][Appendix A, Tables 8-12]{2022ApJS..262...36S}. 
For our analysis we select models with solar metallicity ($Z_\odot$\,=\,0.017) 
and  the \citet{cha03} 
IMF
in the standard mass range from 0.1 and 100 \msun.
Prior to performing the SED fits, the model fluxes are shifted by the redshift of the galaxy, 
$z=0.002905$, and 
reddened by the Galaxy extinction in the direction of NGC\,1386, $E(B-V)= 0.012$ \citep{schafink11}, using the extinction law from \citet{ccm89}.

\subsubsection{Stochastic sampling of the IMF}

The standard approach adopted to compute population synthesis models (e.g.\ \citet{Leitherer+99, bc03}, C\&B) assumes 
that a pool of infinite mass is available to draw stars of all masses within a given range according to a prescribed IMF. 
The models are then renormalised to a finite mass, implying that a fractional (non-integer) number of stars populates each 
mass interval. This approach is not far from reality when the total mass of the stellar population is large ($\ge\,10^4\,M_\odot$).
However, the mass of most clusters in NGC\,1386 are in the  $10^3$--$10^4 $ \msun\ range (Section\,\ref{s:mass}).  In 
this case, stochastic fluctuations in the number of stars of any given mass become important and can leave significant 
imprints 
on
the SEDs and, in consequence, 
on
the physical parameters recovered from SED fitting \citep[e.g.,][and references 
therein]{barbaro+bertelli77, girardi+bica93, cervino+luridiana04, bruzual02, fouesneau+14}.
Therefore, in addition to the \textit{standard} C\&B models, we build a set of alternative models using the same stellar tracks 
and spectral libraries, but sampling the IMF \textit{stochastically}, as described in \citet{bruzual10}, until we fulfil
the 
target mass $M_{\rm pop}$ of the synthetic population. For each realisation, at $t = 0$ $M_{\rm pop}$ is the sum of the mass of 
all individual stars.
We compute 1000 models, corresponding to 50 Monte Carlo simulations for 20 different values of $M_{\rm pop}$ in the range from $500$ 
to $10^5$ \msun, which covers the mass range obtained from the \textit{standard} model.
We then derive via SED 
fitting the parameters \tm, \avm\ and \mm\ for each simulation, and adopt the median of the 
distribution of each parameter as our
stochastic solution. These medians are  denoted by \ts, \avs\ and \ms.  
As shown in Fig.\ \ref{allparam} and Section\,\ref{s:results}, even though the parameters derived from the stochastic models for a given cluster may differ 
from the \textit{standard} solution, the global properties of the 
cluster population as a whole 
do not change.

\section{The circumnuclear star cluster ring}\label{s:results}
\begin{figure*}
\begin{center}
\includegraphics[width=2\columnwidth]{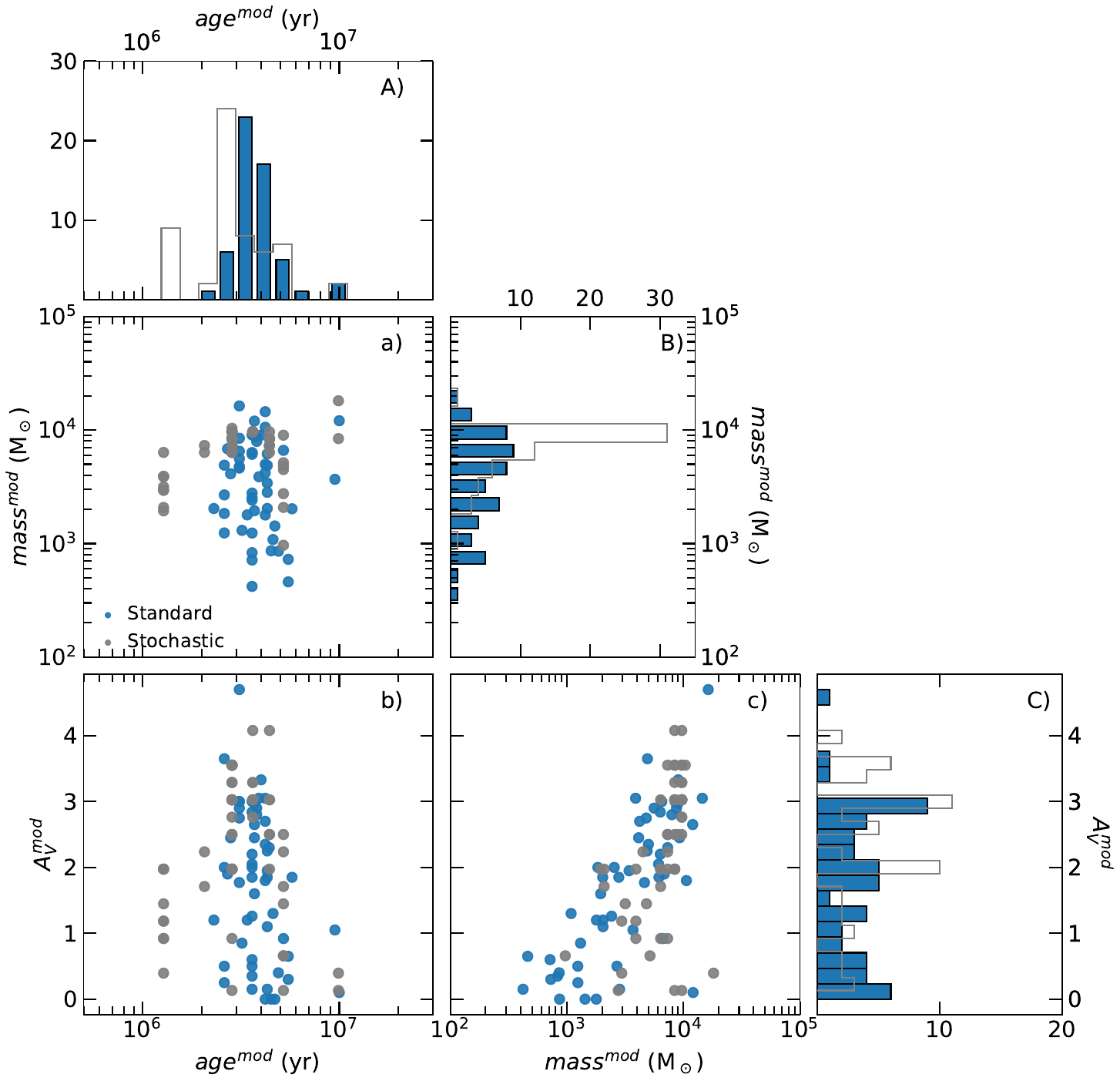} 
 \caption{Age, mass and $A_V$ derived for NGC\,1386 clusters by SED fitting using standard and stochastic C\&B models. The histograms
 show the distribution along each axis (after marginalising over the other two).}
    \label{allparam}
\end{center}
\end{figure*}

This section discusses the results from the spectral fits to the star cluster SEDs. 
For illustrative purposes, the  SEDs of four representative clusters and  
their 
respective fitted model are displayed 
in the insets 
of 
the top panel of Fig.\,\ref{image_V}.
The observed and fitted SEDs for the whole sample are 
given
in Appendix \ref{sec:appendix}.
 Table\,\ref{tab:n1386_fitted_par}  lists \tm, \avm, \mm\ and \lhamod\ for all the clusters with valid 
detections
 in at least three
 photometric bands---less than three fitting points were considered unsuitable for a reliable SED fit. Accordingly,  
clusters \#0, 29, 30, 
46, 47 and 60 were excluded from the analysis. In Fig.\,\ref{allparam} we compare the results obtained using  the C\&B 
models with both  the standard 
(fully populated IMF)
and the stochastically sampled  IMF. 

\subsection{Age}\label{s:age} 

\begin{figure}
\begin{center}
  \includegraphics[width=0.9\columnwidth]{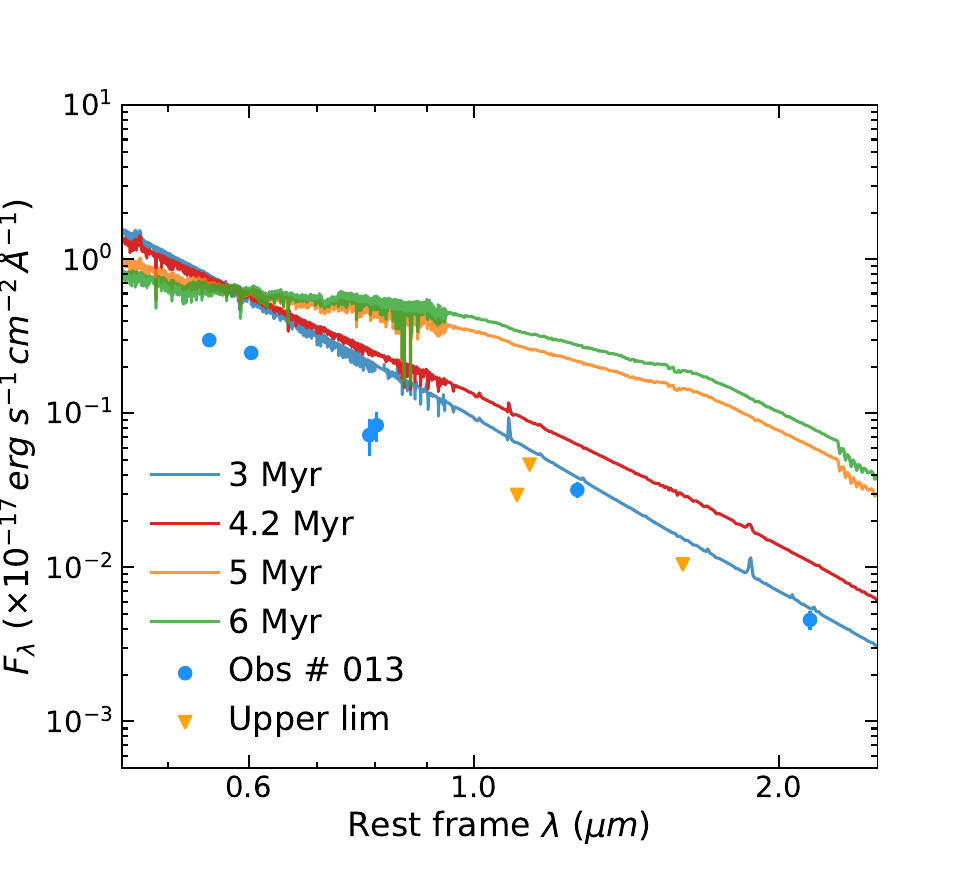}
\end{center}
  \caption{\label{age_evol} 
  C\&B SSP model spectra at ages 3, 4.2, 5 and 6 Myr. For reference, the blue dots and red triangles (upper limits) show the SED of cluster \#13 (\tm = 4.2 Myr).}
\end{figure}

The cluster age distribution displayed in Fig.\,\ref{allparam}(A) shows that most clusters in NGC\,1386 formed roughly 4 Myr ago. 
The median age is 
3.7 Myr, with first and third quartiles at 3 and 4.5 Myr, respectively. 
Figures \ref{allparam} (A, a, b) show that the age distribution derived from the models with 
stochastic sampling of the IMF is similar to 
that derived from the standard models, with a slightly lower median age (2.8 Myr) in the stochastic case.
Regarding the reliability of our age determinations, we remark that  
the lack of 
UV data for these clusters (best age indicator) is compensated by the availability of near-infrared data displaying the rapid evolution of the spectrum in the optical--infrared range at young ages. 
Fig.\,\ref{age_evol}\ shows the C\&B model spectra 
in the 3--6 Myr range.
Below and up to 4.2\,Myr the spectrum is blue, decreasing steeply from the UV to the optical with a weak
contribution in the near infrared. 
Between 4 and 5 Myr the SED suddenly reddens, especially in the \textit{K} band, owing to the appearance of red super giants after the 40\,\msun\ stars leave the main sequence. 
This  reddening acts as a lever, allowing us to determine the age with a relatively small error. 
Also noticeable is the slow evolution of the spectrum below 4 Myr, making the uncertainty in the age determination slightly asymmetric.

\subsection{Extinction}\label{s:extinction}
The clusters in NGC\,1386 are not heavily extinguished, 84\% of them having \avm < 3 mag (Fig.\,\ref{allparam}C). 
Overall, there is a fair agreement between the derived \avm\ and the location of the cluster in the ring. Clusters are 
preferentially detected in relatively 
low-dust or even dust-free regions (see Fig.\ref{image_V}), which may be a selection effect.
The distribution of \avs\ (Fig.\,\ref{allparam}C) is not very different from the one derived from the standard models. 

\begin{figure}
       \includegraphics[width=1.1\columnwidth]{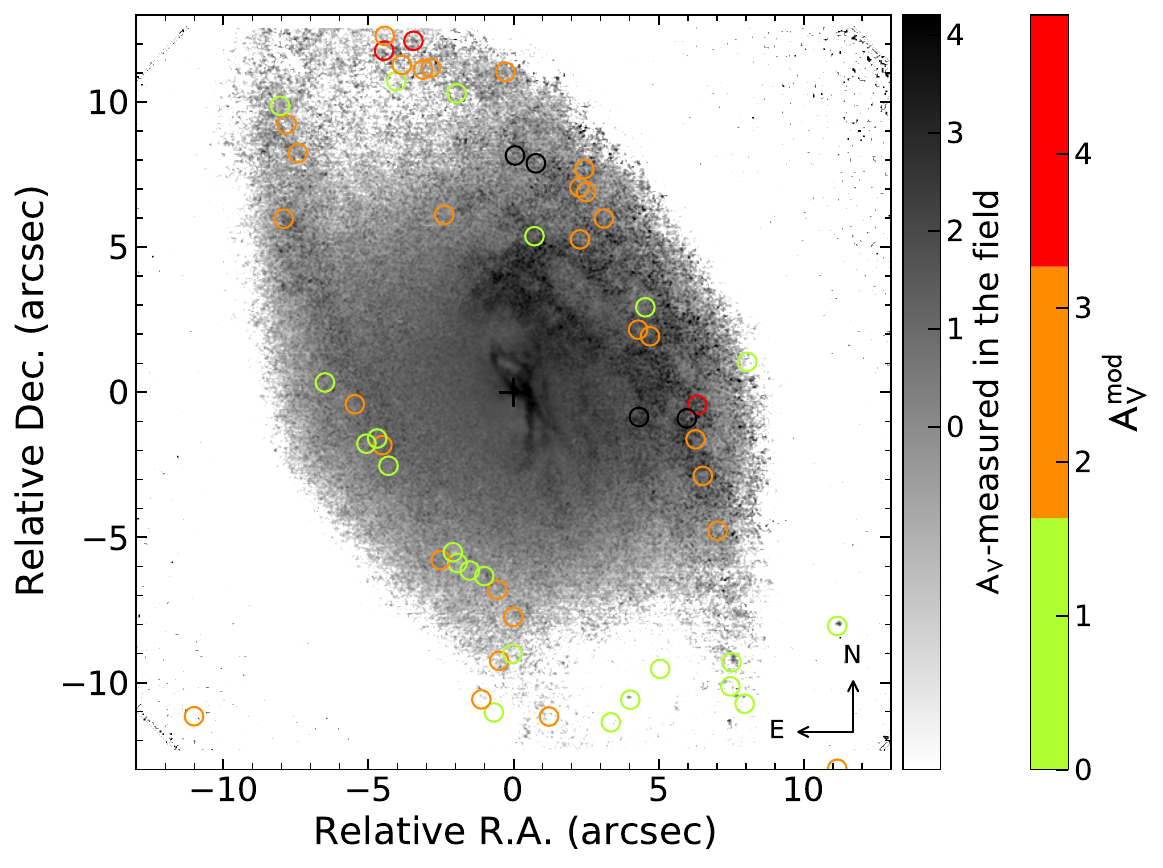}
    \caption{NGC\,1386 (VLT/\textsf{NaCo} \textit{Ks}-band $-$ \textit{HST}/F814W) extinction image. Circles at the detected 
cluster locations are colour-coded 
    according to $A_V^{\rm mod}$. Only clusters detected in at least three bands are shown.
    }
    \label{image_dust}
        \includegraphics[width=0.9\columnwidth]{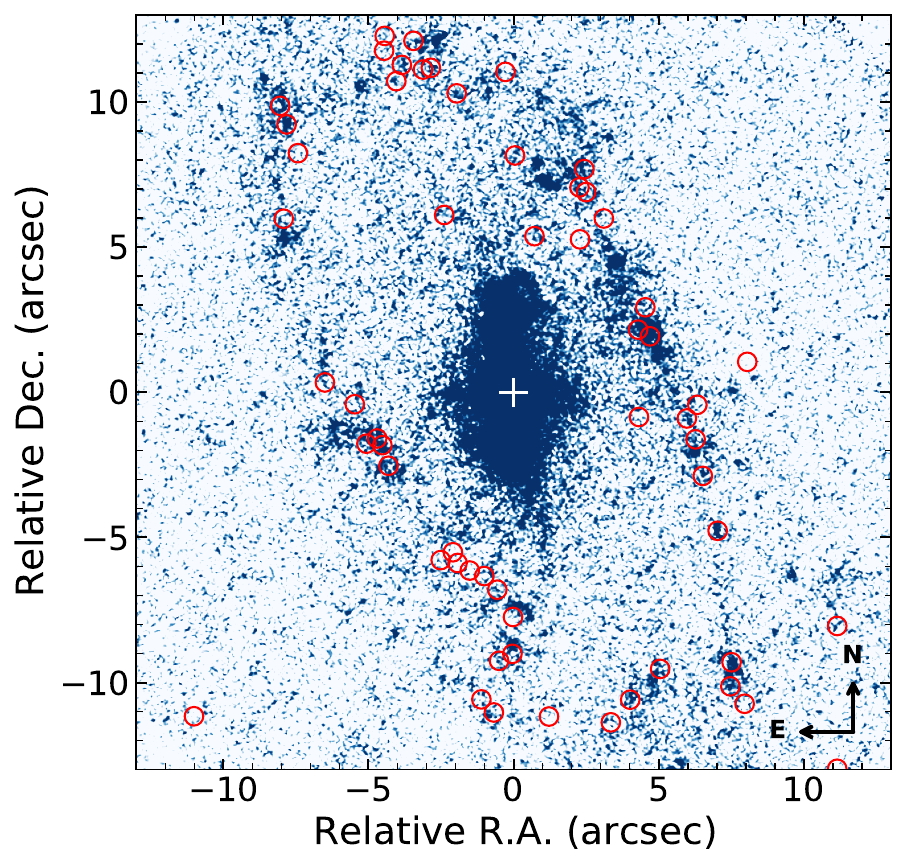}
    \caption{Continuum-subtracted HST / \ha + [\ion{N}{ii}] image. A large number of clusters do not show a counterpart in \ha.}
    \label{image_ha}
\end{figure}

An extinction map\footnote{Even though in this paper we follow the common practice of calling Fig.\,\ref{image_dust} an 
{\it extinction} map, in reality, this is an {\it  attenuation} map: at each point it includes light scattered into the line of sight from all directions in NGC\,1386. The light from each source in this galaxy travels through a different column of dust on its way to us, and is thus attenuated by different amounts.}
of the central kpc of NGC\,1386 is shown in Fig.\,\ref{image_dust}, derived following the procedure used in
 \citet{2014MNRAS.442.2145P}.
Briefly, $A_V$ is derived from the VLT/\textsf{NaCo} \textit{Ks}-band $-$ \textit{HST}/F814W colour image, measuring the 
intrinsic 
colour 
in visually
selected 
dust-free regions.

\begin{figure}
\begin{center}
  \includegraphics[width=0.86\columnwidth]{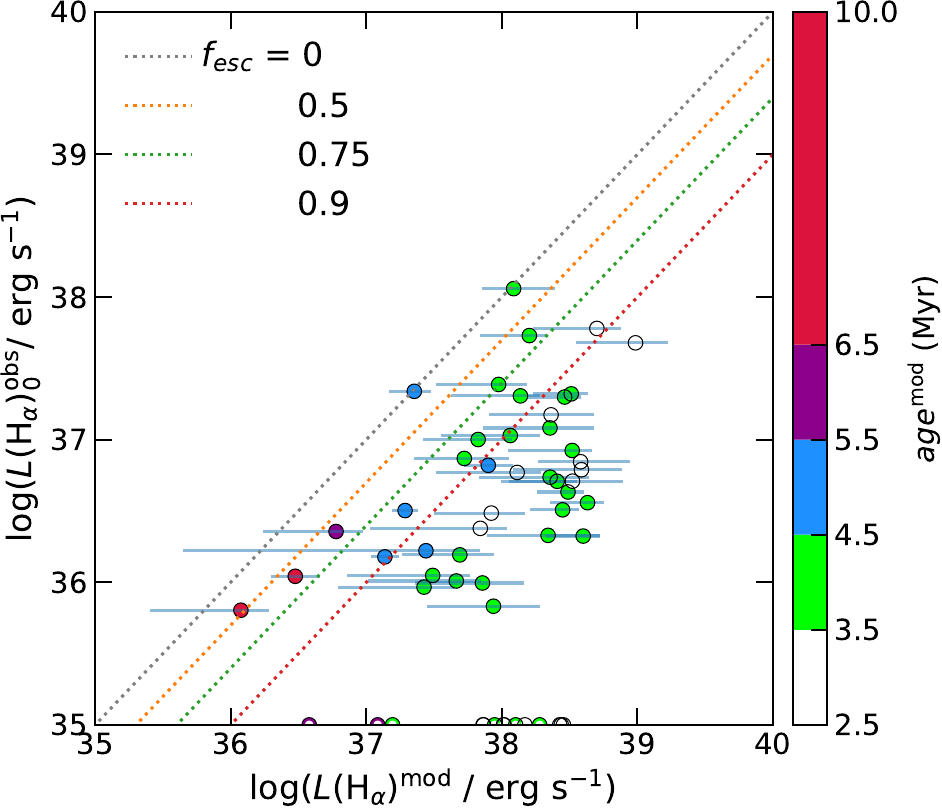}
\end{center}
  \caption{\label{lha_m} Comparison of \lhamod, the \ha\ luminosity derived from optical and NIR SED fits, with the 
corrected observed \lhaobsI.
  Horizontal bars are the 14th and 86th percentiles of the stochastic solution. 
  Circles are colour coded according to \tm. Coloured dotted lines trace the relation corresponding to different escape 
fractions ($f_{\rm esc}$). 
  Upper limits, arbitrarily placed at \lhaobsI\,=\,35, correspond to clusters with no \ha\ detection.}
\end{figure}

\subsection{Mass}\label{s:mass}
The 
clusters in NGC\,1386 are  of  low
mass (Fig.\,\ref{allparam}B). With the standard model  we obtain a mass 
distribution in the range  $4.2\times10^2 < M/M_\odot <1.6\times10^4$, with 
the
median at $4\times10^3 M_\odot$ 
and first and third quartiles at $1.2\times10^3$ and $8.8\times10^3 M_\odot$, respectively.
The mass distribution resulting from the stochastic model  shows a narrow peak near $8\times10^3M_\odot$. 
The  relation between \mm\ and \avm\ is 
shown 
in Figure \ref{allparam}c.  The 
lower-mass 
clusters are in the regions 
of lower $A_V$, this  is more  noticeable in the standard than in the stochastic model fits.  
The absence 
of massive blue stars in some stochastic models 
favours 
lower-extinction solutions. Nonetheless, this effect 
may be due to a selection effect: low mass but highly extinguished clusters are likely to go undetected.

The total atomic plus molecular gas mass in the  clusters can be inferred from \avm\ and the typical cluster size.
$A_V$ is in the range 0.1 to 4 mag. From the standard dust-to-gas ratio,
$N_H/A_V = 2 \times 10^{21}$ cm$^{-2}$ \citep{1979ARAA..17...73S}, 
the inferred H column density is in the range:  $0.2\times10^{21}$\,<\,$N_H$\,<\,$8\times10^{21}$ cm$^{-2}$.  
Using a cluster size of 6.7 pc FWHM (upper limit from Section\,\ref{phot}), the gas mass is in range 
50\,<\,$M_{\rm gas}$\,<\,$2\times10^3$\,\msun,
$\sim$ 10\% of the cluster stellar mass. 
The low gas fraction and young ages indicate that star formation  
probably 
took place in a very short burst. This may 
also explain the observed lower than expected  \ha\  in these clusters (Section\,4.4).

An estimate of the  total stellar mass in the ring, after  summing up the mass of the 61 detected clusters, is $5\times10^5 $ \msun.
Were the size for the clusters in NGC 1386 of the same order as those directly measured spatially  in the nuclear starburst  
NGC\,253 \citep{2009MNRAS.392L..16F} or in M31 \citep{kodaira04}, i.e.\ FWHM $\sim$\,2\,pc, the gas fraction would be reduced to <\,1\% of the stellar mass. 

\subsection{Lyman continuum photon leak} \label{s:halpha}
  
Fig.\,\ref{image_ha} shows the distribution of \ha\ + [\ion{N}{ii}] in the central kpc of the galaxy. It 
comprises
a dominating central emission due to the AGN, and  a ring of point-like sources
identified  with the continuum emission from the detected clusters. 
However, about 17 clusters
(30\% of the sample, Table\,\ref{phot_n1386})
show  
no 
\ha\ + [\ion{N}{ii}] counterpart despite their young age. 
Moreover, a number of  \ha\ + [\ion{N}{ii}]  
sources in the ring --\,some point-like, others more diffuse\,-- do not show a cluster (continuum 
emission) counterpart. We recall that cluster detection is based on their continuum detection in 
our deepest 
\textit{HST}
 F606W image (Fig.\,\ref{image_V} and sect.\,\ref{phot}).
Some of these  orphan sources  could be  
leaked 
\ha\  from  the clusters --\,in some cases, the emission
is 
somewhat extended next to the cluster (e.g. clusters 13, 15), or clusters with continuum emission below our 3 sigma threshold for detection 
in any of the photometry bands.

Table\,\ref{tab:n1386_fitted_par} lists only the clusters for which a reliable model fit was obtained;
it includes all cases 
with continuum detection in at least three photometric bands. The table shows that for most
of the clusters, 
we detect less \ha\,flux than the model predicts, and in 14 cases, no \ha\ 
at all
is detected  despite their young age
 and bluish spectrum. In the cases with lower 
than predicted
\ha , 
the deficit reaches up to an order of 
magnitude in some cases.  The model-derived  \lhamod\ 
in the table results from the integration of 
the Lyman 
ionising 
continuum in the SSP model, under the assumption of     total conversion of 
ionising 
photons to \ha\,flux 
(ionisation-bounded\footnote{In this case absorption and re-emission occur locally in 
the \hii\ region and no photons escape from the 
ionised 
nebula \citep[][]{osterbrock89}.} \hii\ region) 
and a complete coverage of the 
ionising 
source.  \lhaobsI\ in the table is the observed \ha\ luminosity after  
extinction
and stellar absorption
corrections, both   as inferred from the model. 

At the age of these clusters, the contribution of \ha\ in absorption is much less than 1\% of the  total in 
emission, as derived from the SPP models, 
so 
this absorption would not explain the \ha\ deficit. Very 
high $A_V$
($\sim$\,10 mag)
is required to make \ha\ undetectable. Yet, such a high extinction  
departs 
considerably 
from the moderate $A_V\,\lesssim\,4$ mag resulting from the cluster
fits (Fig.\,\ref{allparam}), and the dust 
map of the region (Fig.\,\ref{image_dust}). Moreover, $A_V$ values that high would notably 
depress  the continuum  towards 
the shortest wavelengths ($A_B\sim 1.7 \times A_V$), which is contrary
to the generally flat or blue spectra 
observed (Fig.\,\ref{image_V} and Appendix \ref{sec:appendix}). 
Illustrative examples are the clusters \#2,
 \#7 and \#9 (Table\,\ref{tab:n1386_fitted_par}), which have 
very blue spectra but lower 
than predicted \ha .

Given the relative low emission of these clusters, we 
assess  whether the expected \ha\ emission is 
below 
the level of uncertainty of 
our measurement. This is mostly linked to the choice of the integration aperture size, which we limited 
to a radius equal to the FWHM of the clusters to avoid contamination from neighbouring clusters. This uncertainty
 is estimated 
to be 
 a factor of two after  comparing with fluxes in  
an 
$r = 2~\times $~FWHM. This larger radius 
includes the first Airy ring of the PSF, as judged from the extracted cluster's light profile, but it may also 
include
extra emission from neighbouring regions.

We further   
cross-check  
our 
\textit{HST}/ 
\ha\ + [\ion{N}{ii}] image  with  a  deep VLT-MUSE image of the cluster's ring. Because 
of 
the difference in angular resolution, this comparison could safely be done for  two isolated clusters.  The effective 
FWHM of the clusters in MUSE is 0.8 arcsec, 
a 
factor seven larger than that in  
\textit{HST}, 
and this  introduces confusion 
and light mixing all along the ring. The smearing effect  can be appreciated by comparing with a published MUSE \ha\ image 
in \citet{2019A&A...627A.136I}.  Still, for the selected clusters in the 
MUSE-extracted 
\ha\ image
(clusters 11
and 20), 
their \ha\ fluxes were found 
to be 
within a factor of two  larger in MUSE, in line with  our previously derived aperture 
correction factor. The availability of spectra from the MUSE cube also allowed us to verify two effects: 1) the  
emission spectrum 
inside 
the ring is of LINER type  
([\ion{N}{ii}] 6583 \AA\ is comparable or higher than \ha),
so 
all the gas 
inside 
the ring is entirely associated  with the central AGN; 2) the spectra  in the ring are typical 
\textsc{H\,ii} type, but  the [\ion{N}{ii}] contribution is 
a factor two less than  we assumed on the basis of  
the BPT diagrams (Section\,\ref{phot}). Summing up the contribution of the two effects, aperture correction and [\ion{N}{II}] 
contribution, the reported \ha\ fluxes in Table\,\ref{tab:n1386_fitted_par} had a priory to be increased by a factor of 2.7 at most. We decided 
not to apply this correction explicitly in our photometry  but 
instead to 
retain 
it 
as an uncertainty term. We prefer to 
stick to our \ha\ estimate, as it 
was 
done in a systematic manner for all the clusters in the ring. Still, even if 
accounting for a factor 2.7 increase  in \lhaobsI\, which would yield the maximum possible \ha\ from 
the observations, these remain     under-predicted in some cases by an order of magnitude (Fig.\,\ref{lha_m}).

Thus, we think that  the observed \ha\ deficit is caused by  a significant escape fraction of Lyman continuum 
photons from the HII region. The  escape fraction is estimated after  the comparison of the total 
\lhaobsI\ from  the clusters, and the total predicted from the models. These total values are 
shown 
at the bottom of  
Table\,\ref{tab:n1386_fitted_par}. They differ by factors 
of up to 
10 (7 after applying a factor 2.7 increase 
in \lhaobsI). Thus, the Lyman escape fraction is in the 94\%--85\% range.
As  indicated,  higher $A_V$  
values 
are incompatible with the blue or flat spectrum of most of the clusters. Thus we 
conclude that the  cluster's \hii\ region are  matter-bounded, most 
of the 
 ionizing photons escaping the 
Str\"omgren
sphere, 
this being opposite to 
what is generally believed 
for \hii\ regions  (e.g.\ Osterbrock 1989).

The  leaking photons,
however,
 seem 
to be 
bounded to the ring:
the \ha\ integration over the entire ring on 
both 
the 
\textit{HST} 
\ha\ image (Fig.\,\ref{image_ha}) 
and 
the  MUSE \ha\ image
yields 
a comparable value of  $\sim 10^{40}\,\rm{erg\,s^{-1}}$ --\,after applying a global  extinction 
$A_V$ of 1 mag. 
This is about the total luminosity predicted  from  the models (Table\,\ref{tab:n1386_fitted_par}). Thus, most of the leaking photons at cluster level 
seem to  get trapped in ionised gas in the ring. 
The same result was obtained for the cluster's ring in NGC\,1097.

These  high escape fractions at cluster level are not unprecedented, as 
discussed in the next section.

\begin{table}
	\caption{Results from SED fits to star clusters in NGC\,1386$^a$.}
	\label{tab:n1386_fitted_par}
	\begin{tabular}{p{0.01cm} p{0.02cm} p{0.1cm} p{0.1cm} p{0.1cm} p{0.5cm} p{0.9cm} p{0.8cm}}
			\hline
			& & & & $\mathrm{log}$ & & $\mathrm{log}$ & $\mathrm{log}$\\ 
		\multirow{2}{0.2cm}[0.03cm]{\bf \#} &    
		\multicolumn{2}{c}{Relative} &
		\multicolumn{1}{c}{age} & 
		\multicolumn{1}{c}{mass} & 
		$\mathrm{A_V^{mod}}$ &
        \lhamod &
        \lhaobsI\\ 
		\rule{0pt}{3ex} & \multicolumn{1}{c}{R.A.} & \multicolumn{1}{c}{Dec.} & \multicolumn{1}{c}{[Myr]} & \multicolumn{1}{c}{[\msun]} & [mag] & [$\ergs$] & [$\ergs$] \\ 
		\hline
  1   & \multicolumn{1}{r}{$11$\farcs15}   & \multicolumn{1}{r}{$ -12$\farcs97}   & 3.60  & $  3.44 $ & 1.9 & 38.02  & \dotfill \\
  2   & \multicolumn{1}{r}{$ 3$\farcs35}   & \multicolumn{1}{r}{$ -11$\farcs37}   & 3.60  & $  3.09 $ & 0.5 & 37.66 & 36.01 \\
  3   & \multicolumn{1}{r}{$ 1$\farcs22}   & \multicolumn{1}{r}{$ -11$\farcs17}   & 3.10  & $  3.66 $ & 1.8 & 38.43  & \dotfill \\
  4   & \multicolumn{1}{r}{$-11$\farcs01}  & \multicolumn{1}{r}{$ -11$\farcs16}   & 2.60  & $  3.26 $ & 2.0 & 38.01  & \dotfill \\
  5   & \multicolumn{1}{r}{$-0$\farcs68}   & \multicolumn{1}{r}{$ -11$\farcs03}   & 3.60  & $  2.85 $ & 0.6 & 37.43 & 35.97 \\
  6   & \multicolumn{1}{r}{$ 7$\farcs96}   & \multicolumn{1}{r}{$ -10$\farcs73}   & 5.50  & $  2.66 $ & 0.7 & 36.58  & \dotfill \\
  7   & \multicolumn{1}{r}{$ 4$\farcs02}   & \multicolumn{1}{r}{$ -10$\farcs59}   & 4.50  & $  2.94 $ & 0.0 & 37.29 & 36.50 \\
  8   & \multicolumn{1}{r}{$ -1$\farcs11}  & \multicolumn{1}{r}{$ -10$\farcs58}   & 3.60  & $  3.79 $ & 2.0 & 38.34 & 36.33 \\
  9   & \multicolumn{1}{r}{$ 7$\farcs47}   & \multicolumn{1}{r}{$ -10$\farcs13}   & 3.70  & $  3.29 $ & 1.6 & 37.83 & 37.00 \\
 10   & \multicolumn{1}{r}{$ 5$\farcs05}   & \multicolumn{1}{r}{$ -9$\farcs53 }   & 2.60  & $  3.09 $ & 0.2 & 37.84 & 36.38 \\
 11   & \multicolumn{1}{r}{$ 7$\farcs51}   & \multicolumn{1}{r}{$ -9$\farcs30 }   & 4.60  & $  3.04 $ & 1.3 & 37.35 & 37.34 \\
 12   & \multicolumn{1}{r}{$-0$\farcs50}   & \multicolumn{1}{r}{$ -9$\farcs25 }   & 3.60  & $  3.80 $ & 2.8 & 38.35 & 37.08 \\
 13   & \multicolumn{1}{r}{$-0$\farcs04}   & \multicolumn{1}{r}{$ -9$\farcs01 }   & 4.20  & $  3.25 $ & 0.0 & 37.69 & 36.19 \\
 14   & \multicolumn{1}{r}{$ 11$\farcs15}  & \multicolumn{1}{r}{$ -8$\farcs05 }   & 5.20  & $  3.82 $ & 0.9 & 37.90 & 36.82 \\
 15   & \multicolumn{1}{r}{$-0$\farcs02}   & \multicolumn{1}{r}{$ -7$\farcs74 }   & 3.70  & $  3.95 $ & 3.0 & 38.49 & 36.63 \\
 16   & \multicolumn{1}{r}{$-0$\farcs56}   & \multicolumn{1}{r}{$ -6$\farcs80 }   & 3.10  & $  3.81 $ & 3.0 & 38.59 & 36.79 \\
 17   & \multicolumn{1}{r}{$ -1$\farcs01}  & \multicolumn{1}{r}{$ -6$\farcs33 }   & 10.00 & $  4.08 $ & 0.1 & 36.48 & 36.04 \\
 18   & \multicolumn{1}{r}{$ -1$\farcs51}  & \multicolumn{1}{r}{$ -6$\farcs14 }   & 4.30  & $  3.45 $ & 0.1 & 37.87  & \dotfill \\
 19   & \multicolumn{1}{r}{$ -1$\farcs93}  & \multicolumn{1}{r}{$ -5$\farcs88 }   & 2.60  & $  3.43 $ & 0.5 & 38.17  & \dotfill \\
 20   & \multicolumn{1}{r}{$ -2$\farcs51}  & \multicolumn{1}{r}{$ -5$\farcs78 }   & 4.40  & $  3.86 $ & 2.3 & 37.86 & 35.99 \\
 21   & \multicolumn{1}{r}{$ -2$\farcs09}  & \multicolumn{1}{r}{$ -5$\farcs51 }   & 3.20  & $  3.12 $ & 0.8 & 37.86  & \dotfill \\
 22   & \multicolumn{1}{r}{$ 7$\farcs03}   & \multicolumn{1}{r}{$ -4$\farcs77 }   & 3.60  & $  3.41 $ & 2.0 & 37.98 & 37.39 \\
 23   & \multicolumn{1}{r}{$ 6$\farcs52}   & \multicolumn{1}{r}{$ -2$\farcs88 }   & 4.20  & $  3.70 $ & 2.4 & 38.14 & 37.31 \\
 24   & \multicolumn{1}{r}{$ -4$\farcs31}  & \multicolumn{1}{r}{$ -2$\farcs53 }   & 4.70  & $  3.16 $ & 0.0 & 37.44 & 36.22 \\
 25   & \multicolumn{1}{r}{$ -4$\farcs52}  & \multicolumn{1}{r}{$ -1$\farcs82 }   & 3.10  & $  3.93 $ & 3.0 & 38.70 & 37.78 \\
 26   & \multicolumn{1}{r}{$ -5$\farcs07}  & \multicolumn{1}{r}{$ -1$\farcs77 }   & 9.50  & $  3.57 $ & 1.1 & 36.08 & 35.80 \\
 27   & \multicolumn{1}{r}{$ 6$\farcs27}   & \multicolumn{1}{r}{$ -1$\farcs62 }   & 3.10  & $  3.75 $ & 2.9 & 38.52 & 36.71 \\
 28   & \multicolumn{1}{r}{$ -4$\farcs70}  & \multicolumn{1}{r}{$ -1$\farcs59 }   & 4.30  & $  3.31 $ & 1.1 & 37.72 & 36.87 \\
 31   & \multicolumn{1}{r}{$ 6$\farcs33}   & \multicolumn{1}{r}{$-0$\farcs43  }   & 4.00  & $  3.95 $ & 3.3 & 38.28  & \dotfill \\
 32   & \multicolumn{1}{r}{$ -5$\farcs47}  & \multicolumn{1}{r}{$-0$\farcs41  }   & 4.30  & $  3.53 $ & 1.9 & 37.95  & \dotfill \\
 33   & \multicolumn{1}{r}{$ -6$\farcs50}  & \multicolumn{1}{r}{$ 0$\farcs34  }   & 3.60  & $  2.92 $ & 0.3 & 37.49 & 36.05 \\
 34   & \multicolumn{1}{r}{$ 8$\farcs05}   & \multicolumn{1}{r}{$ 1$\farcs05  }   & 3.60  & $  2.62 $ & 0.1 & 37.19  & \dotfill \\
 35   & \multicolumn{1}{r}{$ 4$\farcs70}   & \multicolumn{1}{r}{$ 1$\farcs93  }   & 4.30  & $  3.79 $ & 1.9 & 38.20 & 37.73 \\
 36   & \multicolumn{1}{r}{$ 4$\farcs29}   & \multicolumn{1}{r}{$ 2$\farcs16  }   & 4.20  & $  4.02 $ & 1.8 & 38.46 & 37.30 \\
 37   & \multicolumn{1}{r}{$ 4$\farcs54}   & \multicolumn{1}{r}{$ 2$\farcs93  }   & 3.60  & $  3.38 $ & 1.3 & 37.94 & 35.83 \\
 38   & \multicolumn{1}{r}{$ 2$\farcs29}   & \multicolumn{1}{r}{$ 5$\farcs27  }   & 3.70  & $  4.08 $ & 2.6 & 38.63 & 36.56 \\
 39   & \multicolumn{1}{r}{$ 0$\farcs72}   & \multicolumn{1}{r}{$ 5$\farcs38  }   & 5.50  & $  2.86 $ & 0.3 & 36.78 & 36.36 \\
 40   & \multicolumn{1}{r}{$ -7$\farcs92}  & \multicolumn{1}{r}{$ 5$\farcs98  }   & 5.75  & $  3.31 $ & 1.9 & 37.08  & \dotfill \\
 41   & \multicolumn{1}{r}{$ 3$\farcs11}   & \multicolumn{1}{r}{$ 5$\farcs99  }   & 3.10  & $  3.68 $ & 2.8 & 38.45  & \dotfill \\
 42   & \multicolumn{1}{r}{$ -2$\farcs39}  & \multicolumn{1}{r}{$ 6$\farcs11  }   & 3.80  & $  3.90 $ & 2.8 & 38.41 & 36.71 \\
 43   & \multicolumn{1}{r}{$ 2$\farcs50}   & \multicolumn{1}{r}{$ 6$\farcs90  }   & 2.70  & $  3.83 $ & 1.9 & 38.58 & 36.84 \\
 44   & \multicolumn{1}{r}{$ 2$\farcs27}   & \multicolumn{1}{r}{$ 7$\farcs06  }   & 3.60  & $  3.80 $ & 2.2 & 38.36 & 36.74 \\
 45   & \multicolumn{1}{r}{$ 2$\farcs44}   & \multicolumn{1}{r}{$ 7$\farcs69  }   & 3.70  & $  3.97 $ & 2.5 & 38.52 & 36.92 \\
 48   & \multicolumn{1}{r}{$ -7$\farcs43}  & \multicolumn{1}{r}{$ 8$\farcs24  }   & 4.20  & $  4.16 $ & 3.0 & 38.60 & 36.32 \\
 49   & \multicolumn{1}{r}{$ -7$\farcs82}  & \multicolumn{1}{r}{$ 9$\farcs23  }   & 3.90  & $  3.59 $ & 3.0 & 38.09 & 38.06 \\
 50   & \multicolumn{1}{r}{$ -8$\farcs04}  & \multicolumn{1}{r}{$ 9$\farcs87  }   & 4.90  & $  2.93 $ & 0.4 & 37.14 & 36.18 \\
 51   & \multicolumn{1}{r}{$ -1$\farcs96}  & \multicolumn{1}{r}{$ 10$\farcs30 }   & 2.30  & $  3.31 $ & 1.2 & 38.11 & 36.77 \\
 52   & \multicolumn{1}{r}{$ -4$\farcs04}  & \multicolumn{1}{r}{$ 10$\farcs72 }   & 3.40  & $  3.25 $ & 1.2 & 37.92 & 36.48 \\
 53   & \multicolumn{1}{r}{$-0$\farcs28}   & \multicolumn{1}{r}{$ 11$\farcs04 }   & 4.30  & $  3.69 $ & 2.2 & 38.10  & \dotfill \\
 54   & \multicolumn{1}{r}{$ -3$\farcs12}  & \multicolumn{1}{r}{$ 11$\farcs12 }   & 3.80  & $  3.94 $ & 2.9 & 38.45 & 36.51 \\
 55   & \multicolumn{1}{r}{$ -2$\farcs85}  & \multicolumn{1}{r}{$ 11$\farcs18 }   & 3.60  & $  3.96 $ & 3.0 & 38.51 & 37.32 \\
 56   & \multicolumn{1}{r}{$ -3$\farcs85}  & \multicolumn{1}{r}{$ 11$\farcs28 }   & 2.80  & $  3.62 $ & 2.5 & 38.36 & 37.17 \\
 57   & \multicolumn{1}{r}{$ -4$\farcs46}  & \multicolumn{1}{r}{$ 11$\farcs77 }   & 2.60  & $  3.69 $ & 3.6 & 38.44  & \dotfill \\
 58   & \multicolumn{1}{r}{$ -3$\farcs45}  & \multicolumn{1}{r}{$ 12$\farcs11 }   & 3.10  & $  4.21 $ & 4.7 & 38.99 & 37.68 \\
 59   & \multicolumn{1}{r}{$ -4$\farcs44}  & \multicolumn{1}{r}{$ 12$\farcs28 }   & 4.20  & $  3.62 $ & 2.7 & 38.06 & 37.03 \\
\hline\\
Total & & & & & &        39.97 &        38.73 \\  
\bottomrule
\multicolumn{8}{l}{{\it Notes.} $^a$Age (column 4), mass (5), $A_V$ (6) and \ha\ luminosity, from fit}\\
\multicolumn{8}{l}{(7) and observed (8). The latter has been corrected for stellar absorption}\\
\multicolumn{8}{l}{and reddening. Clusters \#0, 29, 30, 46, 47 and 60 are not included due to}\\
\multicolumn{8}{l}{poor photometry in most bands.}\\
	\end{tabular}

\end{table}

\subsection{Lyman photon leak in other galaxies}\label{s:halpha_others}

\subsubsection{Observations}

Lyman continuum photon leaks have been reported in, among others, \hii\ regions in the spiral arms of galaxies
\citep{ok97,beckman+00,voges+08, 2022A&A...659A..26B}, and the Milky-Way \citep{2009RvMP...81..969H}
in the cluster population in the 
disc 
of the galaxy M83 \citep{hollyhead+15},   
as well as in circumnuclear rings \citep[e.g. NGC\,1097,][]{prieto+19}. Studies indicate that at least 
30\% of \hii\ regions in nearby galaxies are density-bounded, with a fraction of escaping Lyman continuum 
photons from 10 to 70\% \citep{castellanos+02}. Larger escape fractions are found preferentially in 
low-mass, 
low-luminosity 
\hii\ regions. \citet{xiao+18} derive the properties of 254 \hii\ regions in nearby spiral and 
dwarf galaxies from simultaneously fitting the observed emission lines and the \hb\ equivalent width. They find 
that the escape fraction of 
ionising 
photons evolves with the age of the 
ionising 
cluster, from $\sim$\,90\% at 
1\,Myr to $\sim$\,35\% at 10\,Myr. They estimate that the radius and the mass of the \hii\ regions are in the ranges 
from $\sim$\,0.5 to 10 pc and from $\sim$\,100 to $10^5$\,\msun, respectively.

Recent 
studies 
have mapped individual \hii\ regions at the 10 pc scale with MUSE on the VLT. 
\citet{weilbacher+18} mapped the central region of the Antennae nebula and found that 38 (out of 386) low \ha\ 
luminosity \hii\ regions (\lha\,$\le$\,38.25 dex), comparable to the ones around the NGC\,1386 clusters, have a 
median Lyman continuum photon escape fraction of 83\%, with a standard deviation of 22\%. They estimate a photon 
escape fraction of 7\% when averaged over all the regions. Also with MUSE, \citet{dellabruna+21} find an average 
escape fraction of 67\% for the \hii\ regions in NGC\,7793. \citet{prieto+19} estimate a photon leak 
of 
over 90\% from 
the star clusters in the circumnuclear ring of NGC\,1097.

\subsubsection{Reliability of\ \ \texorpdfstring{\ha}\ \ as a SFR tracer}

A physical explanation for this leak is that the \hii\ regions are optically thin \citep{beckman+00, papaderos+13}, 
or that their gas density is very low, favouring the escape of UV photons from the Str\"omgren sphere \citep{prieto+19}. 
Regarding the fate of these photons, \citet{ferguson+96} and \citet{beckman+00} suggest that the Lyman continuum photons 
leaking from the arms in spirals may give rise to a diffuse 
ionised 
gas component, whose luminosity is estimated to be 
in the $10^{40}\,\ergs$ range.

Modelwise, numerical simulations of  pre-collapse molecular clouds with 
masses 
in the range $10^4$--$10^6$ \msun\ \citep{dale+13} show that the Lyman 
continuum photon escape fraction is closely related to the conditions of the parental cloud, specifically, 
to 
its mass and equilibrium phase (whether the cloud is a bounded or partially bounded system). In particular, 
low-mass
 clusters with mass\,<\,$10^5$\,\msun, have large photon escape fractions, ranging from tens to 
a hundred per cent
\citep[see also][]{dobbs+11}.

The numerical calculations of
\citet{tenoriotagle+99}, 
which consider the mechanical energy deposited by WR and other 
massive stars at the earliest stages of the cluster evolution, imply that once the matter swept
up by the radiation 
front becomes Rayleigh--Taylor unstable and fragments, the region allows not only for the venting of the hot 
matter 
from both wind and supernovae
outwards
but also for the leakage of a large fraction of the 
ionising 
radiation.

An obvious consequence of Lyman  photon leakage is that star formation tracers based on the \hii\ gas  may dramatically underestimate the true SFR. This will equally affect infrared (IR) tracers, as the IR emission comes from dust-reprocessed UV photons. At scales comprising  the entire star-forming region,  the integrated \ha\ over the region may account for the possible  Lyman leakage on local (cluster-level) scales if the gas remains bound to the region,  thus  being representative of a sort of global SFR in the area, but see sect. \ref{s:sfr}.

\begin{figure*}
\begin{center}
    \includegraphics[scale=0.5]{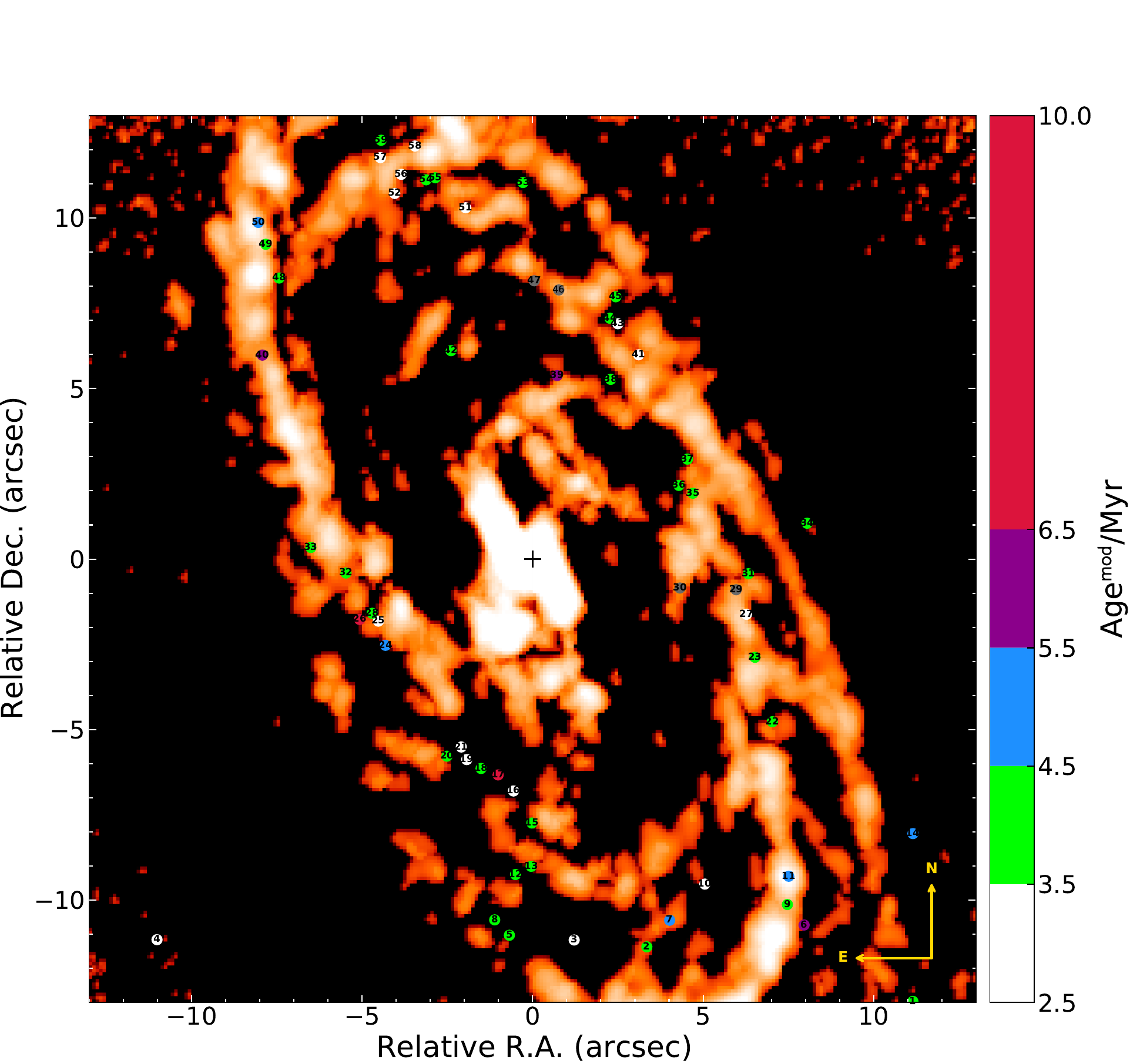}
     \caption{\label{image_CO_large} ALMA/CO$(2-1)$ image. Circles at detected clusters are 
colour-coded
 with model age. In 
grey, 
clusters detected only in two or 
fewer
bands, for which no reliable age can be derived.  Circle size 
is about the angular resolution of our
\textsf{ESO}-\textsf{NaCo}-Adaptive Optics 2\,$\mu\rm{m}$\,image (FWHM\,$\sim\,0.09\,\arcsec$, Table\,\ref{tab_obs}), chosen so  to best illustrate the clusters - molecular gas separation. We take $\sim\,0.09\,\arcsec$ as  an upper limit to the physical  size of the clusters, 6.7 pc FWHM.}

\end{center}
\end{figure*}    

\section{The circumnuclear ring of CO filaments}\label{s:CO}

\subsection{Morphology and mass}\label{s:CO_mass}


The total molecular  mass in the ring, assuming a CO-to-H$_2$ MW conversion factor $X_{\rm{CO}} ~= ~2 \times 10^{20} 
~\rm{{cm}^{-2}~ (K ~km~ s^{-1})^{-1}}$ under the optically thick case \citep[cf.][]{scoville+sanders87, bolatto+13}, 
is $\sim 2\times 10^8$ \msun. No correction for He content
or for the CO$(2-1)$ contribution, usually a factor of 1.3 each, 
has
been applied. The molecular mass 
of 
the central nuclear region  is  comparable
($1.6\times 10^8$ \msun ).

The molecular gas ring resolves into multiple
long and narrow filaments
running parallel to the ring of clusters. 
The whole molecular structure  
defines 
a sort of secondary ring 
partially
enclosing the stellar one (Fig.\,\ref{image_V}).
We distinguish  two main filaments, one runs from the 
north along the 
east 
side of the galaxy to curve 
midway 
towards the nuclear region;
a second one  runs from  
the 
south along the 
west side of the galaxy, to curve 
slightly
above 
midway   
also towards the centre.  In between,  one may count about six or more
shorter filaments 
partially 
connected with the 
main ones. 
Together, all these filaments 
define a kind of nuclear spiral. The filaments resolve in point-like clouds
of
size $\lesssim  40$ pc FWHM (the resolution of the ALMA data (sect. \ref{IR_obs})),  surrounded by diffuse emission. Most of 
the point-like clouds are not associated with visible clusters in the ring;
in turn, most of the star clusters are 
located
in molecular voids. The faintest CO emission in the figure is three sigma 
times
the background level, $\sim 40$ mJy per beam. 

The most conspicuous CO filaments do coincide with equivalent dust filaments seen directly  in the 
\textit{HST}
optical image  (Figs. \ref{image_V} and \ref{CO_contours}). We note two points: 1) the maximum  $A_V$ in the extinction map 
(Fig.\,\ref{image_dust}) is  four 
magnitudes in line with the maximum Av found in the clusters 
and 
2) most clusters are seen at optical wavelengths.  
We 
therefore 
believe that the CO clouds are not hiding   
clusters with ages comparable to  those of the visible ones 
since 
it would be surprising not to see some of them in our near-IR
VLT-Adaptive-Optics or 
\textit{HST} 
images. Instead, we believe that these clouds are   nursing a new generation of clusters  
in the ring, possible  of 
a 
similar nature 
to 
that  of the current ones in the cluster ring. 
Arguments 
in support of
this 
claim 
are put forward in the next section.  
We count about   70 clouds in the CO filaments.
Coincidentally, the number 
of clusters detected in the cluster ring  is similar
(61).

By comparison,   the much richer, in 
both
cluster
number and mass, 
 circumnuclear  ring of NGC\,1097
contains about eleven  protocluster candidates,   each identified  as a point-like source, of size\,<\,40\,pc, 
but 
only at 
the 10 $\mu$m and 20 $\mu$m 
mid-IR 
wavelengths
in the 
VLT diffraction-limited images. These sources are not associated with any 
cluster in the ring, and in turn 
have 
no
optical or near-IR counterpart, as 
is  the case in NGC\,1386. The NGC\,1097 
sources instead coincide with 
the 
CO or HCN emission peaks of almost any molecular cloud in the ring \citep[][their fig.\,1]{prieto+19}.
Similar detailed mid-IR--CO source associations are starting to be seen  in  
\textit{JWST} images, e.g.\ some of the mid-IR sources 
discovered by 
the \textit{JWST} 
in NGC\,1365 \citep[e.g.][their fig. 3]{whitmore+23} are  
seen to
coincide
with CO emission peaks and could be 
protoclusters. Unfortunately,
 equivalent mid-IR information for the CO point-like clouds in NGC\,1386  is still lacking, 
pending 
data from
the \textit{JWST} and  VLT-VISIR.

The mass of the point-like clouds in NGC\,1386 is in the $10^6$ \msun\ range 
(based on their CO peak emission and optically thick luminosity conversion). This mass turns to be of the same order 
of the virial mass if taking as  their virial velocity the  
dispersion  velocity 
(line width)
measured  in the CO maps
($\sim$ 20 $\rm{km~ s^{-1}}$ on average). 
On this
basis, taking  the cloud upper 
limiting
 radius
(20 pc), 
the cloud masses are found 
to be 
of the same order as that inferred from the optically thick mass
($\lesssim 2.5\times 10^6$ \msun ).  
In comparison, the still more massive molecular clouds,  $10^7$ \msun\, in the circumnuclear 
cluster ring of NGC\, 1097  are also virialized \citep{hsieh+11}, 
as
is 
widely
 the case  in   molecular clouds in the  Galaxy \citep[e.g.][and references therein]{bolatto+13}.

The mass of the clusters in NGC\,1386 is in the $10^3$--$10^4$ \msun\ range. 
Where 
the molecular clouds in the filaments   give 
birth to a cluster population with masses  similar  to those of the current ones, the  star formation efficiency, 
cluster-mass/gas-mass would be  0.1--1\%. In global terms, the ratio of the total mass in clusters in the  ring over the 
total mass of the molecular ring is $ 5\times 10^5$ \msun\,/ $1.6\times 10^8$ \msun, also about 0.1\%. These efficiencies  are just 
within 
the     range  measured  in  the Galaxy and  in 
nearby 
 stellar clusters \citep[e.g.][and references therein]{krumholz+14}. 

To summarise,   the    circumnuclear rings of  molecular gas and  star clusters in this  
early-type 
spiral share  similar properties  
in terms of molecular cloud
masses, cloud
stabilisation 
(virial)  
and star formation efficiency, as found in the disc of the 
Galaxy.

\subsection{Clusters in molecular voids}\label{s:COvoids}
Figure \ref{image_CO_large} shows the relative  distribution of the star clusters  and  
CO filaments in  NGC\,1386. The 
clusters' 
location is marked with circles whose size is the nominal upper limit,
FWHM < 0.09 arcsec, in our AO-VLT/K-band image. In this 
way  the cluster--filament relative location  can  be better  assessed. It can be noticed that most of the clusters are 
located 
away from 
the molecular gas. The separation between the filaments and the clusters  varies along  the ring, presumably 
because of 
projection effects, 
but it is persistent and can be seen at any point in the ring.

Typical cluster--molecular filament offsets are 40 to 60\,pc, in projection. The galaxy inclination is 65 degrees \citep[2MASS 
catalogue and  CO kinematics,][]{ramakrishnan+19}, the near side of the ring is 
to 
the 
west and the dust filaments are seen 
in the foreground. 
When correcting for the galaxy inclination the offsets amount 
to
90--140\, pc. Some clusters are found in CO voids, 
e.g.\  the 
arch-like rim of clusters \#3  to \#10, 
in
the 
southern part
 of the ring, whereas other are in strings running 
tangentially 
to the CO 
filaments, 
e.g.\  the stream of clusters \#24 to \#33 
on
the 
eastern side of the ring,  or the
clusters \#15 to \#20 
to
the  
south.  A few clusters  
coincide with   CO depressions in the filaments, e.g..\ clusters \#40, \#51, \#23 and \#9.
Regardless of their location, no age differences or other properties are found among the clusters. Equally,  
there are no 
signs of distortion 
in  the filaments closer to a cluster that 
might suggest 
feed-back effects.

Equivalent spatial offsets between molecular cloud peaks  and sites of star formation
are regularly  seen in the spiral arms of galaxies, in general with the 
star-forming 
regions siting at the outer edge of the
 molecular arms 
(e.g.\ the prototype case M51, \citealt{tully74ii}, \citealt{rand+kulkarni90}, \citealt{scoville+01}; NGC\,4254, \citealt{egusa+04} and 
\citealt{tamburro+08} and references therein; 
M100, \citealt{castillo-morales+07}).
Still,  this spatial separation  is not  seen at all scales in the spiral arms
 
These displacements  are in general interpreted in the context of galactic shock-wave theory \citep{fujimoto68, roberts69, tully74iii, 
tamburro+08}
and could be used to characterise the various  phases of star formation \citep[e.g.][and references therein]{kim+23}. 

In 
this study,
 we see  an equivalent spatial offset but  in a circumnuclear ring
(more precisely,  between the clusters  continuum emission  and the molecular gas).
The resolution of  a few parsecs in these
data allows us to identify with high precision  the  
location of the clusters' 
continuum emission.  
Thus,   the 
quoted spatial separations are with respect to the  true star-forming location 
 as measured from the  
continuum light, and not with respect to the \hii\ gas as 
is usually measured in spirals 
(c.f.\ above references).
Offsets with respect the \hii\ gas  may be misleading when estimating  star formation time scales, 
because 
\hii\ gas is by itself subject 
to displacements
produced by  stellar winds or
simply
caused by the gas escaping from the cluster boundaries if  the \hii\ region 
is optically thin (sect.\ \ref{s:halpha}).

 Moreover,  the 
positions 
of most of the clusters  in NGC\,1386 usually 
coincide
with regions of low $A_V$ in the  dust extinction 
map  (Fig.\,\ref{image_dust}). The dust in the region also resolves 
into 
filaments  (sect.\ \ref{s:CO_mass} and Fig.\ \ref{image_V}), 
and these  coincide in morphology and location  with those in CO,  as seen in  
Figure \ref{CO_contours}. 
The
clusters are 
therefore 
equally offset from the dust  filaments as from the molecular ones.

Putting NGC 1386 ring  in the context of  the galaxy scale  
shock-wave 
theory, it is interesting to note the curvature of the 
molecular/dust filaments towards the nuclear region,  reminiscent of that followed by spiral arms 
on 
galactic  scales, but  
more concentric in the ring.  Were  the filaments  the consequence of the passage of a density wave through the disc of the galaxy, 
as proposed for spiral arms,  leading  to the compression of  molecular gas in the filaments, 
an estimate of the delay between the peak 
compression and the peak of star formation could be inferred from the current clusters location,
on the assumption that the  cluster ring arises from a previous shock compression.  This delay would be 
less than 2--3  Myr, the  lowest age   dated in the ring.
On a speculative 
basis, a typical cluster--filament separation of 100 pc (deprojected), implies a differential velocity with respect to the gas 
in the disc of $\sim 50\ {\rm km~s^{-1}}$.  The CO kinematics  is consistent with rotation \citep{ramakrishnan+19}  but there are 
structural residual velocities 
as large as $40 - 60\ {\rm km~s^{-1}}$,
after subtracting the rotation component (fig.\ 2 in \citet{ramakrishnan+19}).
These residuals   could be indicative of the level of streaming motions expected in the density-wave scenario
 \citep{roberts69, tully74iii, tamburro+08}.  

On the other hand, the age of the youngest clusters and their ubication  in 
free CO regions, already imply a  very high star formation efficiency or a short time scale for the removal of the parental molecular gas, 
of $ < $ 3 Myr.  In the later case,   the evacuation of this   gas had to be driven by the first stellar winds,  prior to the supernova phase 
\citep{krumholz+14}. 
A slightly longer time scale, $\sim$\,5\,Myr, is estimated  for the clearing 
of the molecular gas
in \hii\ regions in spiral arms 
\citep{kruijssen+19Nat}. 
In this case, this is measured from the molecular  spatial offset   with respect to the \hii\ gas, whereas we measure it with respect 
to the  location of the 
ionising 
star cluster, which as just mentioned may lead to larger time scales in the former.

\subsection{The connection: SFR in the cluster
ring,  free-fall time of the clouds molecular ring}\label{s:sfr}

The  peak of cluster formation in the ring is in the 3--4.5\,Myr range (Fig.\,\ref{allparam}). Clusters with these ages  are found at any location in the 
ring, meaning that star  formation  occurred in a synchronised manner over the 2 kpc ring diameter. This might suggest  
a large-scale event in the 
disc 
triggering star formation. The star formation surface  density is very different from place to place in the 
ring, with 
some sectors 
being 
 devoid of clusters  
and 
others  showing  a high cluster density
(Fig.\ \ref{image_V}).  This variation is presumably due to the availability of  molecular clouds  of suitable  mass and density. 
As discussed in Section\,\ref{s:CO_mass}, the absence of clusters in certain areas, even at near-IR wavelengths, is presumably not caused 
by dust. 
$A_V$ values of tens of magnitudes  are required to obscure clusters 
such as those currently seen  in the $K$\,band.

Accordingly, a  global estimate of 
the 
star formation rate in the ring, i.e.\ normalised to the surface density,   may not be  
representative. 
Instead,  we estimate the SFR in the  ring directly from  the clusters individually. 
If we sum up the total mass in star clusters
($5\times 10^5$ \msun )
and use a burst  duration of 2\,Myr, as per the width of 
the peak of 
the age histogram in Fig.\,\ref{allparam},
this  yields a global ${\rm SFR_{ring}\,\sim\,0.25\,M_\odot\,{\rm yr}^{-1}}$.
Were this value normalised to the ring area, the SFR surface
density would be factor two larger
($\sim$\,0.4\,M$_\odot\,{\rm yr^{-1} kpc^{-2}}$)
which is  unrepresentative given the sparse   distribution of clusters in the ring.

Conversely,  the SFR at the cluster level  should be   more representative of the event as a whole. Using the average cluster mass
$\sim 5 \times 10^3$ \msun )
and  the same burst time duration, the SFR is obviosuly much lower
($ {\rm SFR_{cluster} \sim 2.5 \times 10^{-3}  ~ M_\odot yr^{-1}} $) -- two orders of magnitude 
below  
the  global one.

In comparison,  the SFR at the cluster level in the circumnuclear cluster 
ring of NGC\,1097
(SFR$_{\rm cluster}$)
is an order of
 magnitude 
higher. This is driven by the   average cluster mass in NGC\, 1097, 
which is
  an order of magnitude higher. The  burst duration, 
sharply defined   
in this case,  is   also $\sim$\,2\,Myr \citep{prieto+19}. 

Putting these numbers in context, the Milky
Way  SFR 
throughout 
the disc is  much higher,  1.65--1.9 ~ ${\rm M_\odot yr^{-1}} 
$\citep{chomiuk+povich11, licquia+newman15}, presumably because of the 
widely varying 
scale of the regions  considered:  the Milky
Way disc 
compared with  the kpc ring in NGC\,1386 or NGC\,1097. Yet, when comparing with the SFR 
in
individual Milky
Way \hii\ regions, this  
is in line with the SFR  at the cluster level in the circumnuclear rings: 
in the 
$10^{-3}$--$10^{-2}~{\rm M_\odot\,yr^{-1}} $ range
for  cluster
mass in the few $10^3$--$10^4$ \msun\
range, as in NGC\,1386 and NGC\,1097, respectively \citep[e.g. 
table 2 in][]{chomiuk+povich11}.   
Thus, in  a rather different environment, as  
is the case 
for
these circumnuclear 
star-forming 
rings, the SFR at the cluster level  
is the same 
as 
for 
MW \hii\ regions for the same  stellar mass. 
The SFR at the cluster level may even decrease  with  cluster  mass, were that to follow the trend shown by   \hii\  regions in \citet{chomiuk+povich11}.

Focusing on the molecular ring, at the rate of the SFR$_{\rm cluster}$,    and assuming that   the clusters had as a progenitor a molecular cloud of similar properties as those  of the point-like clouds  in the molecular ring,
 the depletion time of  any of these clouds, typical  mass of $10^6 $  \msun\ 
would be 
$\sim 4 \times 10^8\,{\rm yr}$, 
of the order of the average depletion time in galaxies, including
the Milky
Way  (e.g.\ \citealt{bigiel+08} and references therein; 
\citealt{evans+20}).

The stability criteria of the disc, quantified by the Toomre  parameter $Q$ \citep{toomre64}, yields a value  of about  
unity \citep[following 
the thick disc aproximation,][]{behrendt+15}, indicating that the disc is prone to fragmentation with a time scale in the range 1--10 Myr.  In estimating Q, we consider 
 a  $10^8$ \msun\ disc  of  1 kpc radius, rotating  with 
${\rm V \sim 200 ~km~ s^{-1}}$  
\citep{ramakrishnan+19}, surface density $\sim 10^2~\mathrm{M}_\odot ~ \mathrm{pc}^{-2}$  and  
sound speed
$10~\mathrm{km~s}^{-1}$.


In comparison with the 
clouds' 
free-fall time, this is also about 10 Myr  for  a cloud 
of
radius <$\sim$ 20 pc (sect. \ref{s:CO_mass}). 
Interestingly, these time 
scales compare  with   the  orbital time of  the  ring
($\sim 20$ Myr   
for  ${\rm V \sim 200 ~km~ s^{-1}}$).  
 One could think that    
 in about an orbital time,  the disc may have time to fragment, perhaps due to the passage of a density wave,   and even collapse 
into  
 clouds,  as those   seen regularly distributed in the ring,  to     eventually  start  delivering at locations somewhat offset from the peak compression, a new cluster population  in a synchronised manner in the ring.  The clouds in the ring have similar properties
in terms of    mass,  size
and kinematics - sigma dispersion.  One may 
expect the new cluster population 
 also
to 
 span a narrow range of properties in terms of  ages, masses
and 
SFR, as 
appears to be the  case of the 
current cluster
ring.   Given that the current burst  had
its peak $\sim$ 4.5 Myr ago,   in  a   time  frame of 
10 Myr
one may expect a new   cluster
 ring  in the next 5 Myr or so.  

In anticipating 
a 
future 
generation of clusters,
it is logical to ask for the detection of previous ones, 
i.e., older clusters  
in the region. 
We 
detect 
no such 
population. Still, we detected two clusters,  dated  10 Myr, which are among 
the most massive, and 
obviously brightest,
in the sample (Fig.\,\ref{allparam}).  Three additional clusters, only detected in the  $J$
and $K$
bands, and thus not dated 
(Table\,\ref{phot_n1386}),  may also 
signal
old ages 
since 
the infrared contribution increases with age (Fig.\,\ref{age_evol}). The issue  in detecting older generations  is the foreseen 
 low mass of the  
clusters,
  together with a dominant contribution of the bulge light  in the infrared.   For comparison, in the circumnuclear 
ring of NGC\,1097,  several 
cluster
generations
 spanning a 4--100 Myr period are distinguishable, yet
the clusters are  
an 
order of magnitude 
more massive.

\section{Overview and Conclusions}\label{s:conclusion}
This paper 
focuses on the spatial and temporal star formation process in the circumnuclear 
star-forming ring  
in an 
early-type 
spiral galaxy. The   target under 
study, NGC\,1386, was selected because of   its moderately populated
nuclear cluster
ring, with 
approximately
61 clusters   spatially 
isolated  
at 
optical and near-IR wavelengths. Equally,  its associated   molecular ring is resolved 
into 
strings of  point-like clouds, 
of which 
 $\sim$ 70 are counted in CO(2-1).
Observations 
on 
scales of several  parsecs 
enable us to  temporally separate 
some  of the evolutionary
 phases of star formation at 
the 
star cluster
and  molecular cloud
level.
\\

{\it The cluster
ring}
\\
The  cluster
population is  homogeneous in terms of  mass
($\sim 5 \times 10^3$ \msun ), age
(peak $\sim$  4.5 Myr),  
size
($\lesssim 8$ pc FWHM), 
 moderate extinction
($A_V \sim 0$--3) and  SFR
($\sim  2.5\times 10^{-3}~ \rm{M}_\odot yr ^{-1}$,
comparable to that of \hii\  regions in the 
Milky Way; Fig.\,\ref{allparam}, sect. \ref{s:sfr}). 

The analysis of the Lyman photon budget of the clusters  unveils  a high 
Lyman leakage, in the 
80\%--90\% 
range for most of the
 population, 
which implies
that the associated \hii\ regions are optically thin. 
High Lyman photon 
leakage
has 
been  reported in the literature 
for 
different environments,
from \hii\ regions in  spiral 
arms to  star  clusters in 
galaxies (sect. \ref{s:halpha_others}),  and in circumnuclear 
star-forming 
rings  as the case of NGC\,1097, or as 
or as reported in this paper for
NGC\,1386. 
The leaking 
may be more common than previously thought
and set 
as 
a warning for the  use of \hii\ gas as a tracer of  
star formation 
on  
scales of at least several hundred 
parsecs. 
\\

{\it The molecular ring}
\\
The molecular gas is  resolved 
into  
long, collimated  filaments that 
circle around and eventually spiral into
the centre. The whole structure  
defines a  circumnuclear ring    somewhat detached from the  cluster
ring. The filaments   resolve into clouds
$<40~\rm{pc}$ FWHM  in size
of
mass
$\sim 10^6$ \msun,    comparable to those  of the Milky
Way.

There is a close correspondence
in morphology and 
spatial location
between the  
CO filaments and the equivalent dust filaments 
in the region (Fig.\,\ref{CO_contours}). The  clusters are 
spatially  displaced   from these 
filaments, most being located 
in   CO voids or at the edges
of the filaments.   
This displacement, seen at any 
location in the disc, argues for a net spatial separation rather than a projection effect. 

The  filamentary
structure   appear  as a   
well-defined  
entity,   as judged from  their  preserved 
collimated morphology over 
kpc scale. 
The average offset  between the clusters ring and the filament
ring, peak to peak,   is $\sim$ 100 pc (deprojected).  
We 
suggest 
the possibility 
that these offsets  could be due to the  occasional drift produced after the passage of a density wave through the disc, 
as postulated in the
density-wave 
theory for similar molecular--\hii\ gas offsets  in  spirals (sect.\ \ref{s:COvoids}).  Speculating on this 
possibility,
a 100 pc offset
over the time scale of a cluster age of 3 Myr  yields  a   differential velocity for the drift of 
$\sim 50~ \rm km~ s^{-1}$,  which, coincidentally, is   of the order of the residual, not rotational, velocities inferred 
from the CO kinematics in the 
 disc (sect. \ref{s:COvoids}).

Because of observational limits, the molecular  ring 
is currently 
traced only  in  the central kpc. 
However, its 
spatially
associated
dust
is  
followed  
out 
to the edge of the optical disc (Fig.\,\ref{image_dust} and \cite{2014MNRAS.442.2145P}), which  argues for 
their origin 
in
the outer 
part of the galaxy or 
even outside of it. As possible driver of the material to the centre, we do  not see any evidence  for a bar in this galaxy,    and \citet{2015ApJS..217...32B} classify it as SA. Nor there are indications for a companion galaxy as potential donor of the material falling into the center. Yet,  NGC 1386 is in the Fornax cluster, and the high density  environment in which it resides \citep{2019A&A...627A.136I} could be a possible source, or cause, of the infalling material.

 Independently 
 of the cause of the molecular--cluster spatial offset, the fact that some of the youngest clusters are  
in CO   
dust-free  
regions, 
  points towards 
a rapid parental-cloud evacuation phase
on time scales 
of 
less than 2--3 Myr. Thus, the  evacuation  has to 
be driven by the collective 
effect of winds from 
first-generation
 stars.
\\

{\it Their connection}
\\
We interpret the 
point-like 
CO 
clouds in the filaments as  
locations where a new generation of clusters is  in the 
making. The stability 
of the $10^8$ \msun\  molecular disc against collapse, measured by the Toomre 
$Q$ 
parameter, is close to unity. This sets a   
 time scale for  
fragmentation    in the 1--10 Myr range.  Subsequent collapse of the $10^6$ \msun\ mass  clouds by  free
fall implies a
  time scale 
of   $\sim$ 10 Myr;
interestingly, 
these 
 time scales are 
of 
the order of  the orbital time of the CO ring (sect. \ref{s:sfr}). 
Thus, we postulate that this  time scale may set the clock for the molecular gas
to fragment and  collapse into the molecular clouds 
that are 
currently 
seen
in the filaments, and that share similar properties.
Given the relatively periodic distribution of clouds in the molecular ring, it is expected that the 
anticipated 
new cluster
 population would equally 
be distributed 
in a ring, 
the clusters 
having
similar 
ages and presumably similar 
masses at any location in the ring.
The synchronised nature and homogeneous
properties of the current  cluster ring could be explained by a 
similar 
previous 
event.
Within the time frame of 10 Myr,
 and given  that the peak of star formation  was  $\sim 4$ Myr ago,
 a time delivery  
for  
the new generation of clusters in the region 
may be within 
the next 5 Myr.

We note that 
the inferred 
free-fall- and ring orbital- time, are longer that those 
inferred from the 
molecular--cluster spatial offset  if interpreted in the 
density-wave 
scenario (sect. \ref{s:COvoids}). The combination of 
multiple phenomena seen in the region implies a more complex scenario than that inferred 
from 
any one 
time scale. We still believe that a 
major event in the disc may have caused the 
onset of cluster formation  simultaneously in the ring. If that is a density 
wave, this  will 
naturally lead to  gas compression
and star formation at a location left behind in the disc after the 
passage of the wave.
 The induced compression could have accelerated 
star formation to limits bellow those suggested
by
free-fall.
\\

{\it Putting this ring in context} 
\\
Compared
with the prototype circumnuclear ring of  NGC\,1097, both NGC\,1386 and NGC\,1097 are  
early-type 
spirals and  
star formation 
prevails mainly
in  the central region. Their star formation history have many aspects in common.
 In the   kpc ring of NGC\,1386   a  4.5 Myr old single burst
is detected,  to be possibly joined by  a  second one in  
a few Myr.
The  kpc ring of  NGC\,1097 shows  a long burst (lasting 100 Myr), 
with  multiples bursts  separated by  quiescent periods of 10--20 Myr. 
 The current burst is 4 Myr old, an incoming  burst from its molecular ring
may be expected at
any time.
  
In 
a global context,  NGC\,1386 appears 
to be
a scaled
 down version of NGC\,1097, with a factor of four 
fewer 
clusters, in line with its much less 
massive molecular disc
($10^8$\,\msun\, vs. $10^9$\,\msun\,in NGC\,1097).
Accordingly, NGC\,1386 has a total mass in clusters and a SFR at 
the cluster level
an 
order of magnitude lower than NGC\,1097. Perhaps the difference in mass may be due to a very effective  driving mechanism, a bar, in the later. 
  Yet
in both cases their SFRs are in line with those in 
Milky Way \hii\ regions of similar mass. 
  Interestingly, 
 the star formation efficiency
(cluster-to-gas mass)
 in all cases
is roughly similar, in the  0.1-1\% range.

\section{Acknowledgements}\label{s:acknow}

We are thankful to J. Beckman
and 
B. Elmegreen for suggestions, and to M. Barendth for the Toomre stability calculations. AP thanks the CAST group of the Ludwig Maximilians University of M\"unchen  for comments, the Max-Planck Institute f\"ur Extraterrestrische Physik for their hospitality. Funding support: the National Autonomous University of M\'exico (UNAM) through grants DGAPA/PAPIIT IG100319 and BG100622 for GB; Spain I+D+i PID2020-114092GB-I00 for AP; the Spanish Ministry of Science and Innovation (MCIN/AEI/10.13039/501100011033), `ERDF A way of making Europe' and ``European Union NextGenerationEU/PRTR'' through the grants PID2021-124918NB-C44 and CNS2023-145339, and the European Union -- NextGenerationEU through the Recovery and Resilience Facility project ICTS-MRR-2021-03-CEFCA, for JAFO; the Excellence Cluster ORIGINS, funded by the German Research Foundation, under Germany's Excellence Strategy - EXC-2094-390783311, for AB and AP. 
Based on  ESO - VLT programs 076.B-0493, 084.B-0568 and 070.B-0409 and HST 5479, 6419, 7278 and 7458. This paper makes use of the following ALMA data: ADS/JAO.ALMA\#2016.1.01279.S. ALMA is a partnership of ESO (representing its member states), NSF (USA) and NINS (Japan), together with NRC (Canada), NSTC and ASIAA (Taiwan), and KASI (Republic of Korea), in cooperation with the Republic of Chile. The Joint ALMA Observatory is operated by ESO, AUI/NRAO and NAOJ

\section*{Data Availability}\label{s:datav}
The data will be made available upon request to the authors.

\bibliographystyle{mnras}
\bibliography{references}

\appendix

\section{SED fitting and parameters of NGC 1386 star clusters} \label{sec:appendix}

In this Appendix we present the results from our spectral fits, as well as the parameters determined for each cluster in NGC\,1386. Fig.\,\ref{fit_all} contains the observed SED of each cluster together with the best fitting model corresponding to the minimum \ch\ solution. The inferred values of age, mass, $A_V$ and 
\lha\ are listed inside each panel.
Clusters with the smallest error\footnote{We measure the error of a property as the semi-difference between percentiles 84th and 16th of its marginalised posterior distribution.} 
 in their properties are \#7, 8, 12, 14, 15, 17, 18, 42, 44, 53, with error(age) < 0.5 Myr, error(mass)/mass < 0.5 and error($A_V$) < 0.4.
Clusters \#29, 30, 46, 47 and 60 are not included since they have valid detections in only two bands, not enough for a reliable fit and parameter determination. 
\begin{figure*}
     \centering
     \subfigure{	\includegraphics[width=0.35\textwidth]{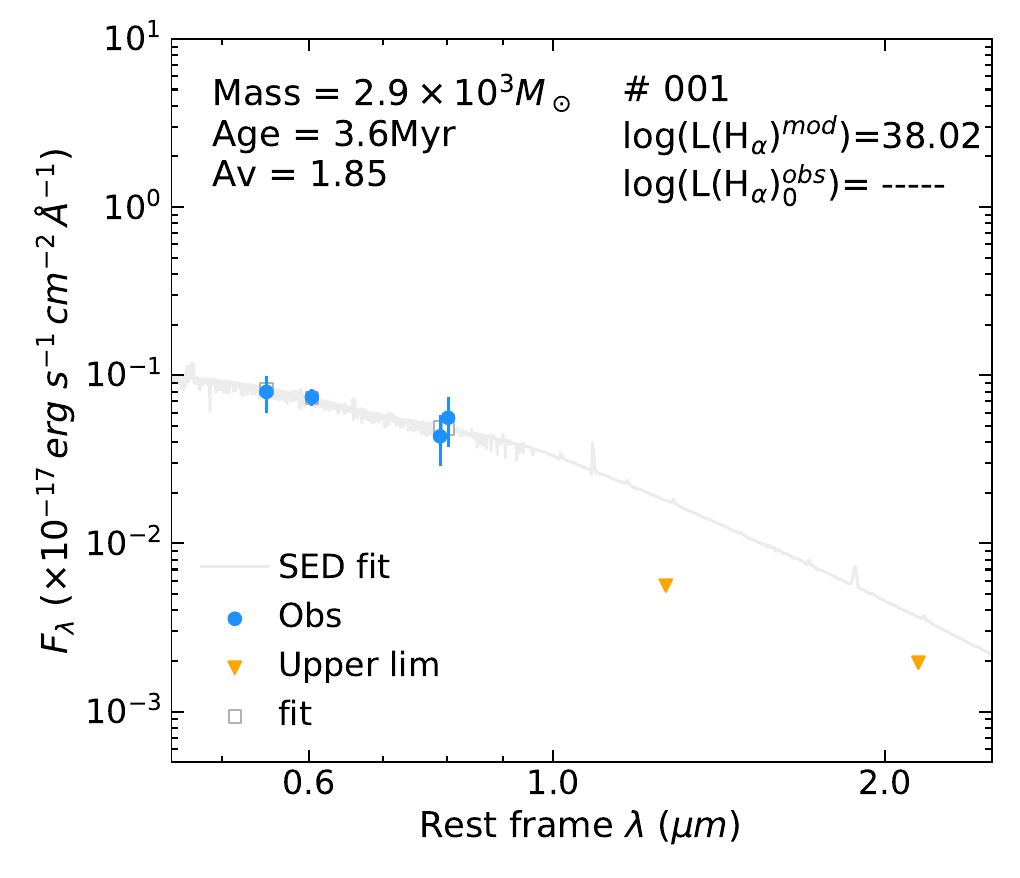}
     			\includegraphics[width=0.35\textwidth]{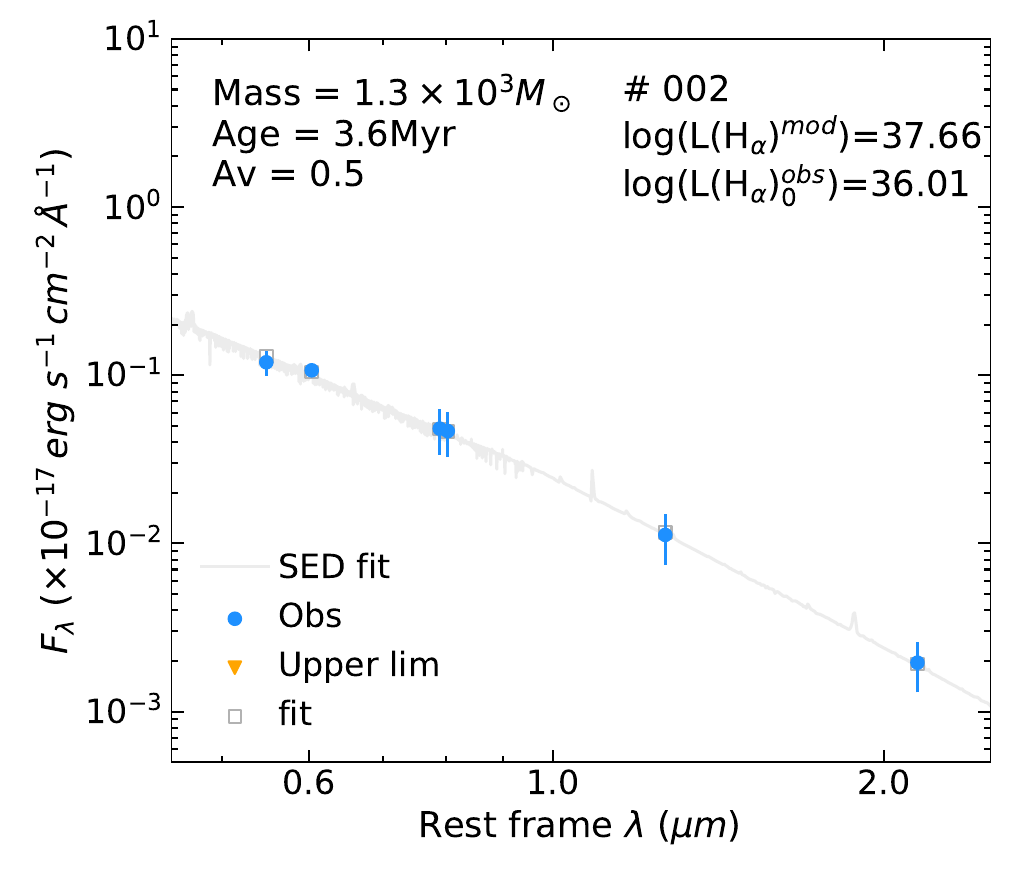}
			\includegraphics[width=0.35\textwidth]{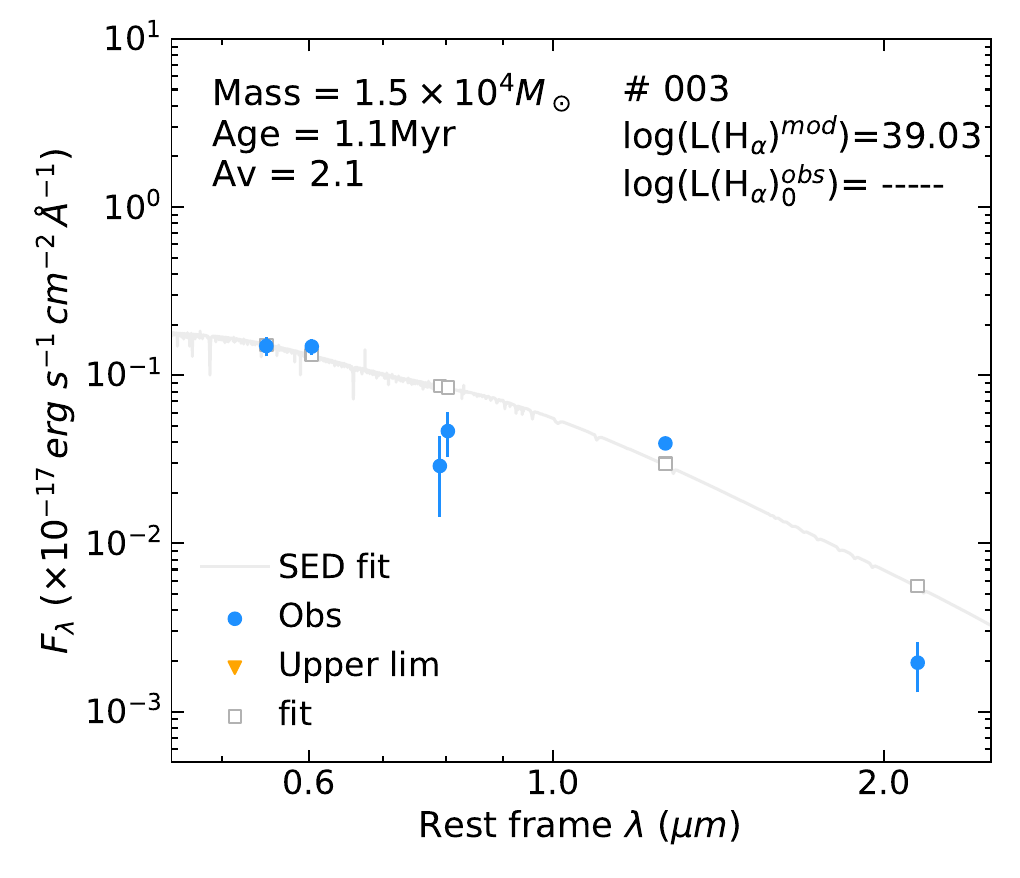}}
     \subfigure{	\includegraphics[width=0.35\textwidth]{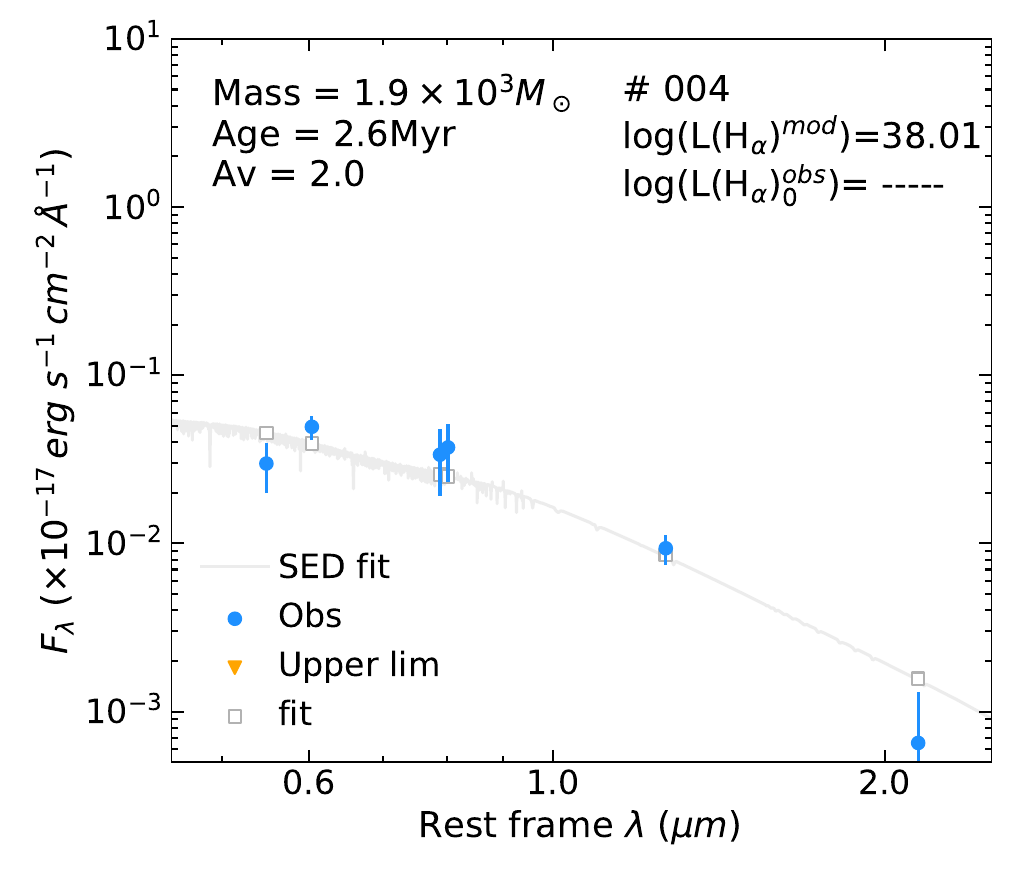}
     			\includegraphics[width=0.35\textwidth]{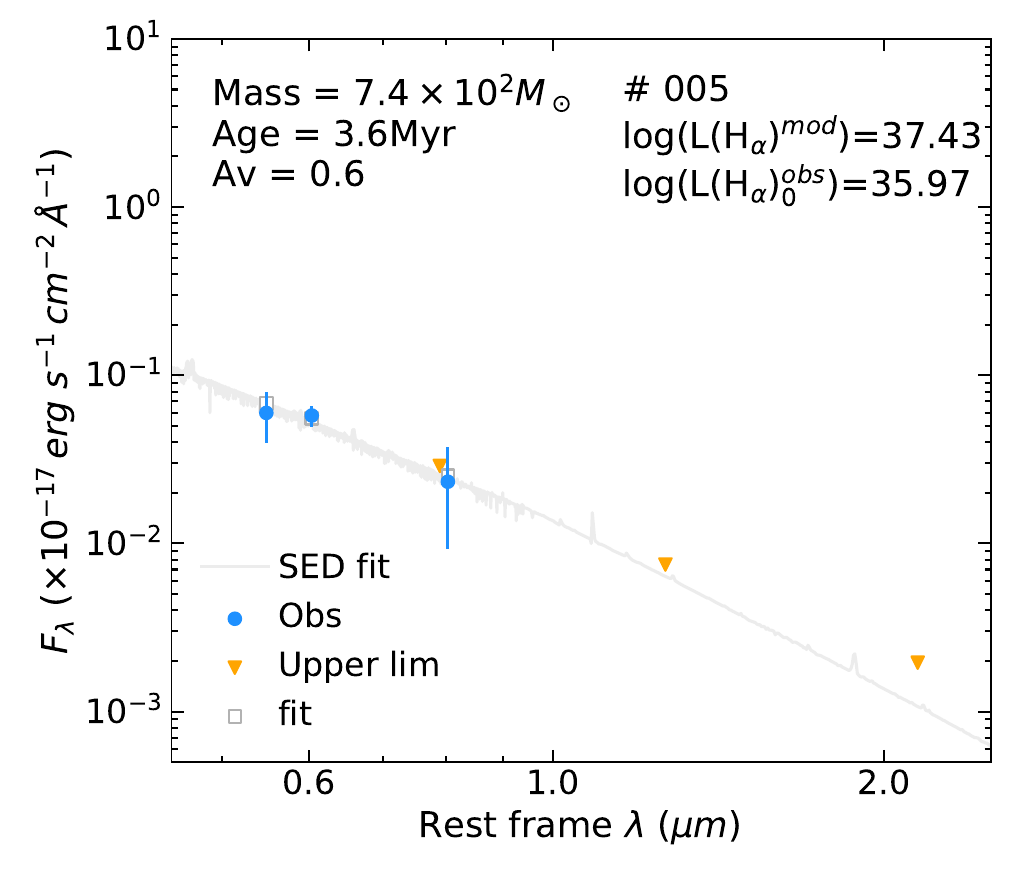}
     			\includegraphics[width=0.35\textwidth]{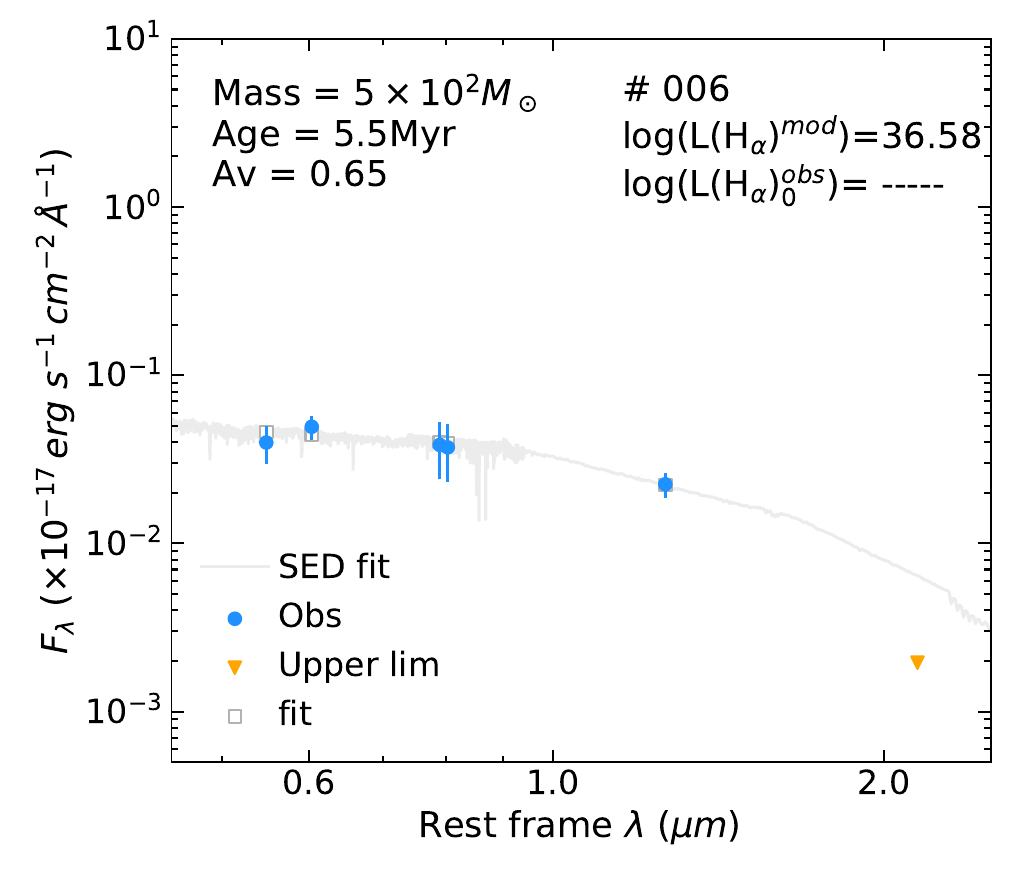}}
     \subfigure{	\includegraphics[width=0.35\textwidth]{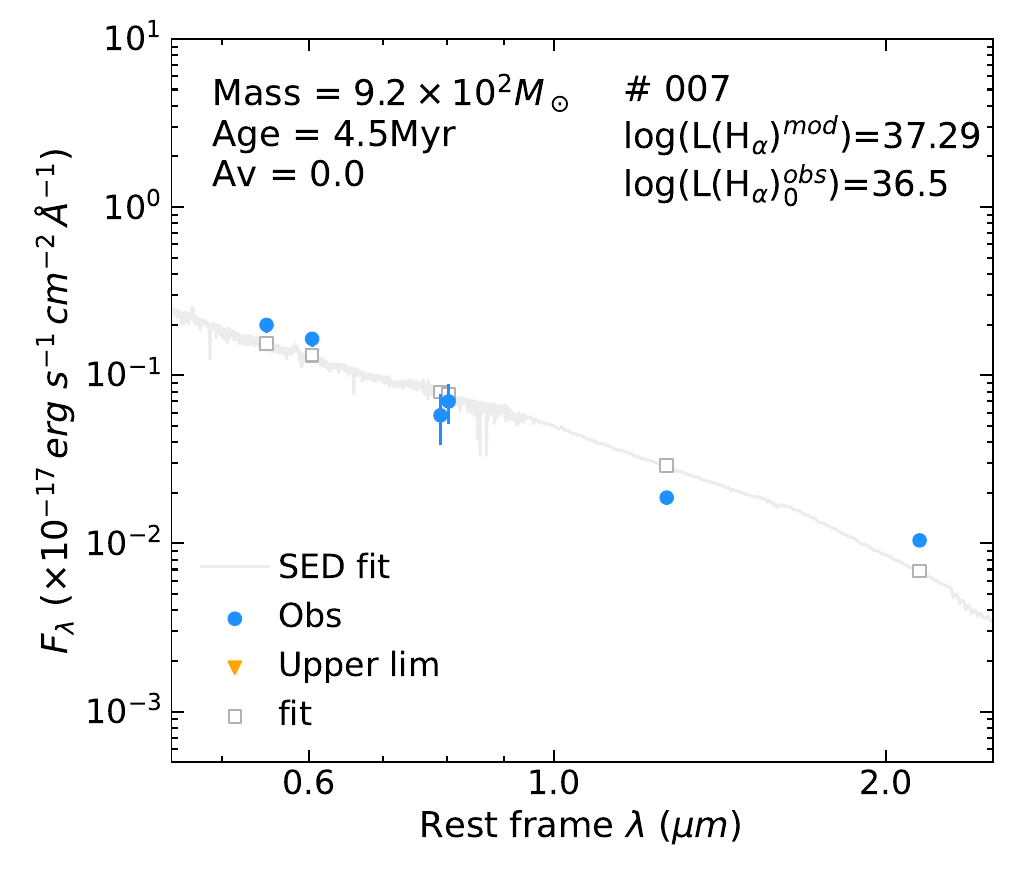}
     			\includegraphics[width=0.35\textwidth]{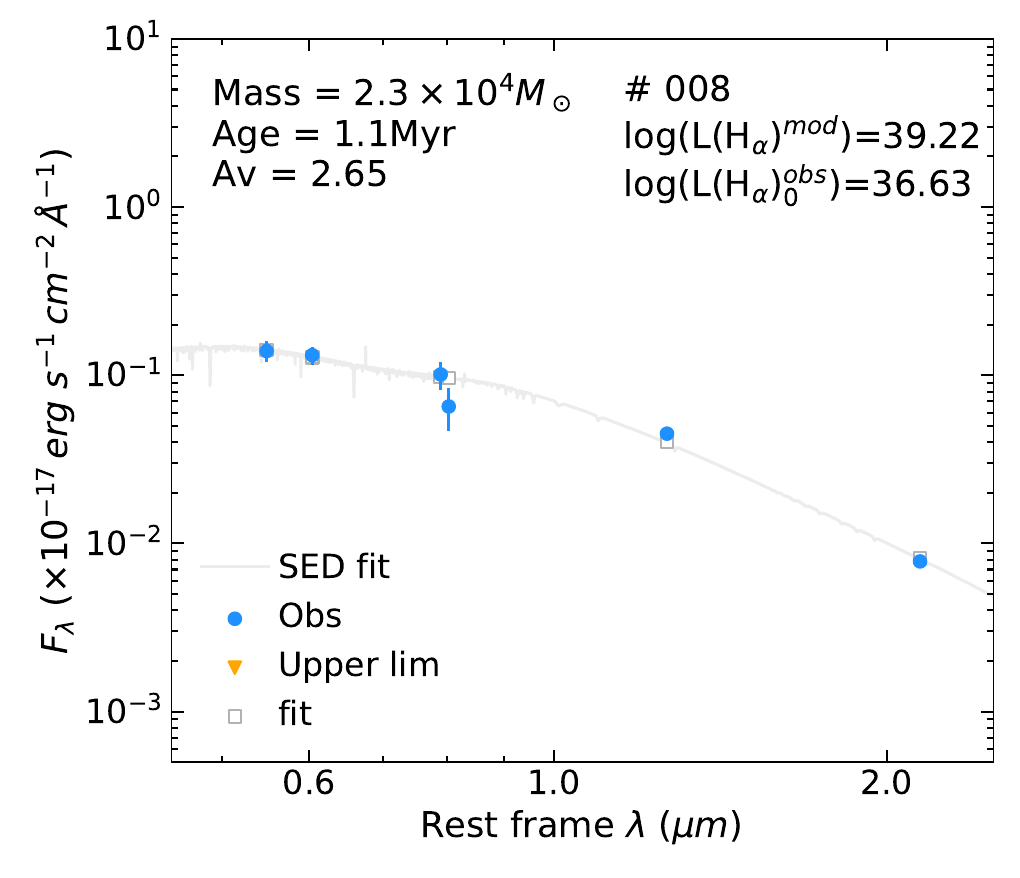}
			\includegraphics[width=0.35\textwidth]{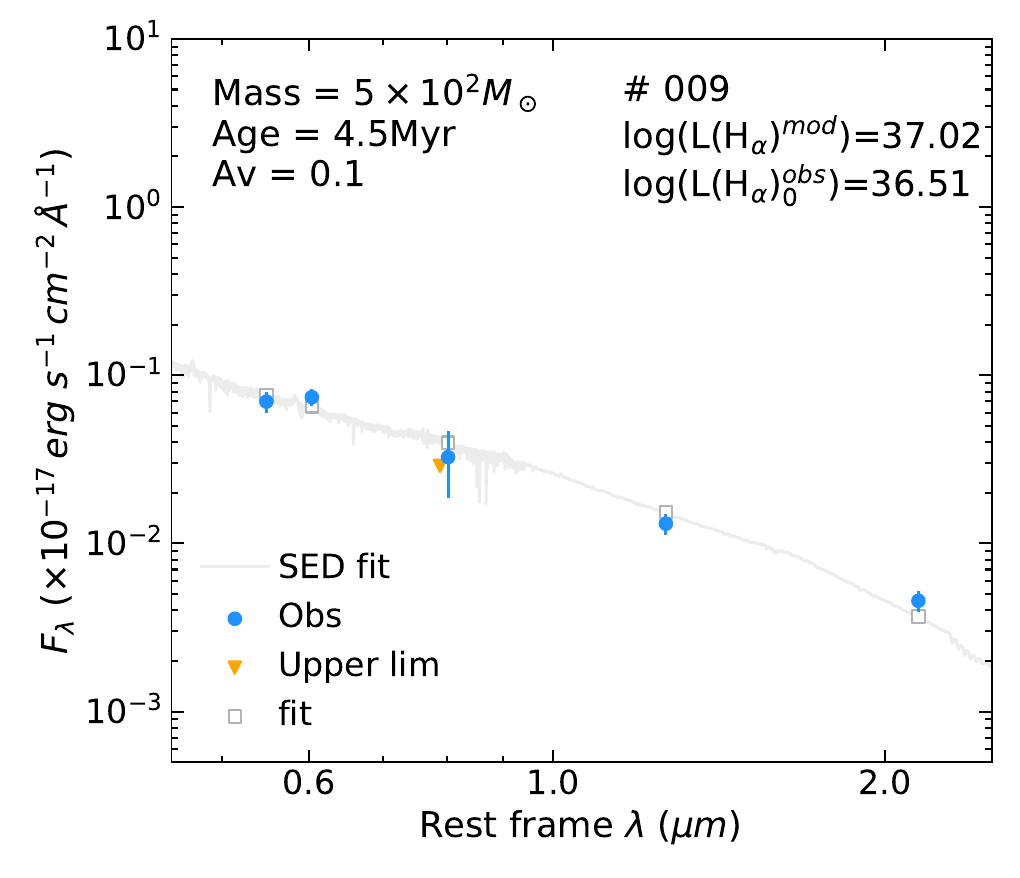}}
     \subfigure{	\includegraphics[width=0.35\textwidth]{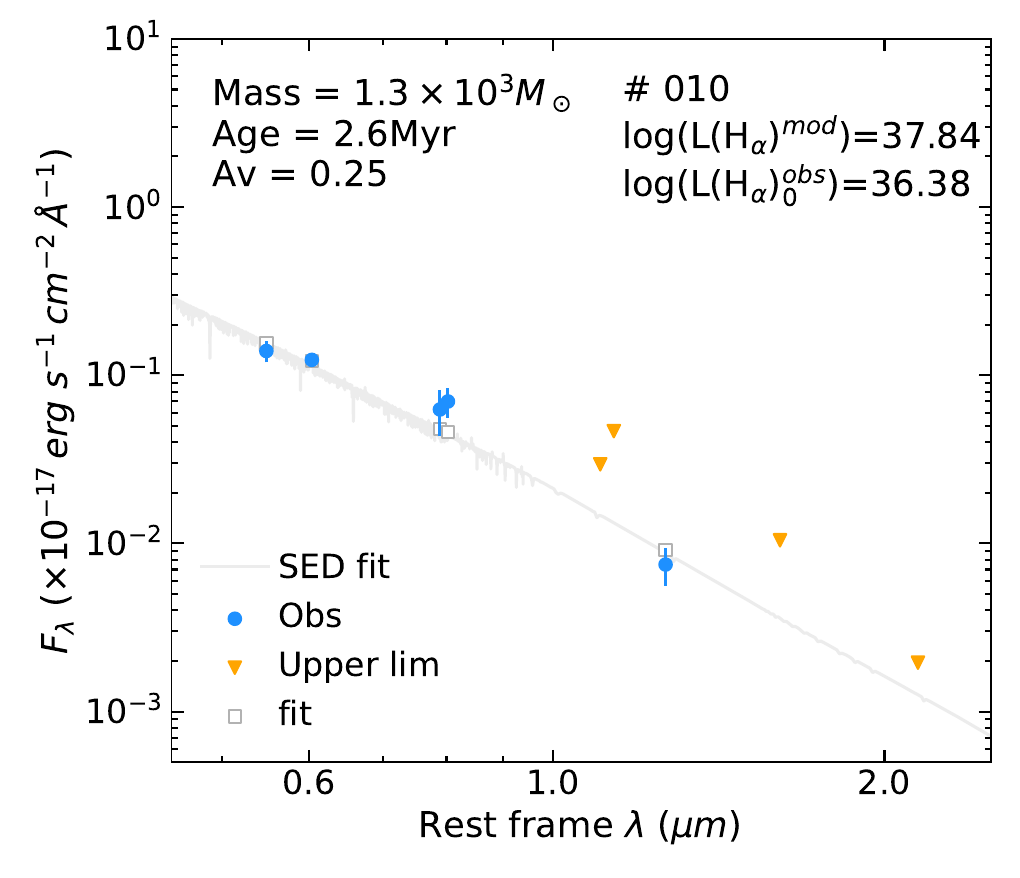}
     			\includegraphics[width=0.35\textwidth]{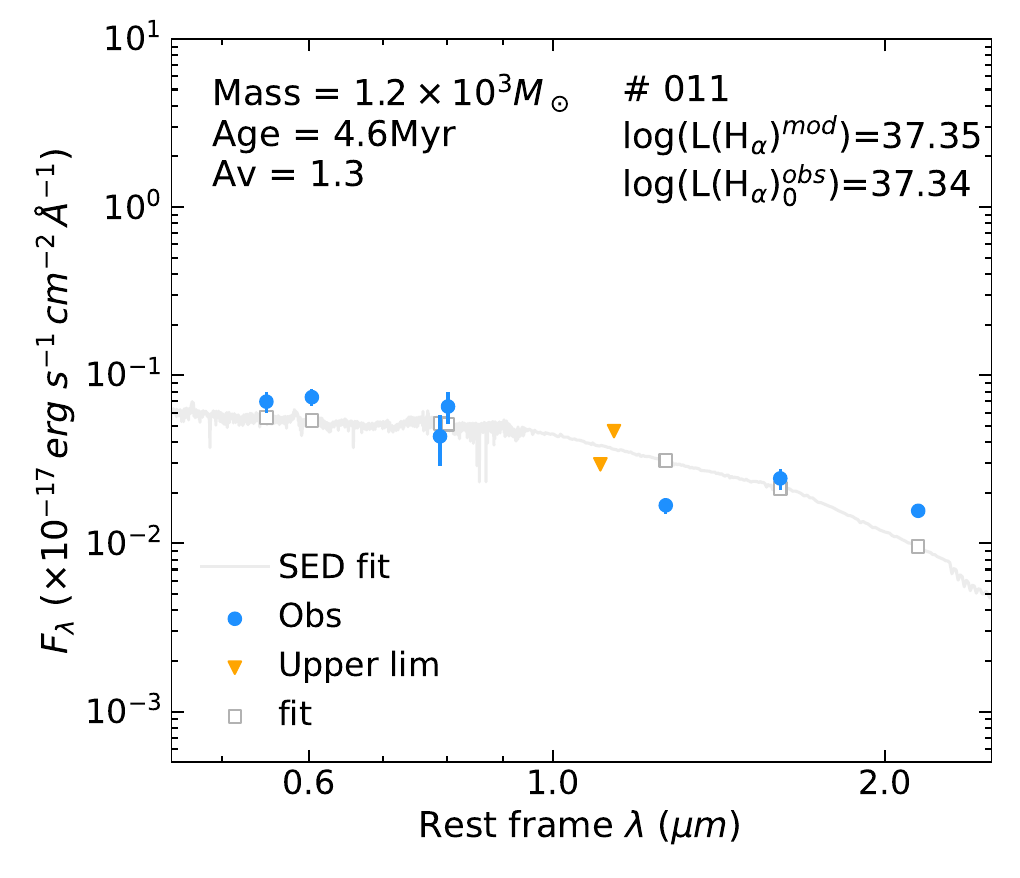}
			\includegraphics[width=0.35\textwidth]{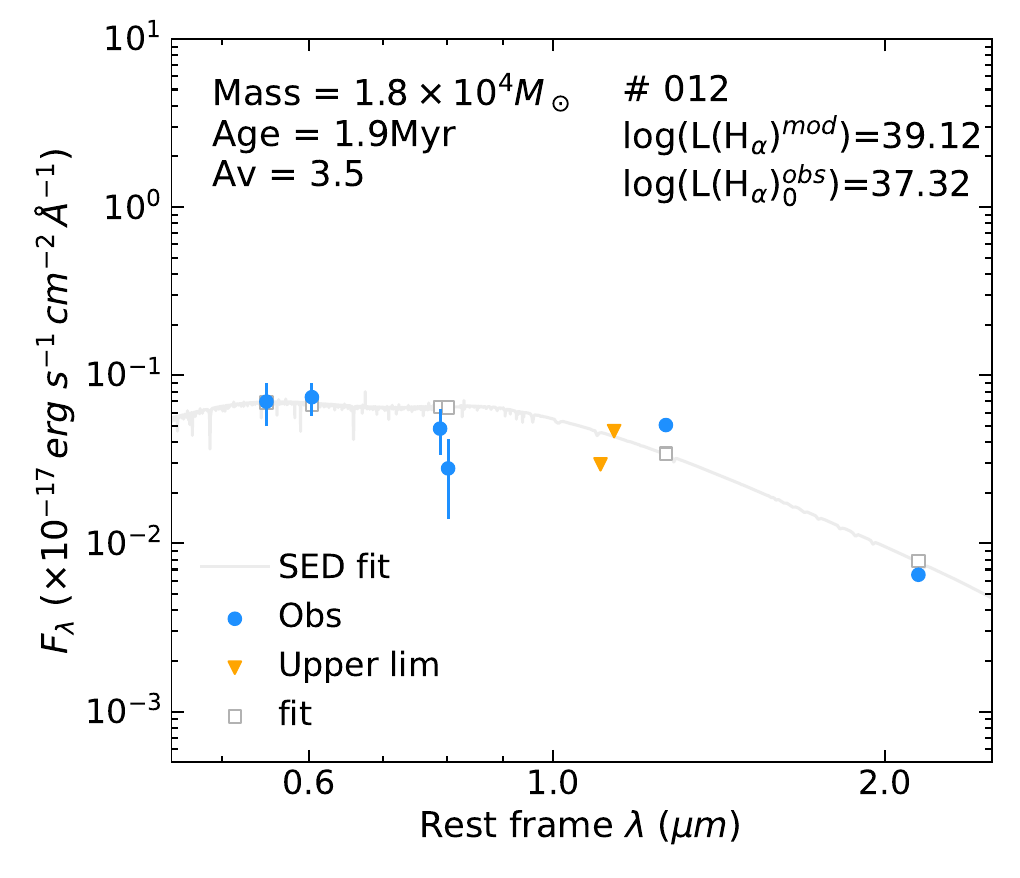}}
\caption{Results of the SED fit for each cluster. In the left panel, blue circles indicate the observed fluxes and orange triangles upper limit detections.
The gray open squares and the gray solid line indicate the flux at the observed bands and the spectrum of the best-fitting model, respectively, 
Model and observed \ha\ luminosity are listed in the panel.  
}
\label{fit_all}
\end{figure*}     

\begin{figure*}
     \centering
      \setcounter{subfigure}{5} 
     \subfigure{	\includegraphics[width=0.35\textwidth]{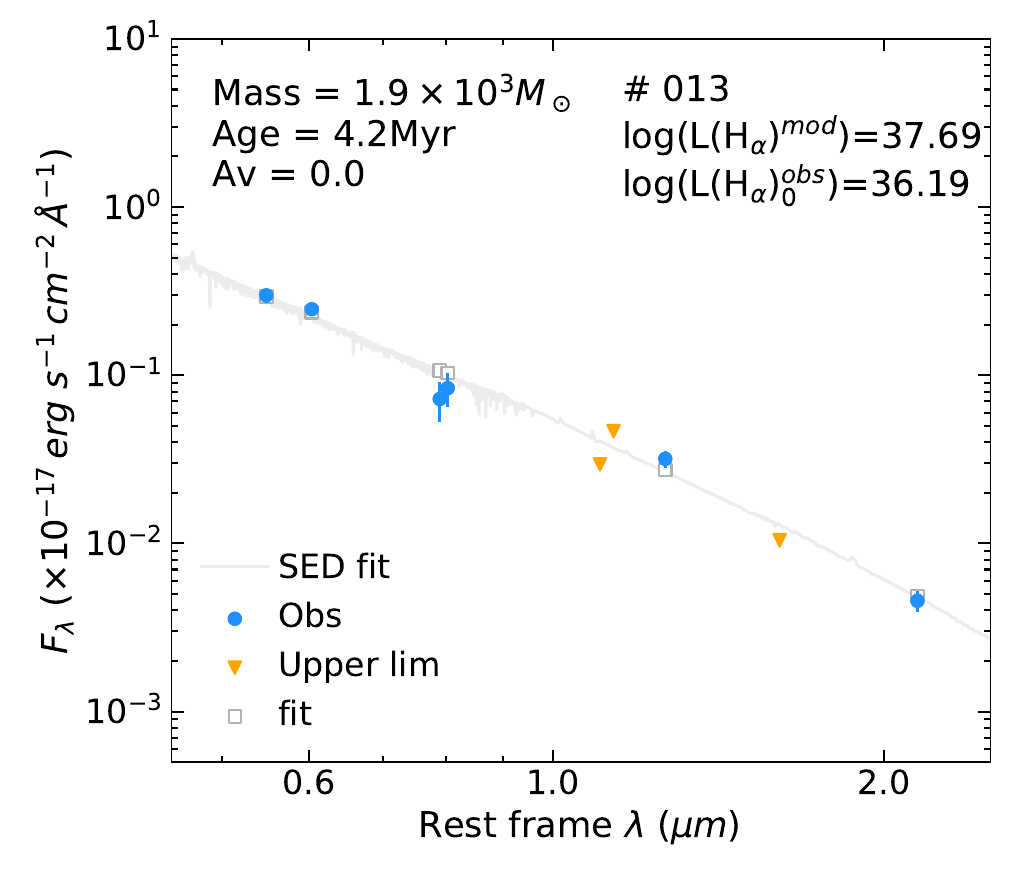}
      			\includegraphics[width=0.35\textwidth]{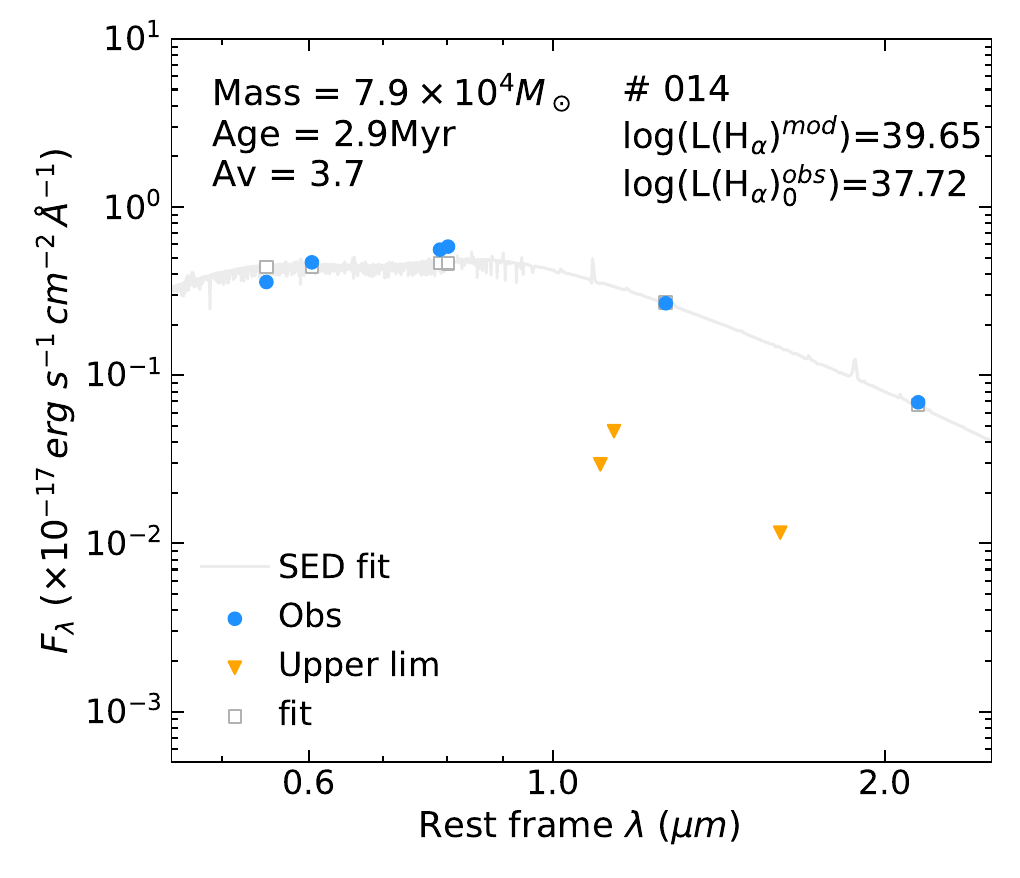}
			\includegraphics[width=0.35\textwidth]{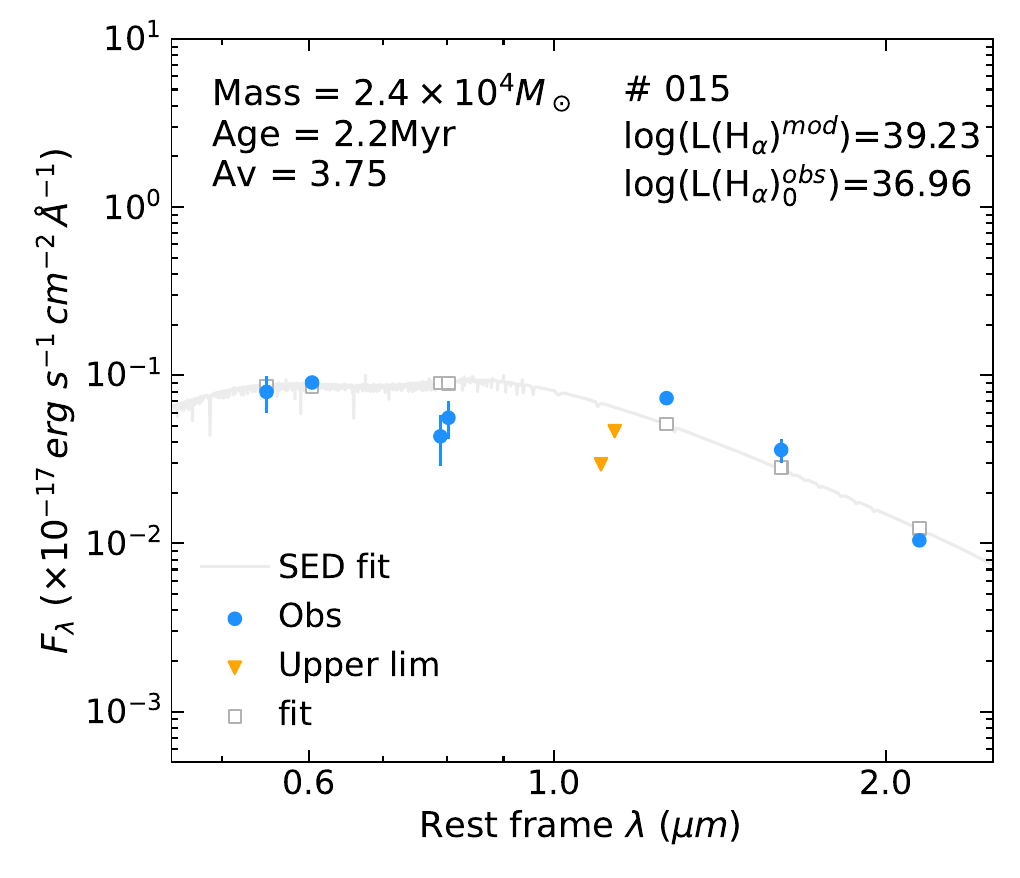}}
     \subfigure{	\includegraphics[width=0.35\textwidth]{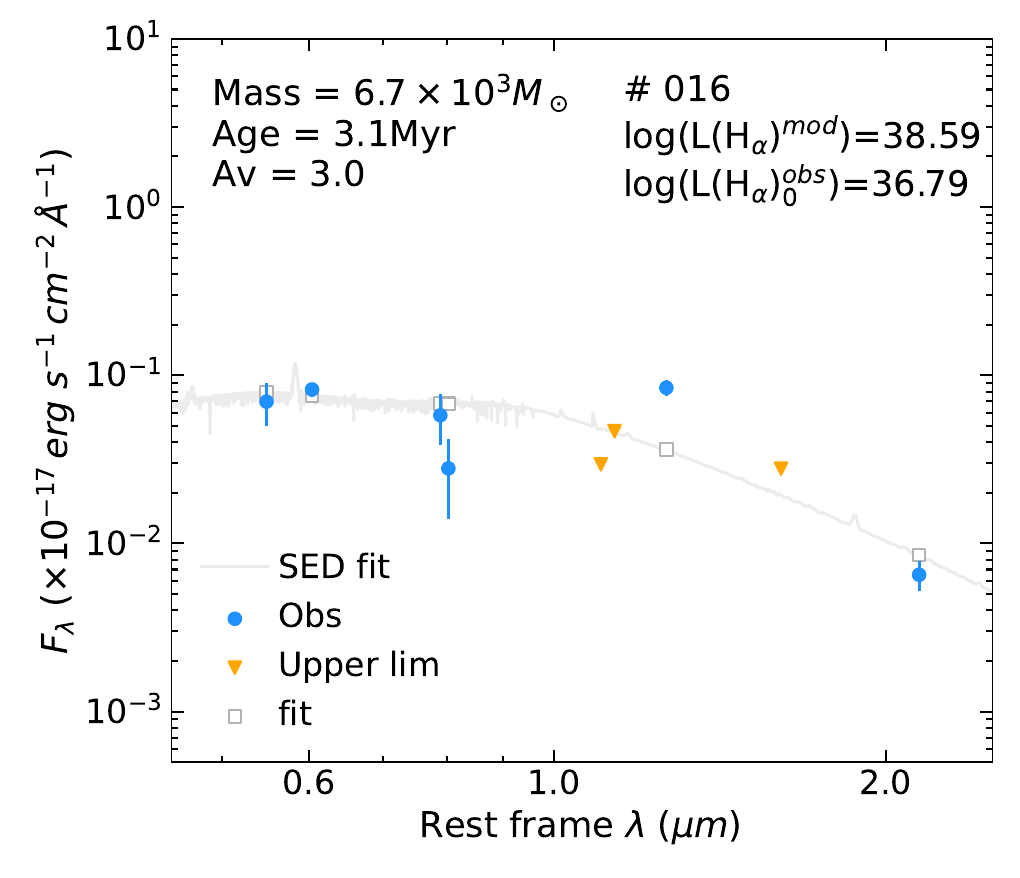}
      			\includegraphics[width=0.35\textwidth]{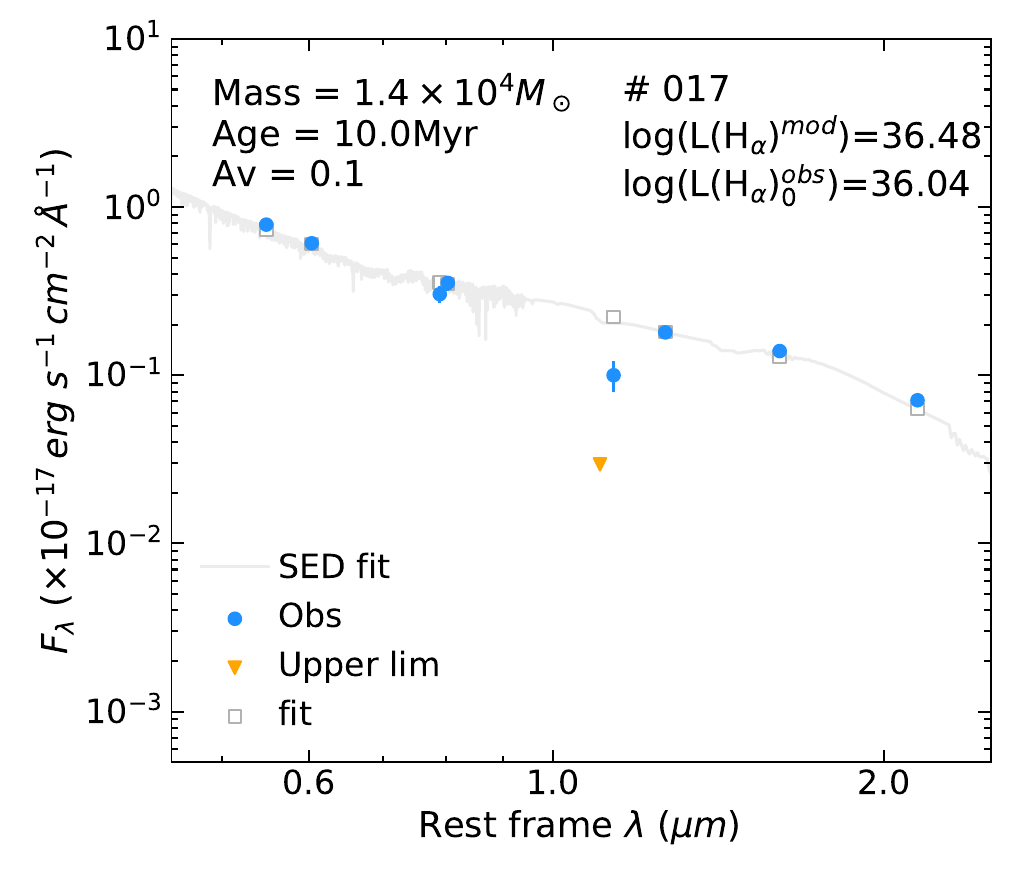}
			\includegraphics[width=0.35\textwidth]{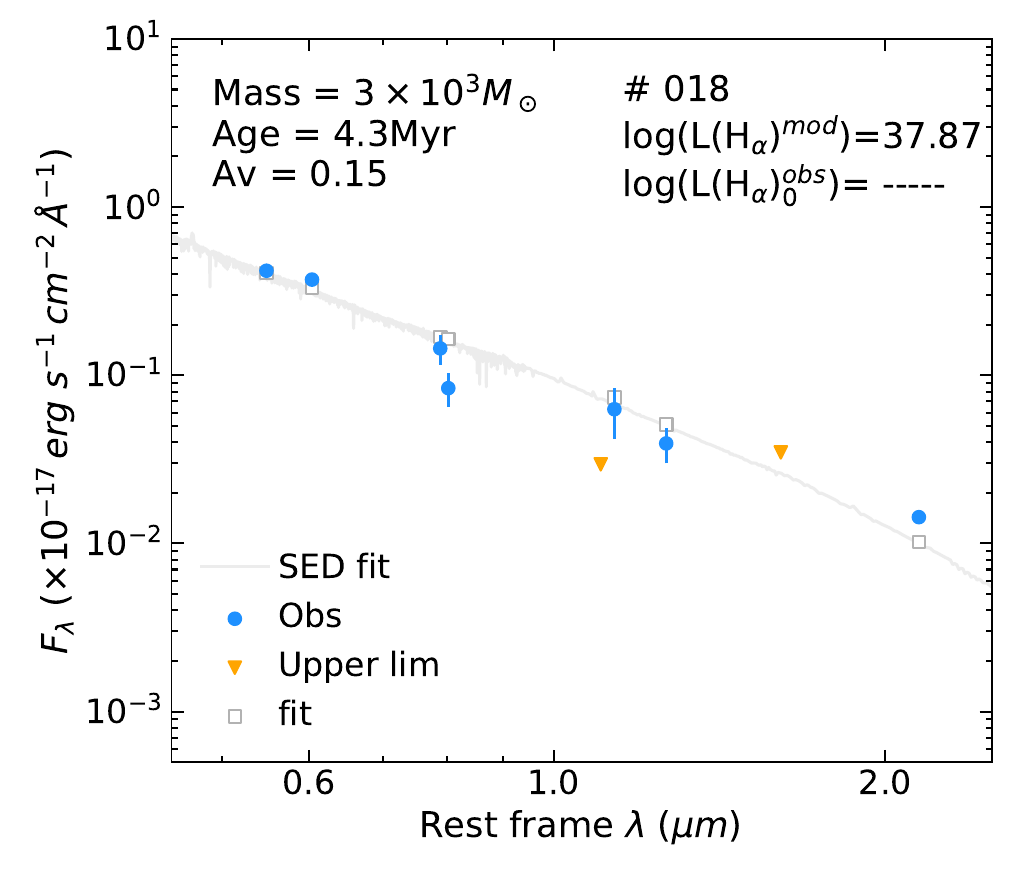}}
     \subfigure{	\includegraphics[width=0.35\textwidth]{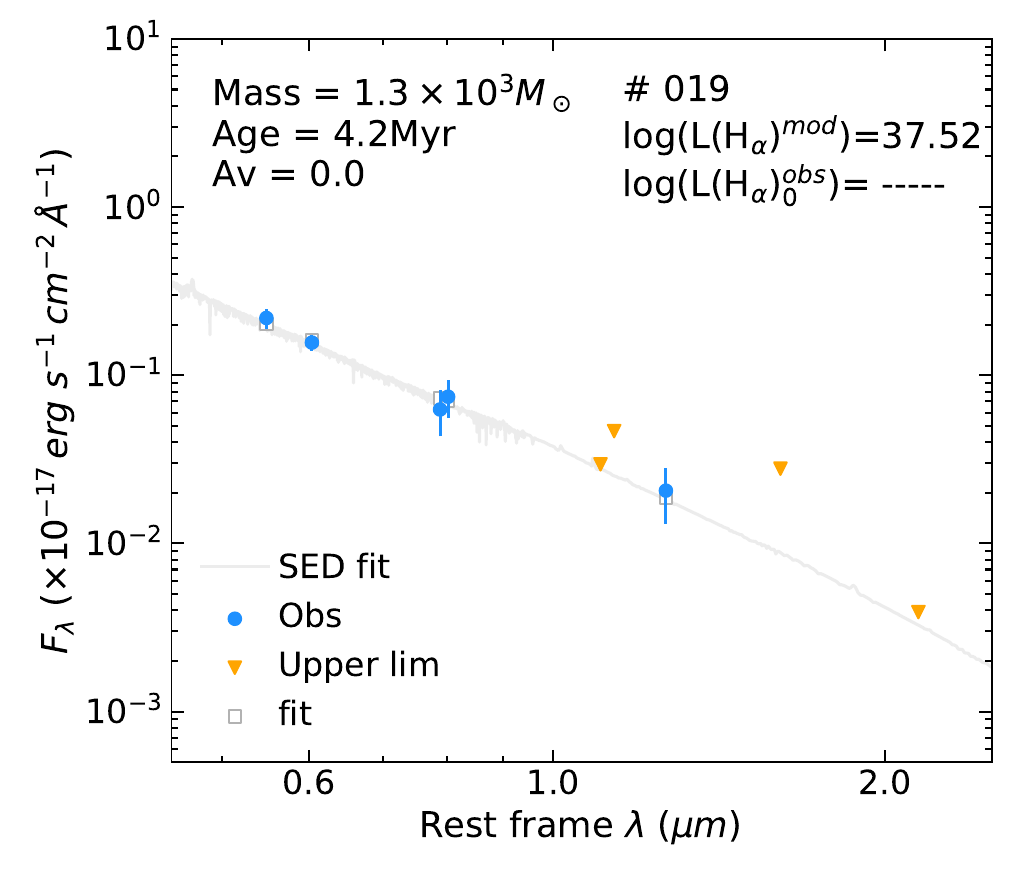}
      			\includegraphics[width=0.35\textwidth]{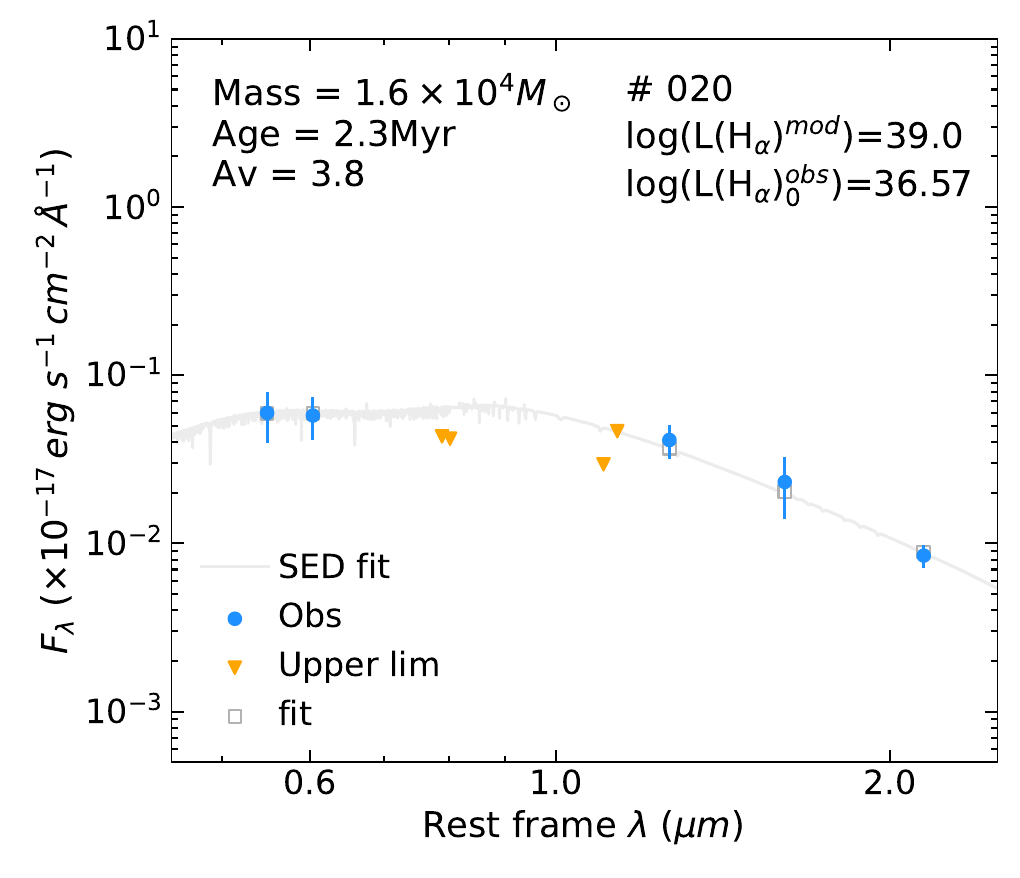}
			\includegraphics[width=0.35\textwidth]{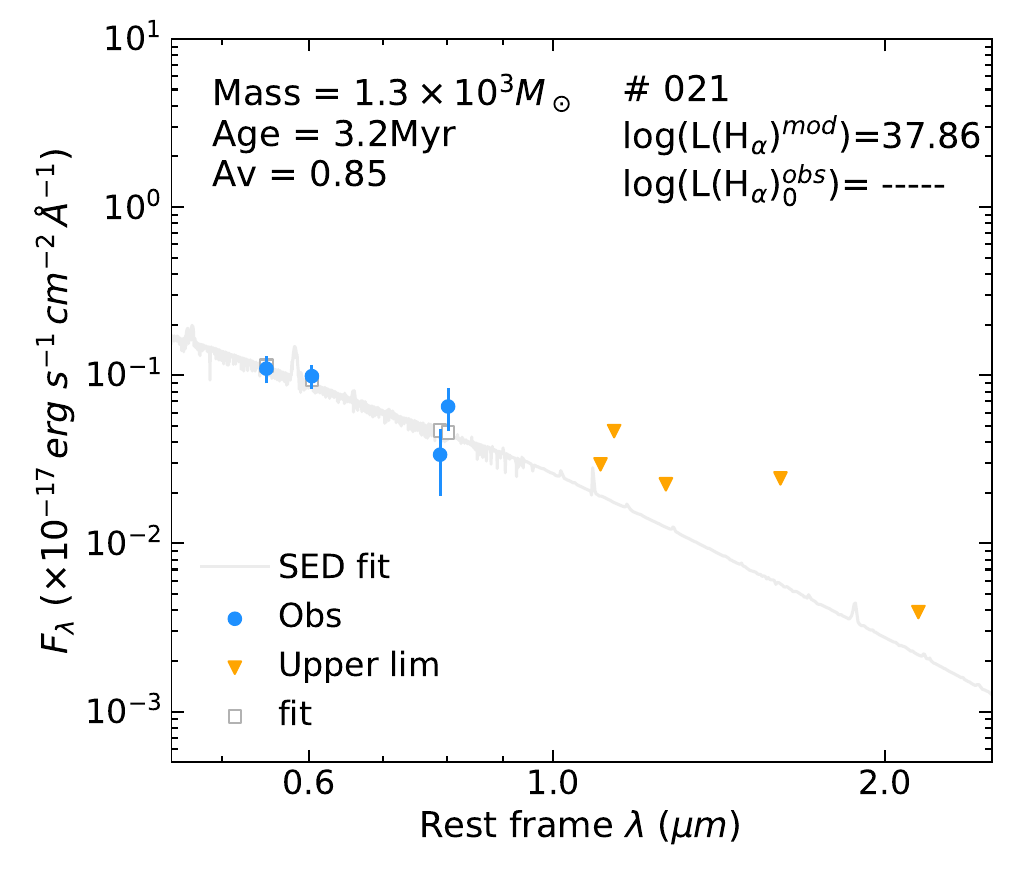}}
     \subfigure{	\includegraphics[width=0.35\textwidth]{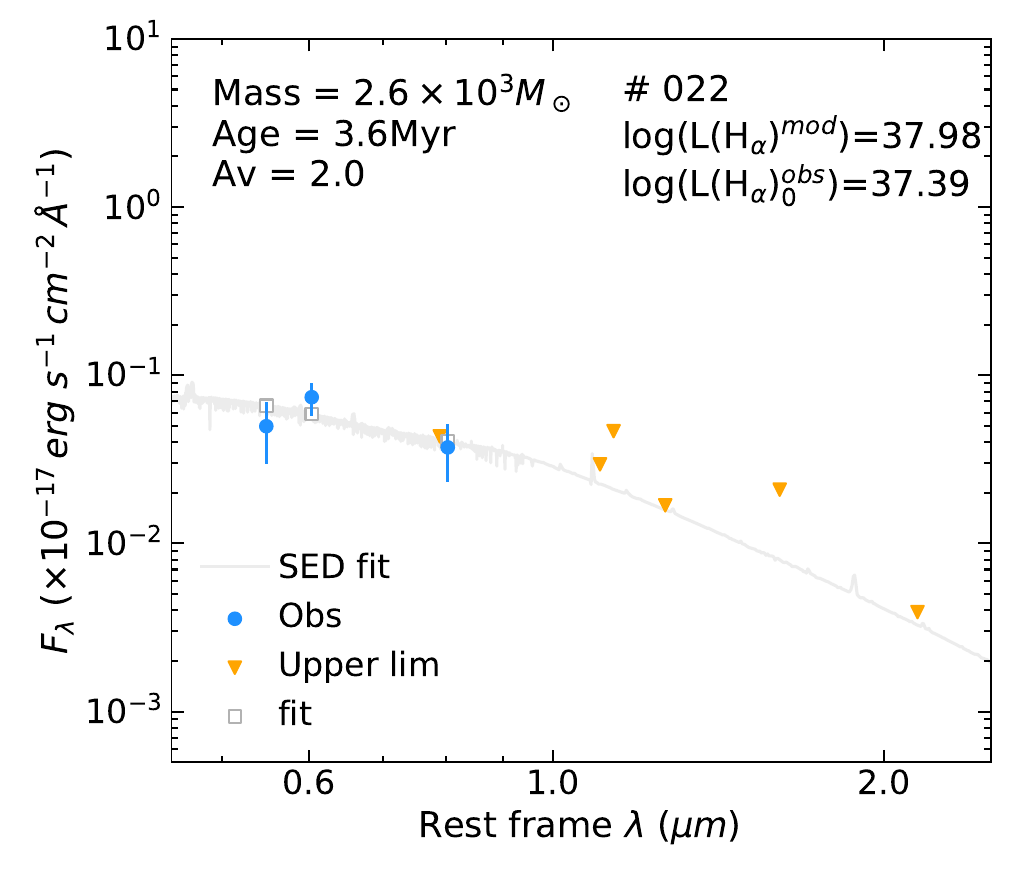}
      			\includegraphics[width=0.35\textwidth]{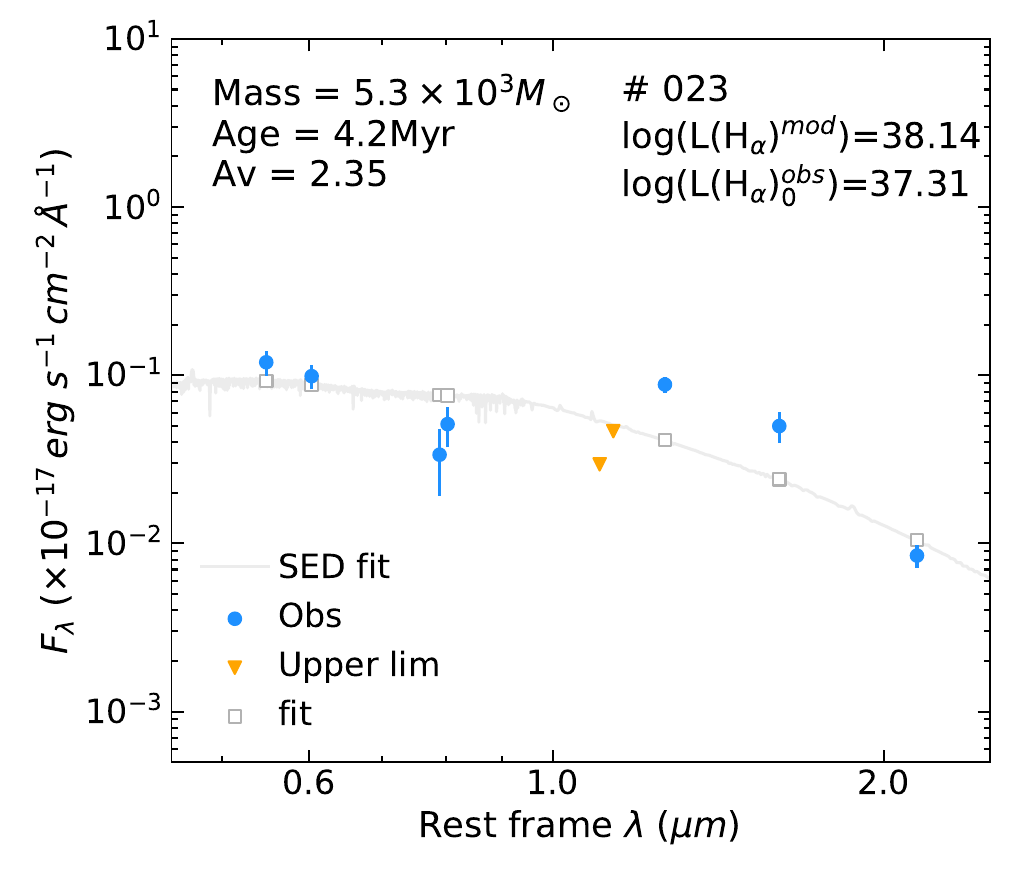}
			\includegraphics[width=0.35\textwidth]{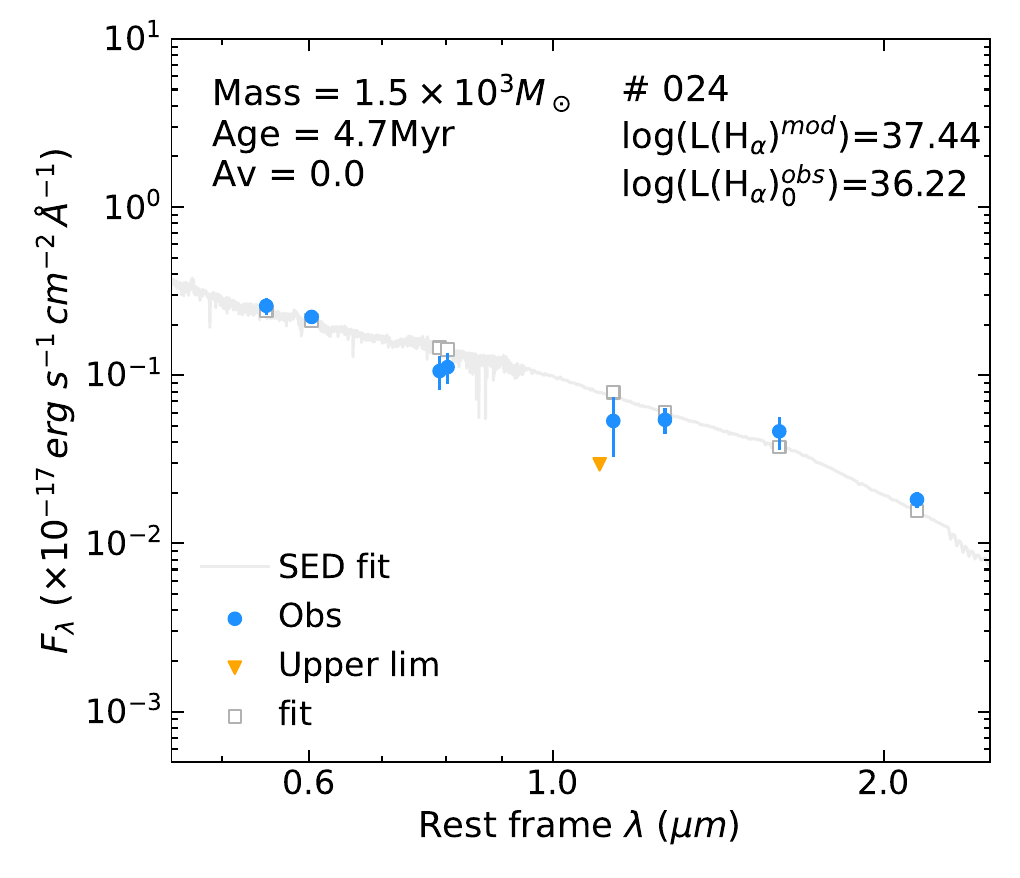}}
\contcaption{Results of the SED fit for each cluster.}
\end{figure*}     

\begin{figure*}
     \centering
     \subfigure{	\includegraphics[width=0.35\textwidth]{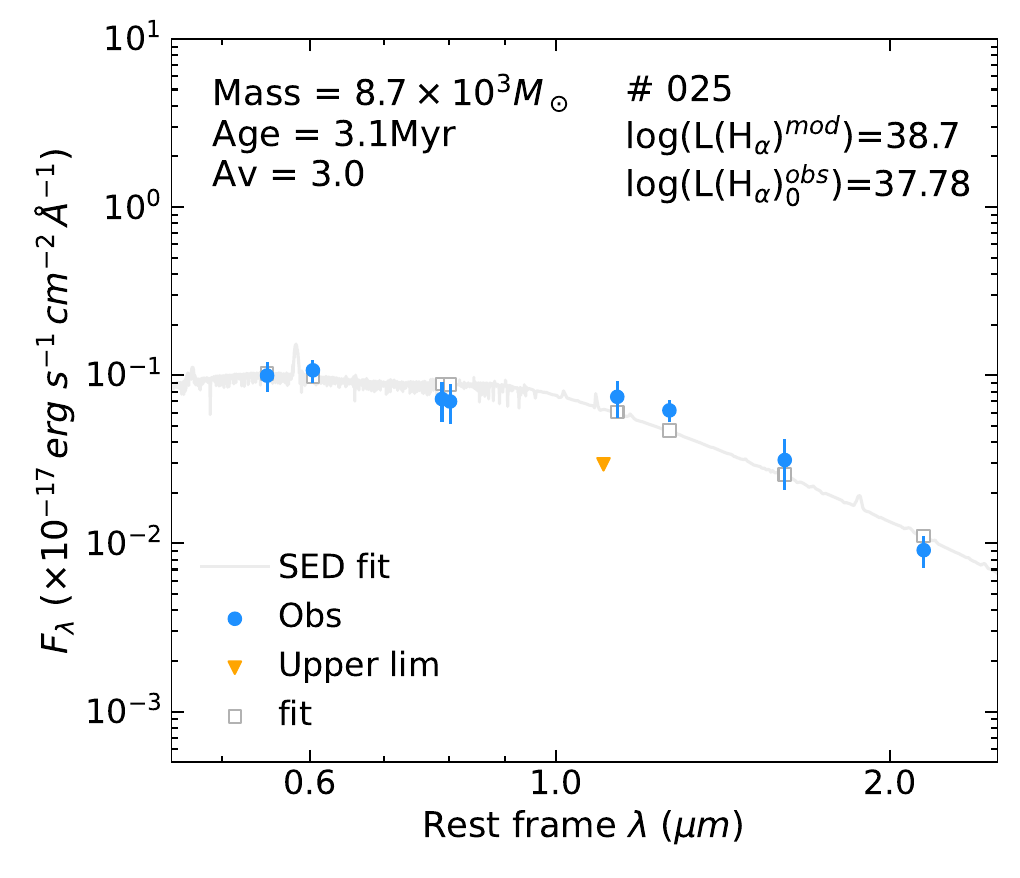}
      			\includegraphics[width=0.35\textwidth]{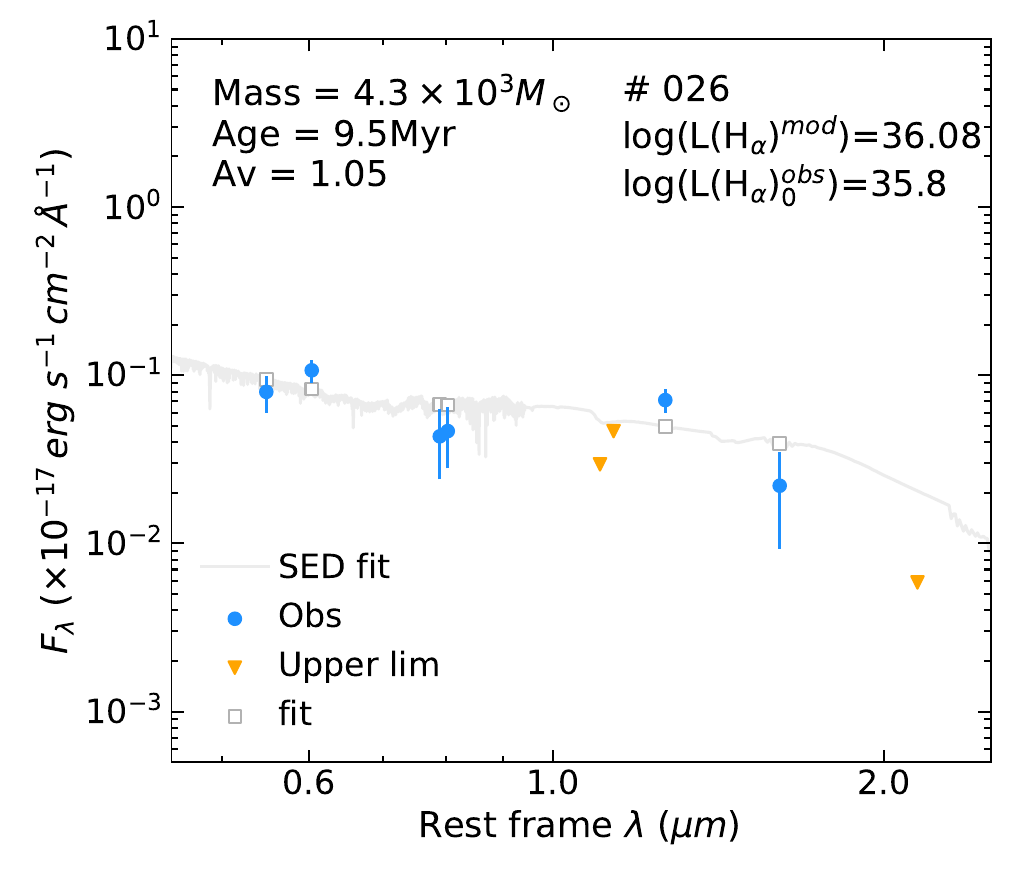}
			\includegraphics[width=0.35\textwidth]{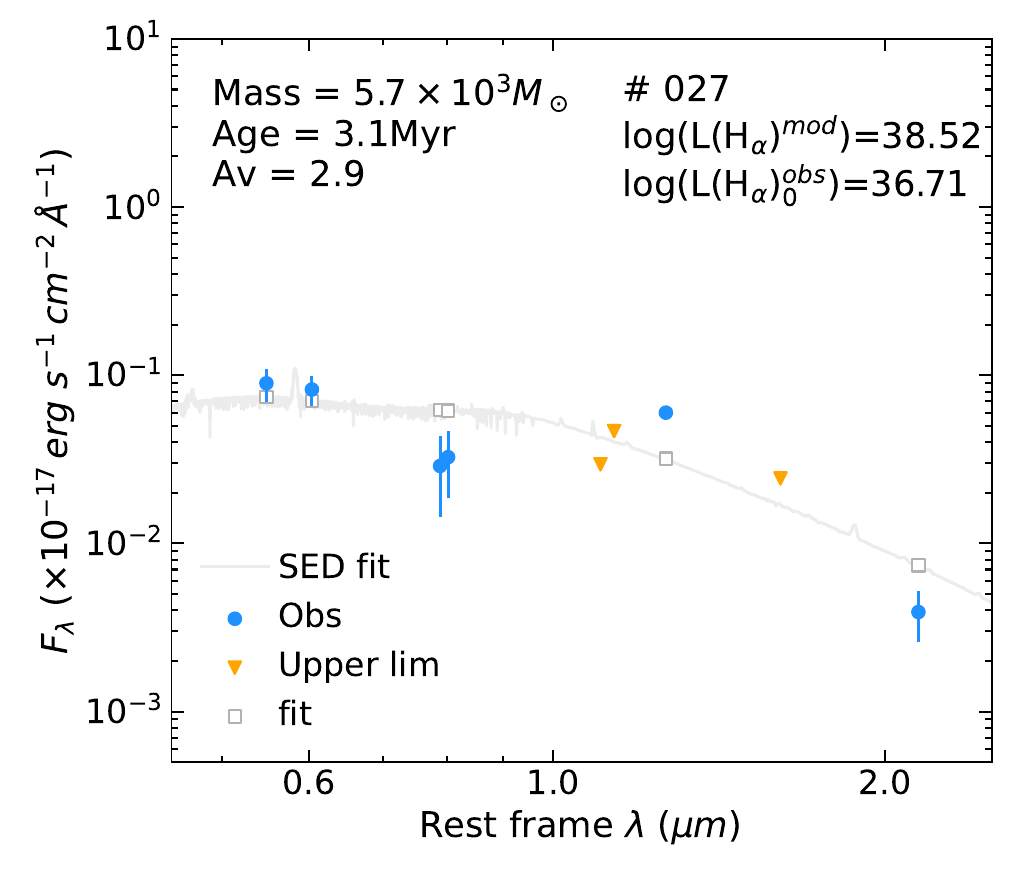}}
     \subfigure{	\includegraphics[width=0.35\textwidth]{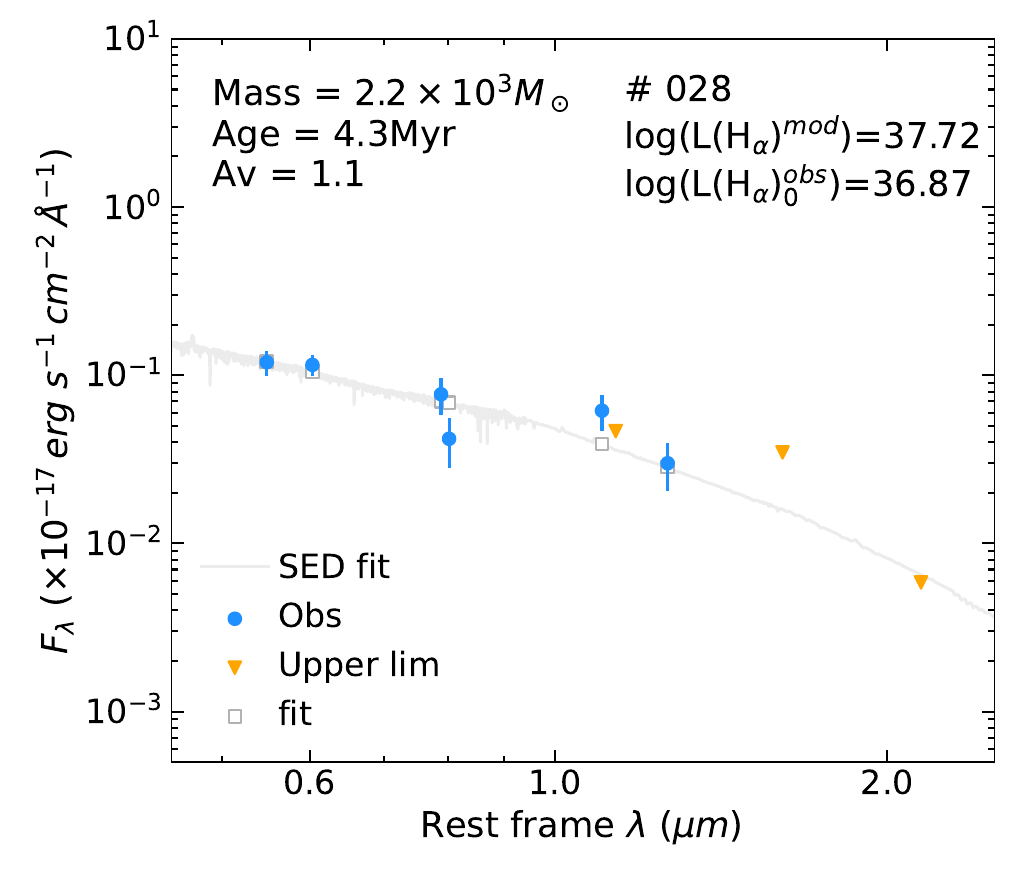}
			\includegraphics[width=0.35\textwidth]{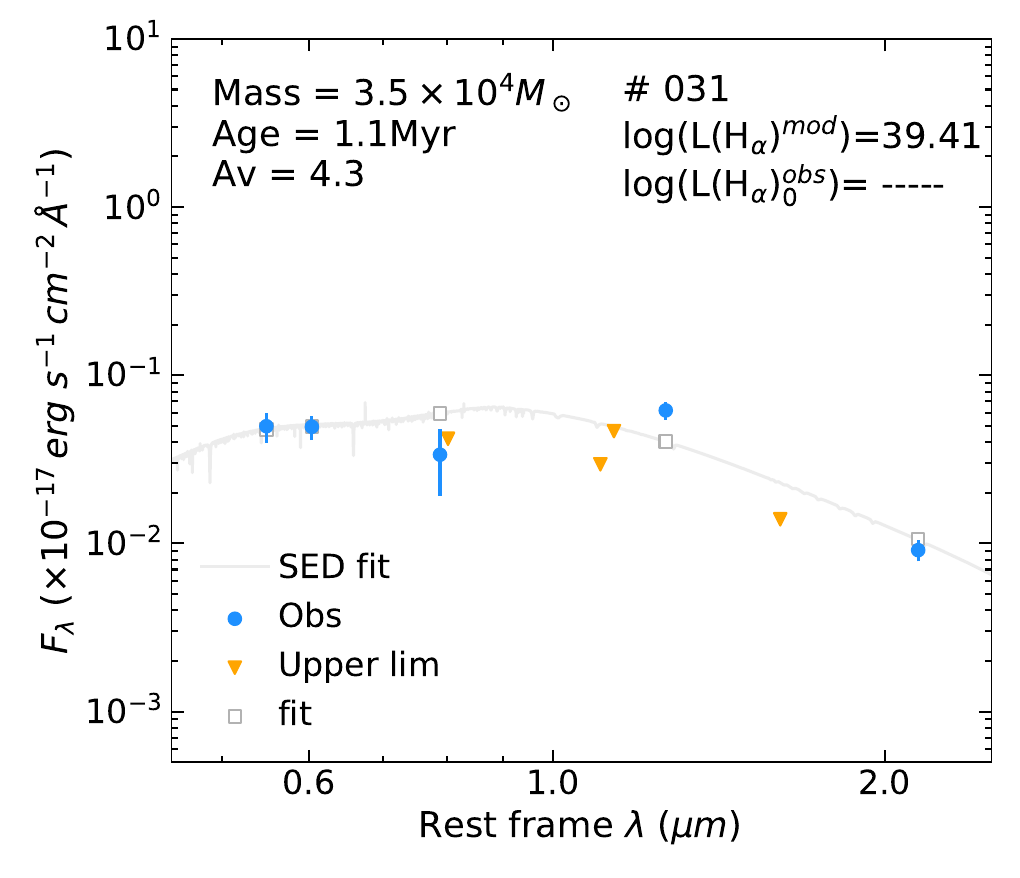}
			\includegraphics[width=0.35\textwidth]{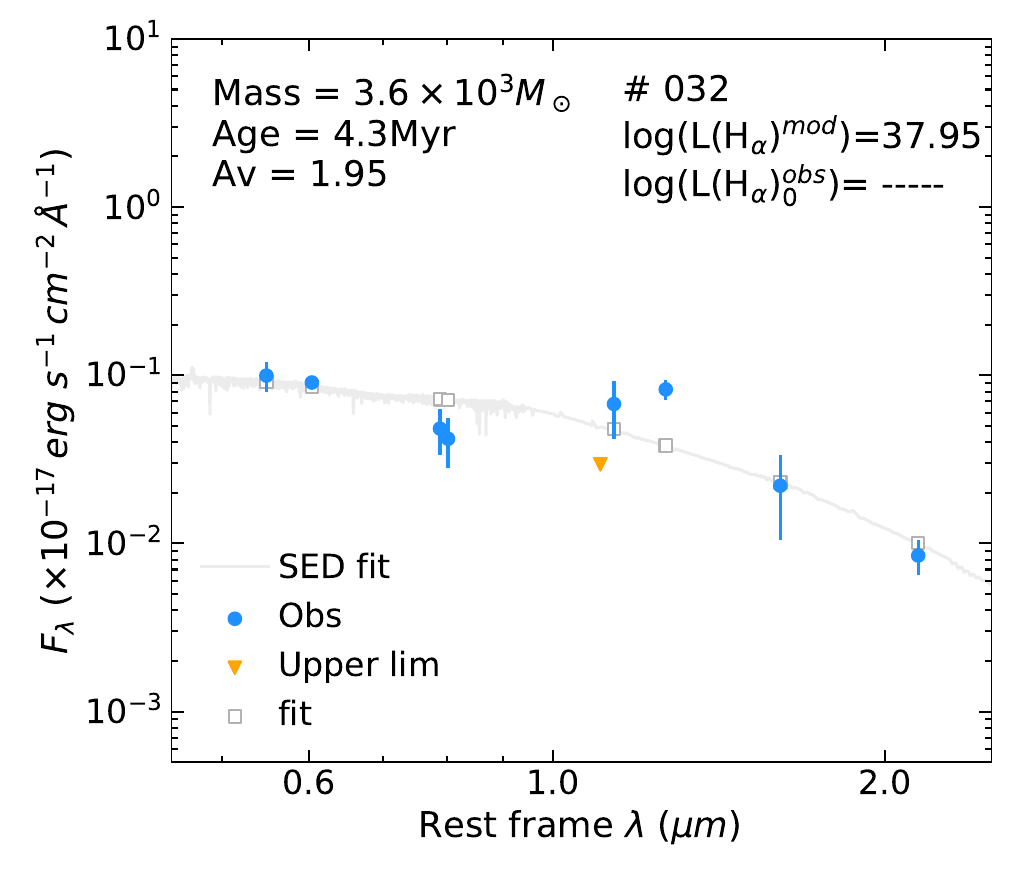}}
     \subfigure{	\includegraphics[width=0.35\textwidth]{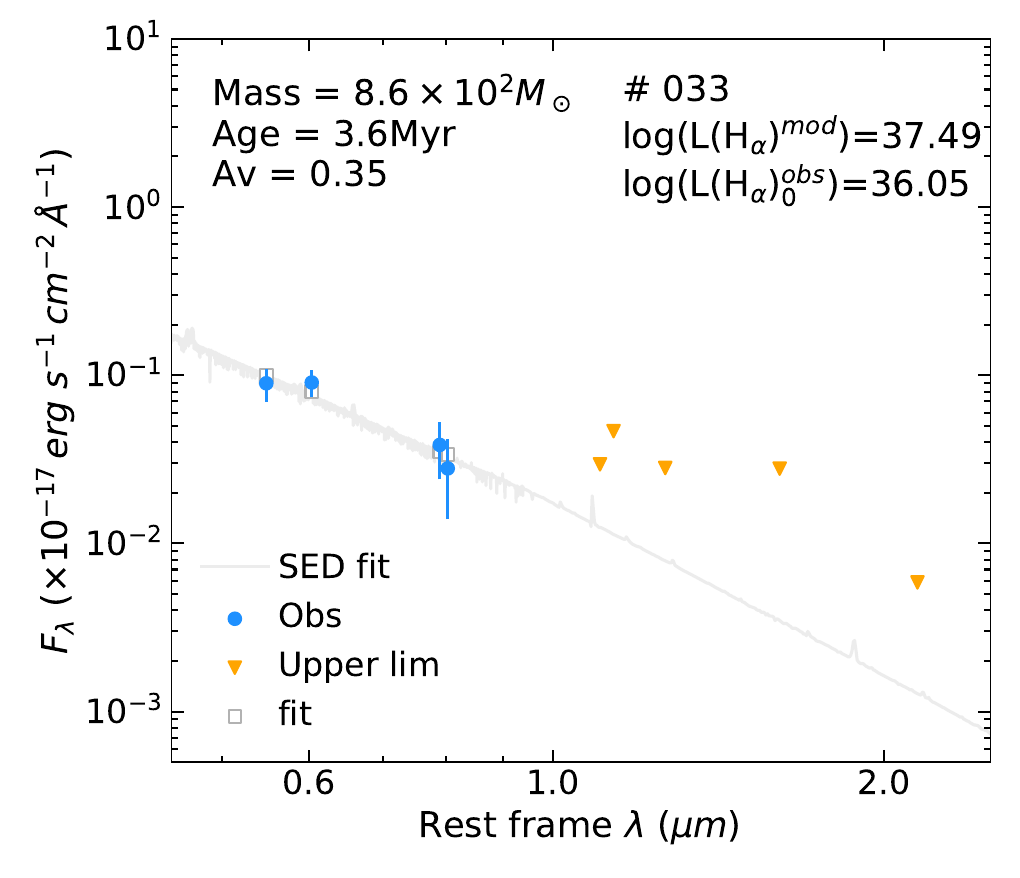}
      			\includegraphics[width=0.35\textwidth]{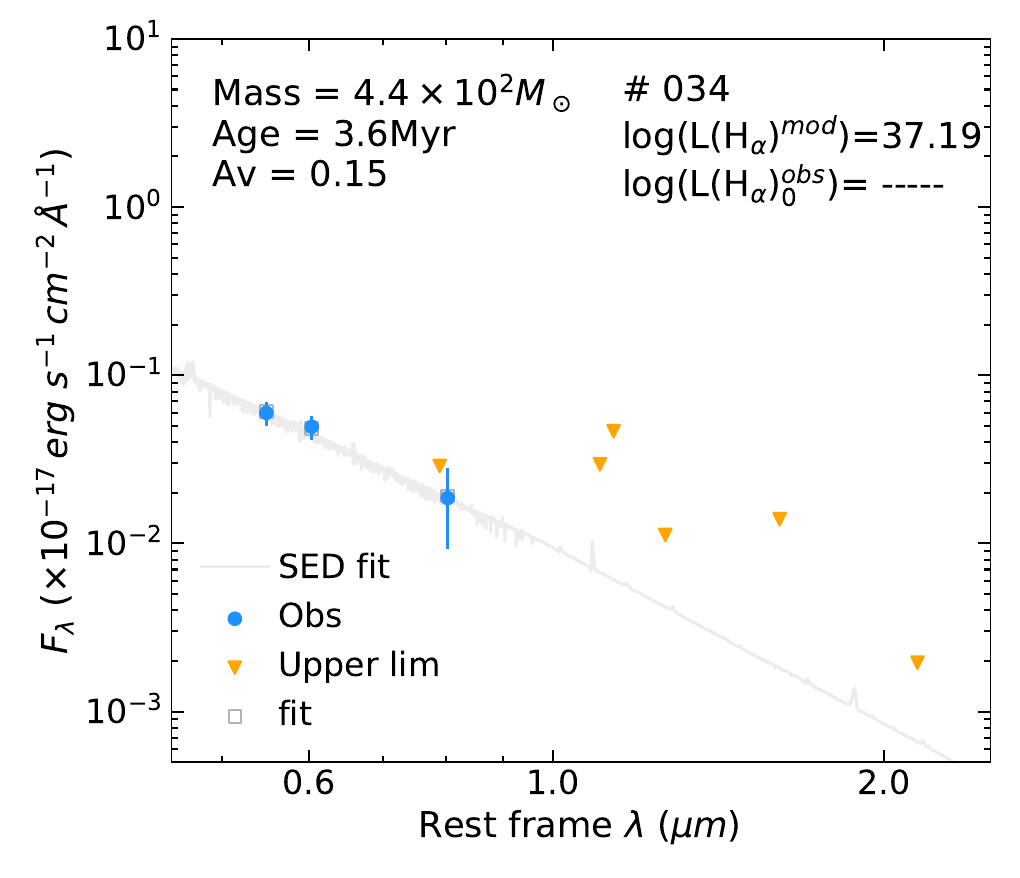}
			\includegraphics[width=0.35\textwidth]{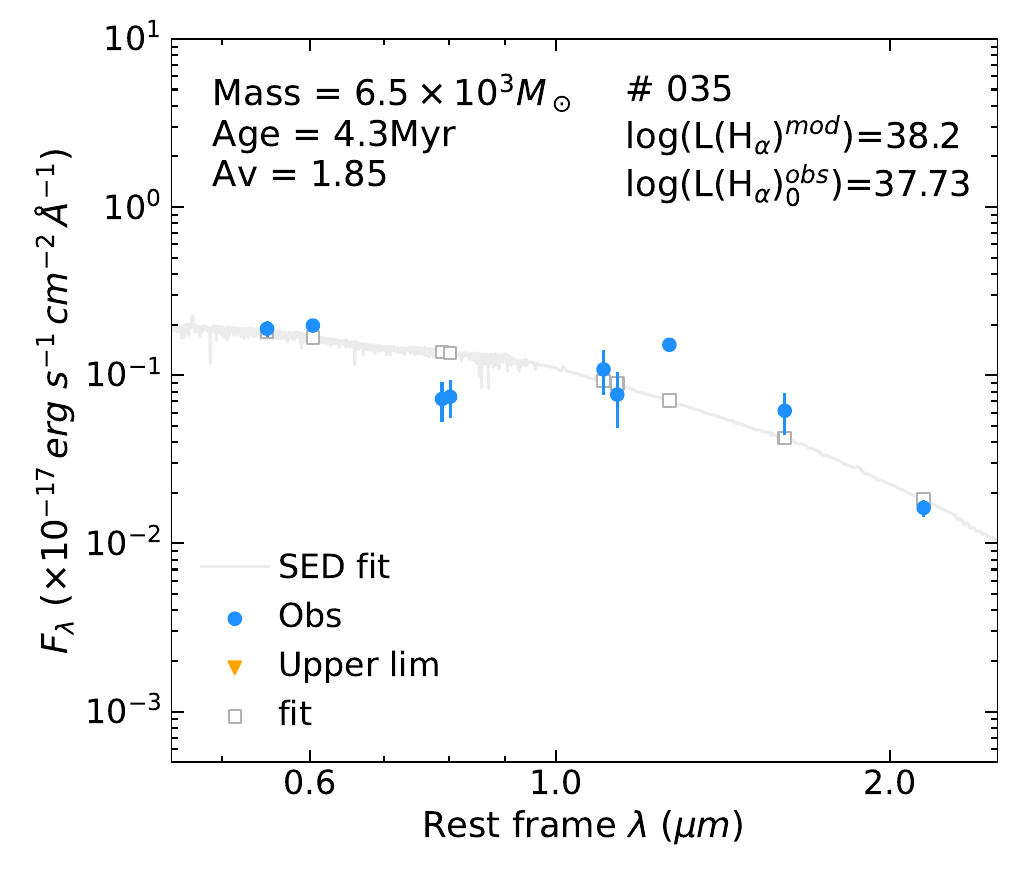}}
     \subfigure{	\includegraphics[width=0.35\textwidth]{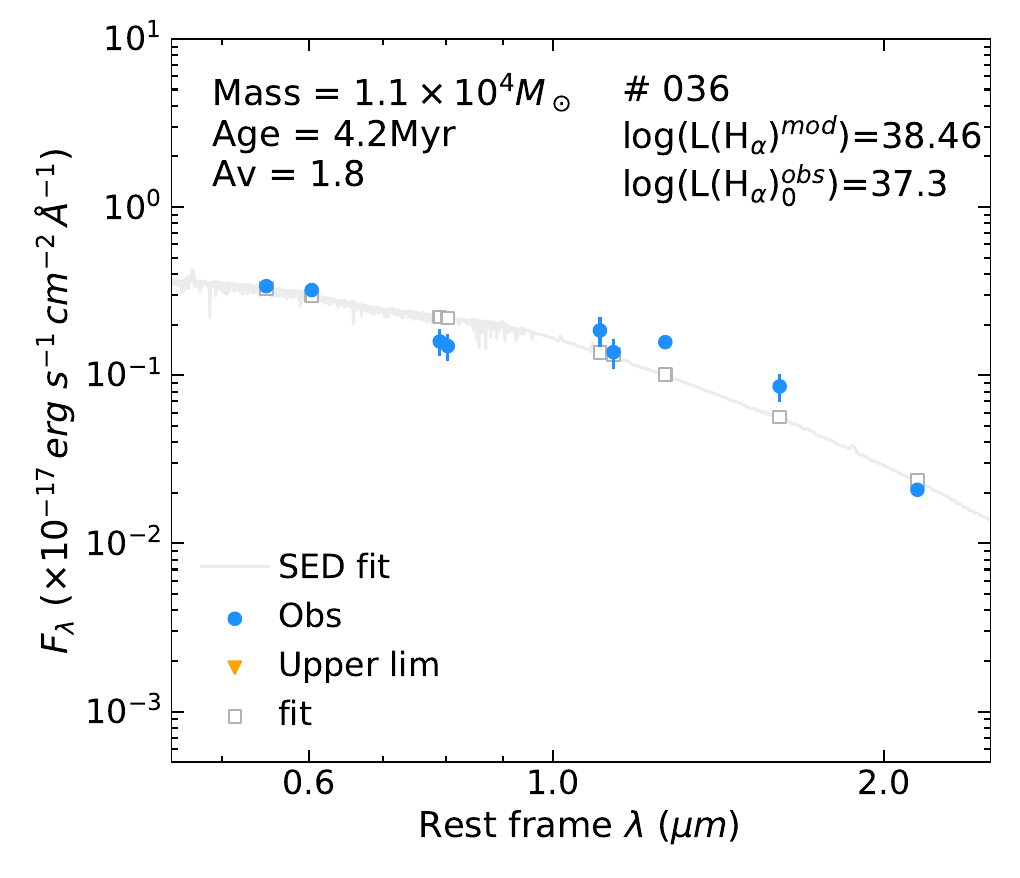}
      			\includegraphics[width=0.35\textwidth]{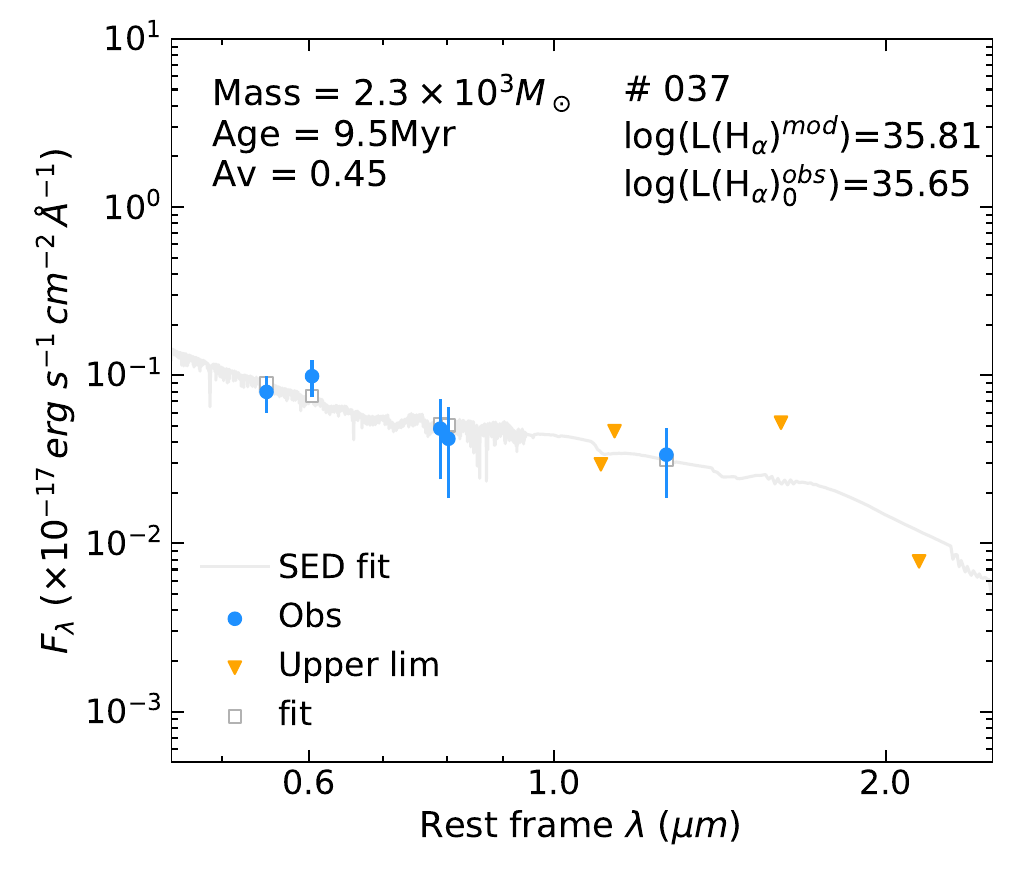}
			\includegraphics[width=0.35\textwidth]{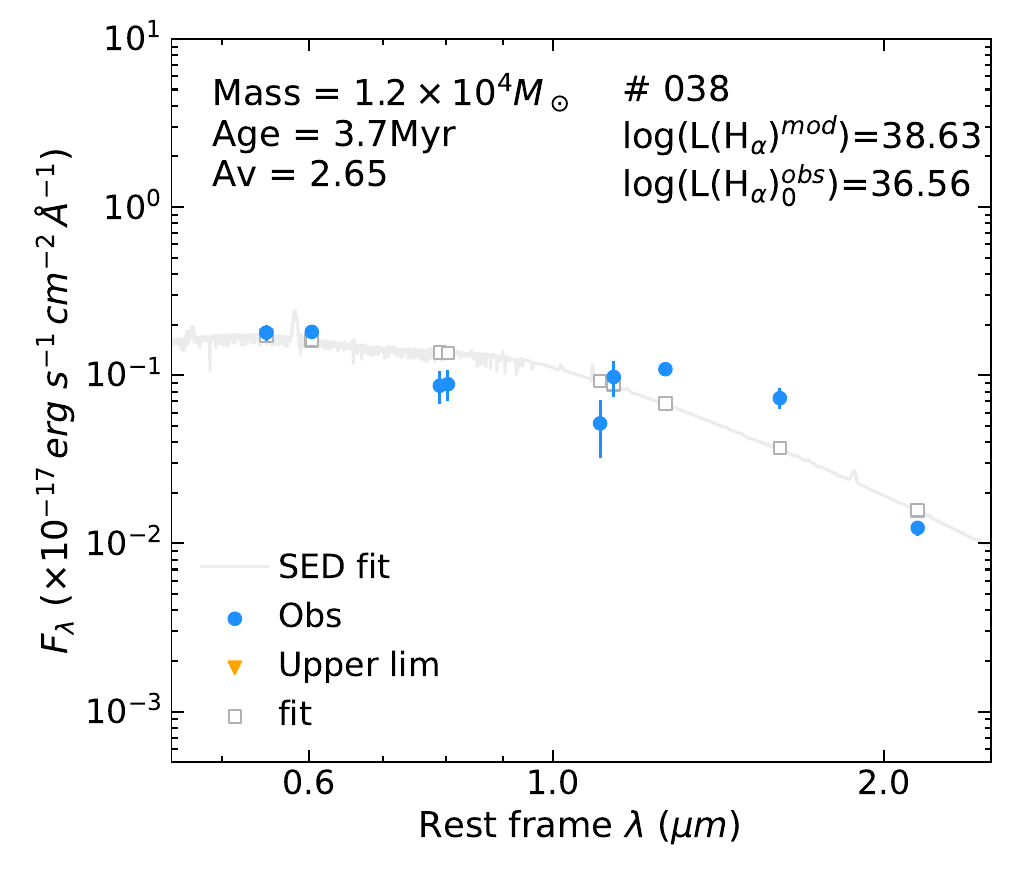}}
\contcaption{Results of the SED fit for each cluster.}
\end{figure*}     

\begin{figure*}
     \centering
     \subfigure{	\includegraphics[width=0.35\textwidth]{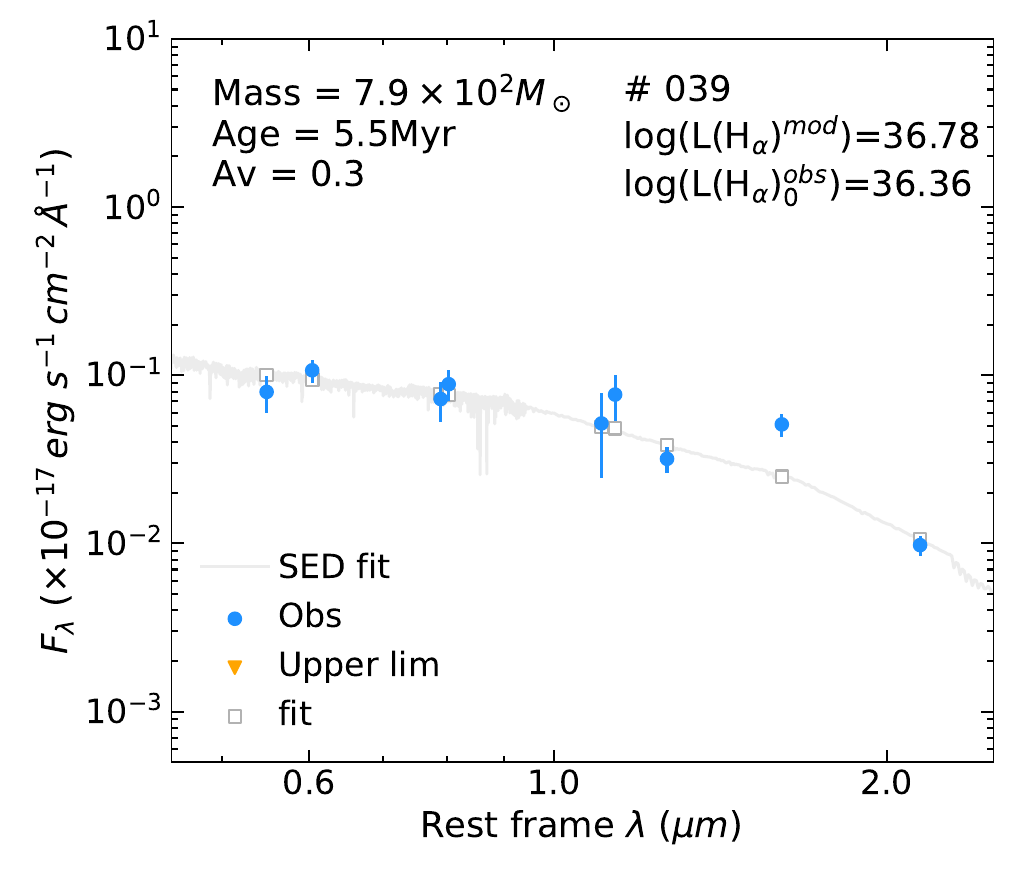}
      			\includegraphics[width=0.35\textwidth]{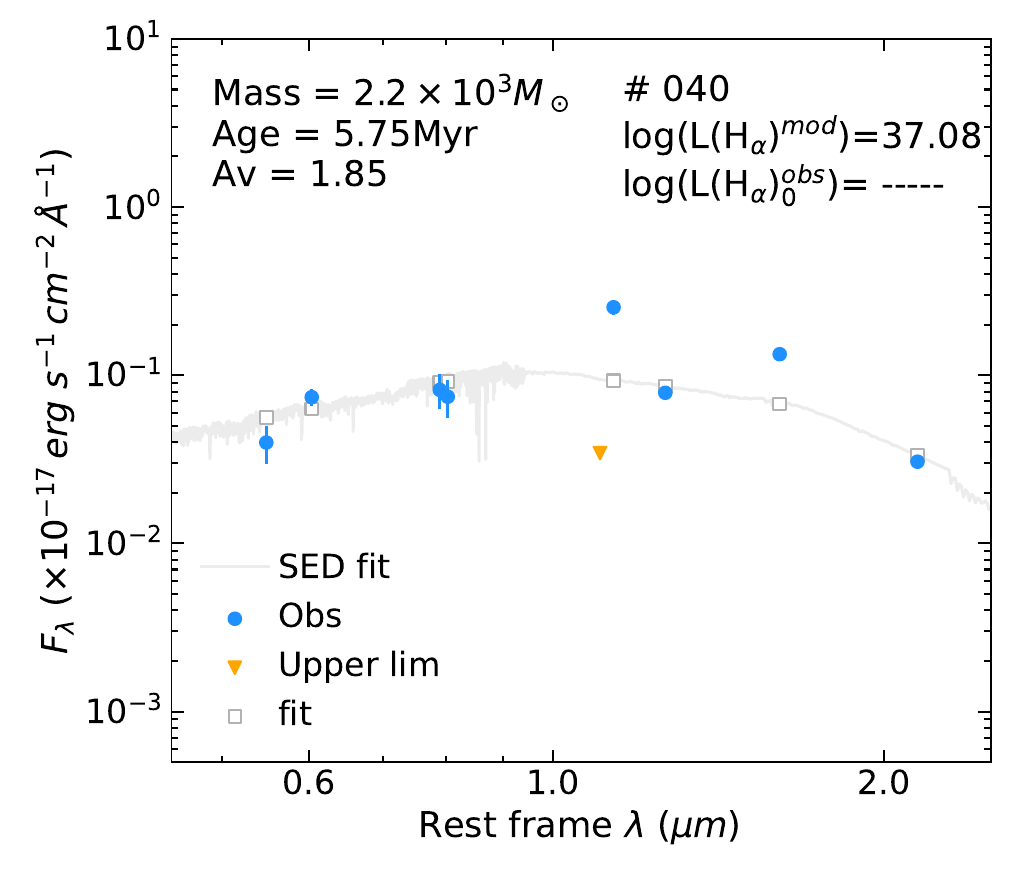}
			\includegraphics[width=0.35\textwidth]{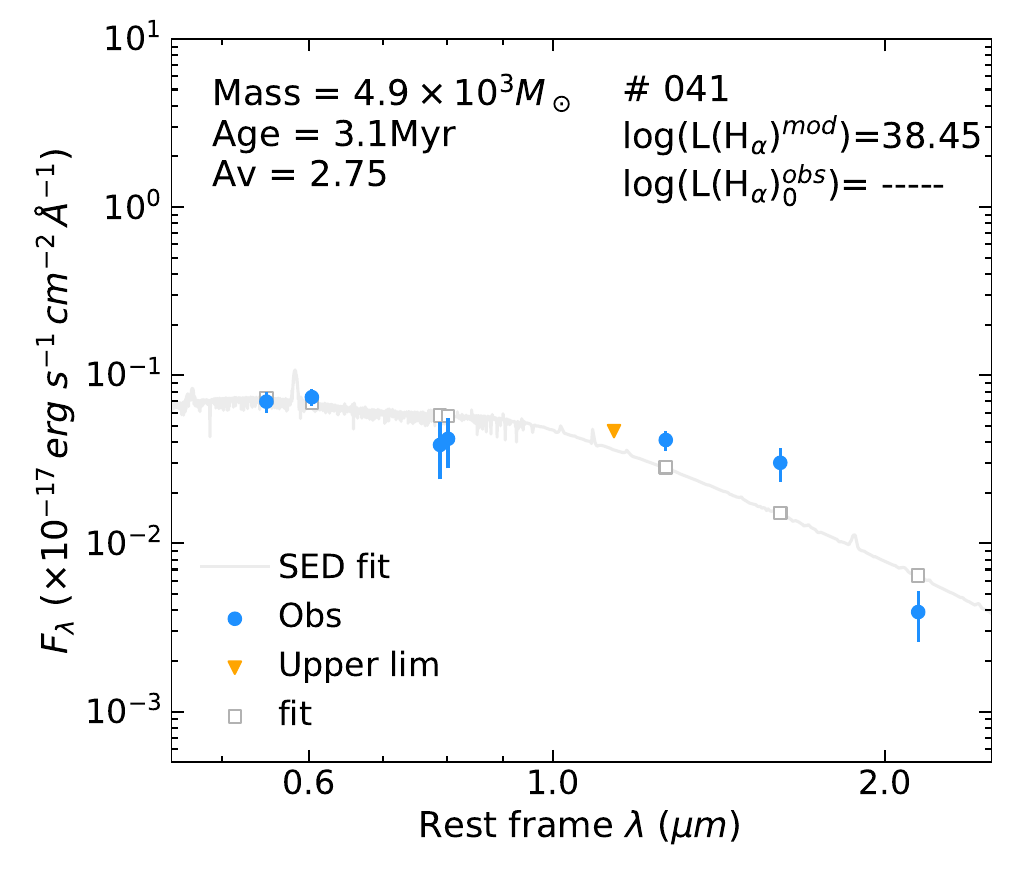}}
     \subfigure{	\includegraphics[width=0.35\textwidth]{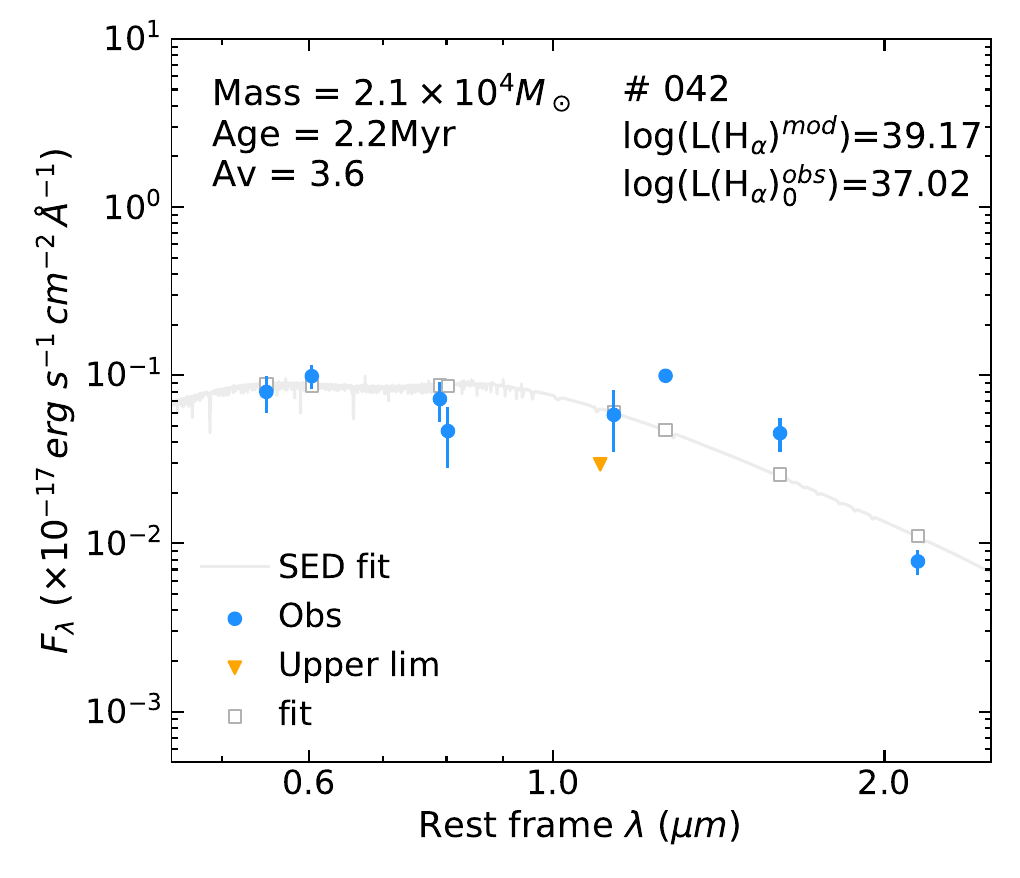}
      			\includegraphics[width=0.35\textwidth]{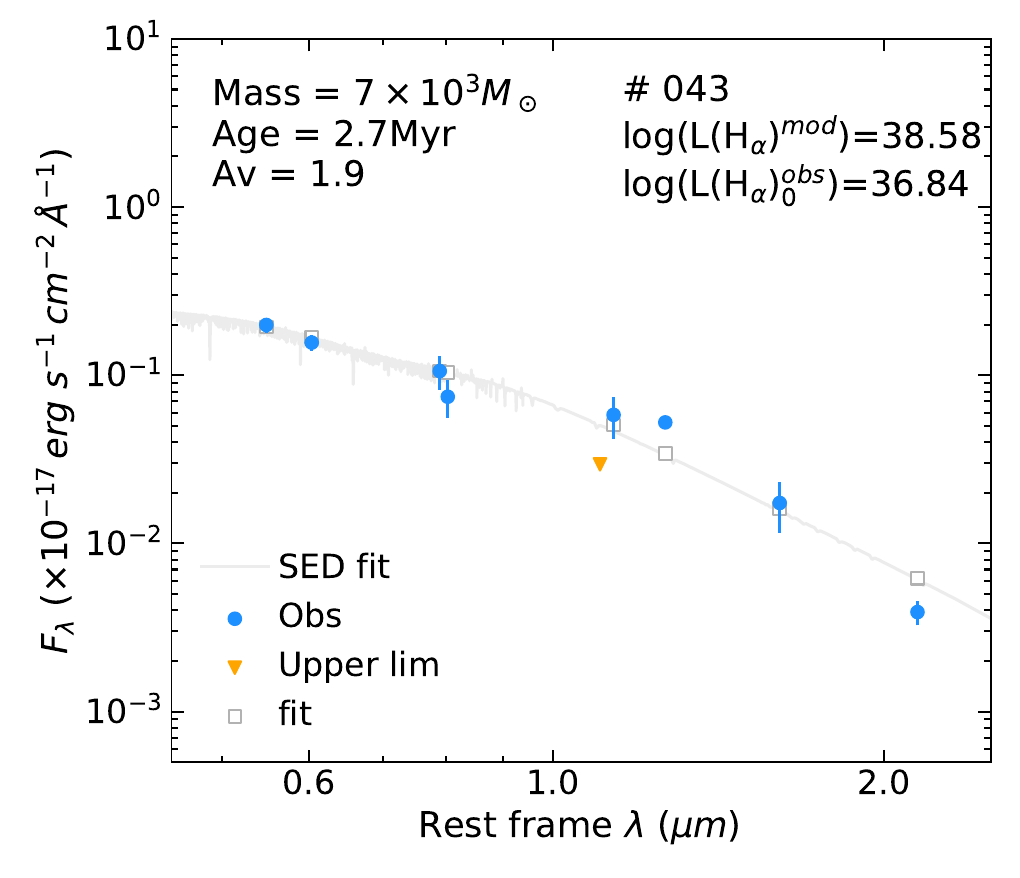}
			\includegraphics[width=0.35\textwidth]{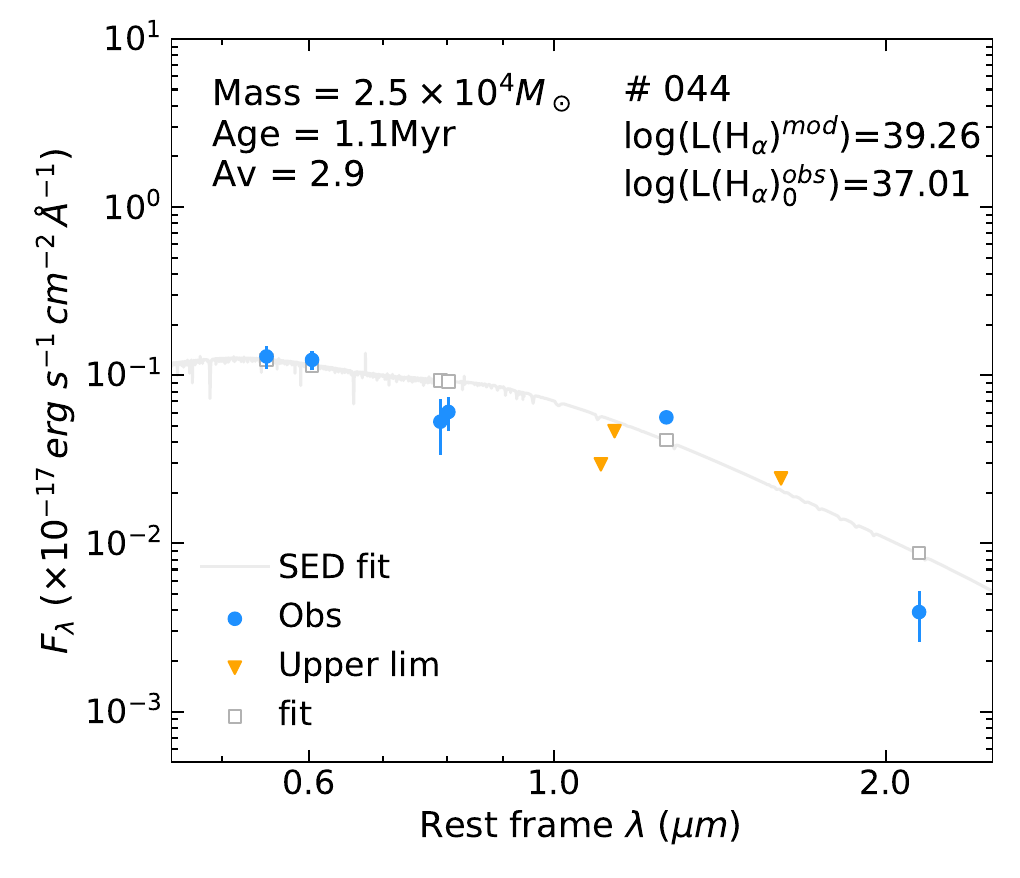}}
     \subfigure{	\includegraphics[width=0.35\textwidth]{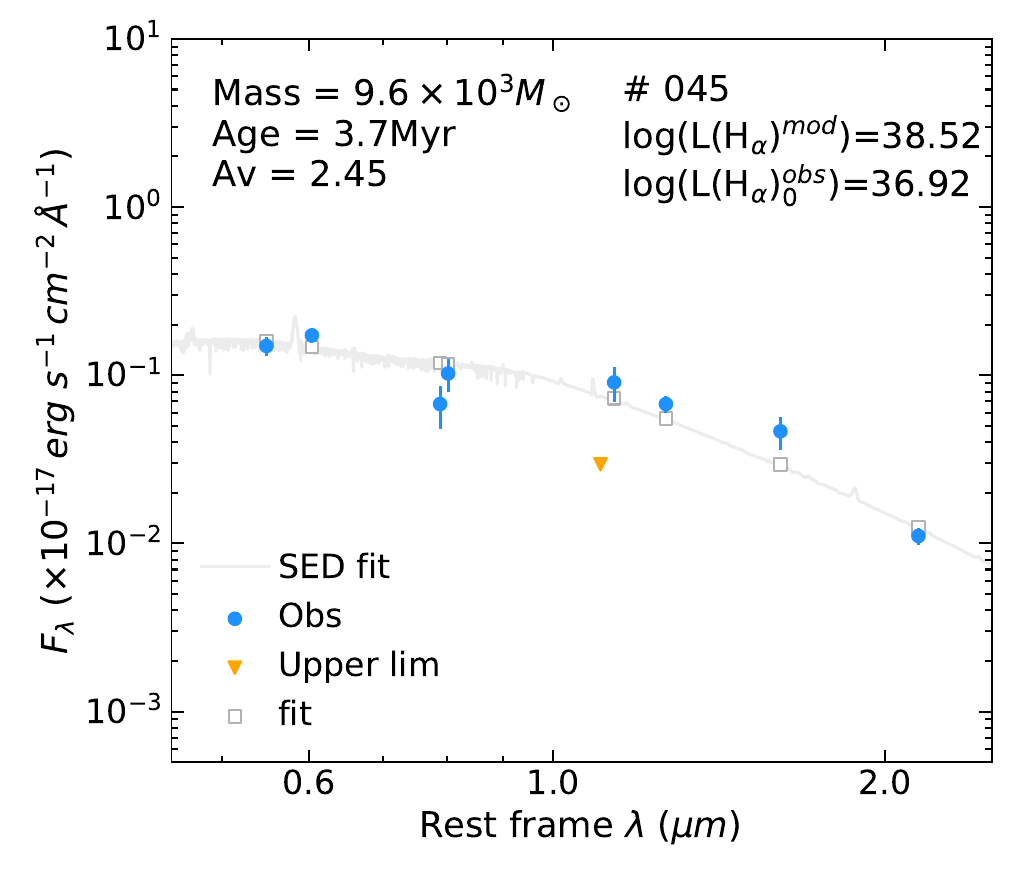}
      			\includegraphics[width=0.35\textwidth]{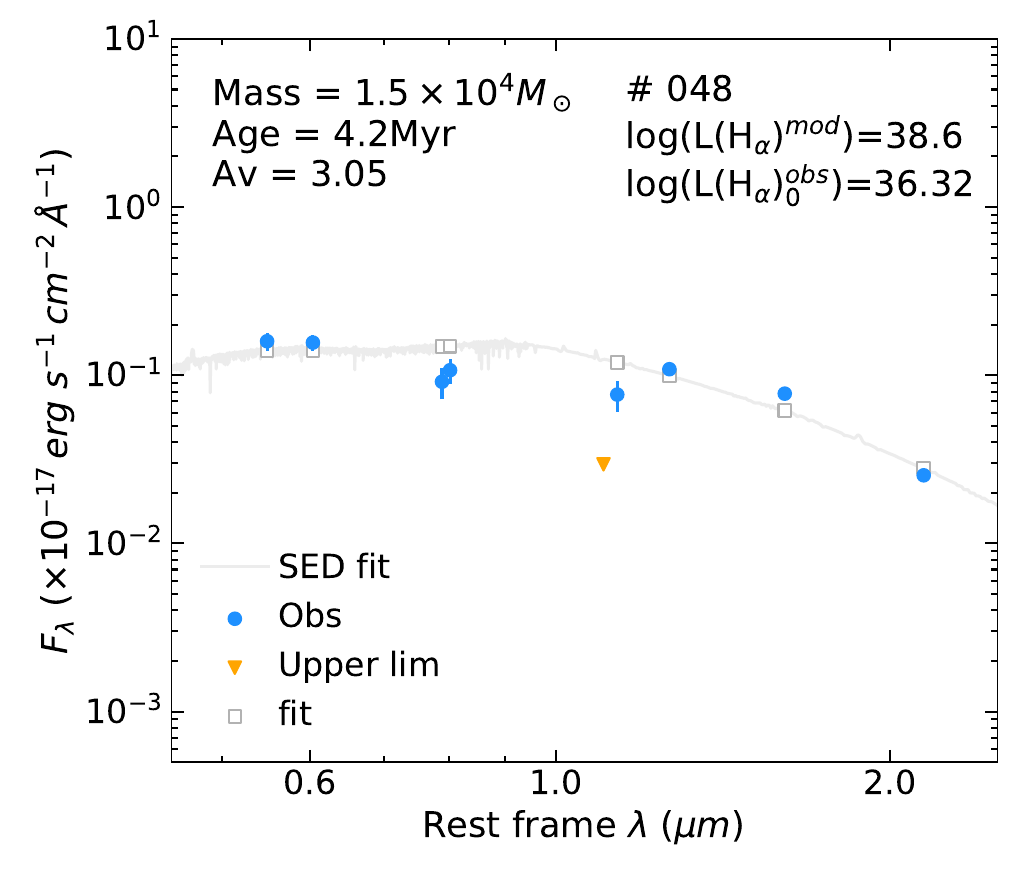}
			\includegraphics[width=0.35\textwidth]{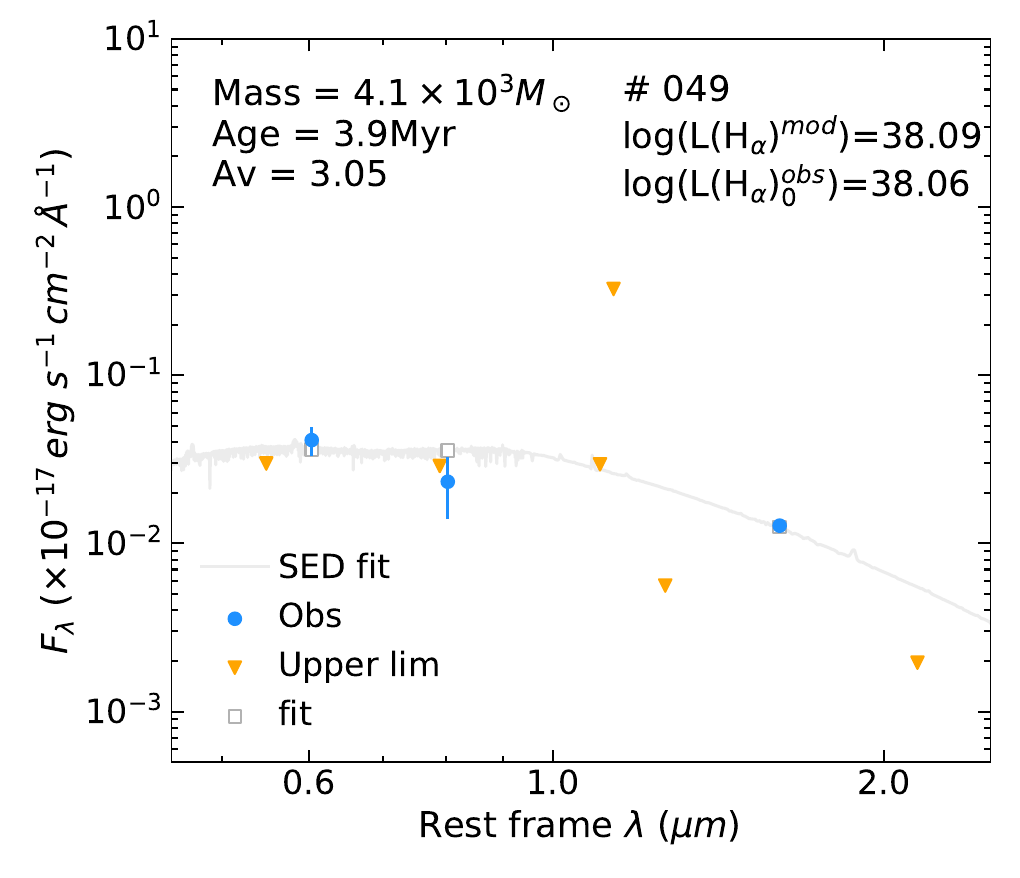}}
     \subfigure{	\includegraphics[width=0.35\textwidth]{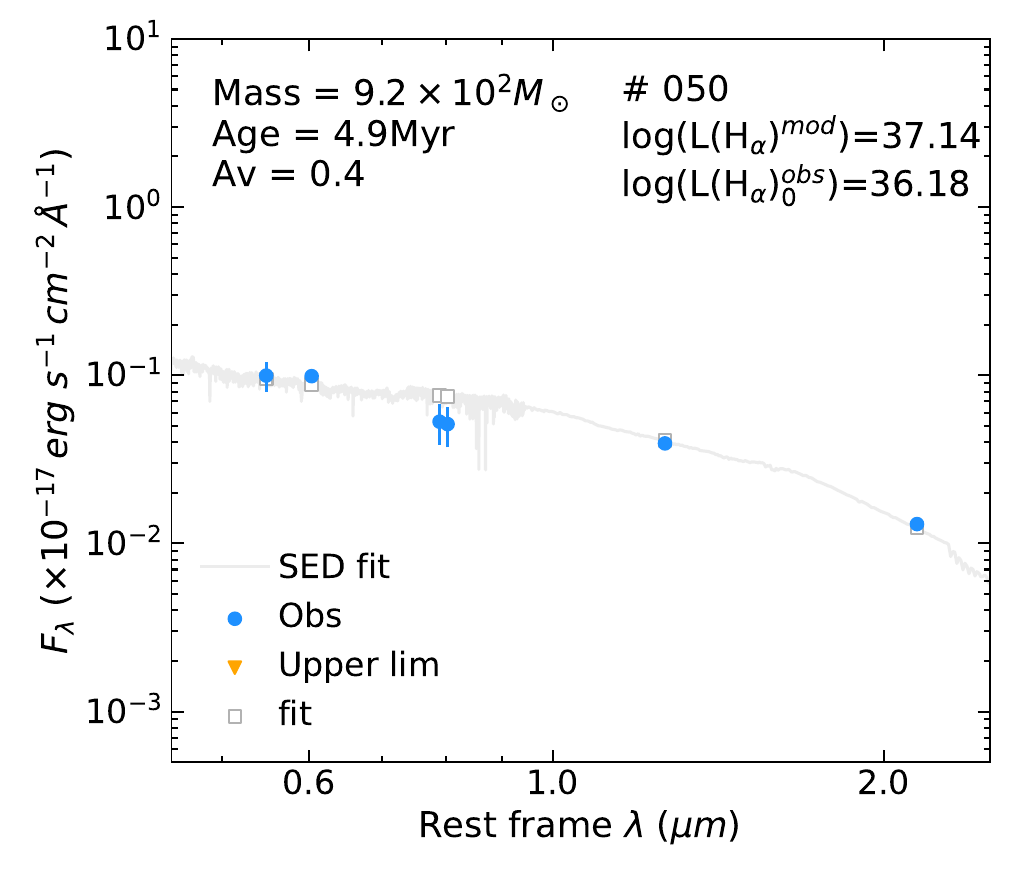}
      			\includegraphics[width=0.35\textwidth]{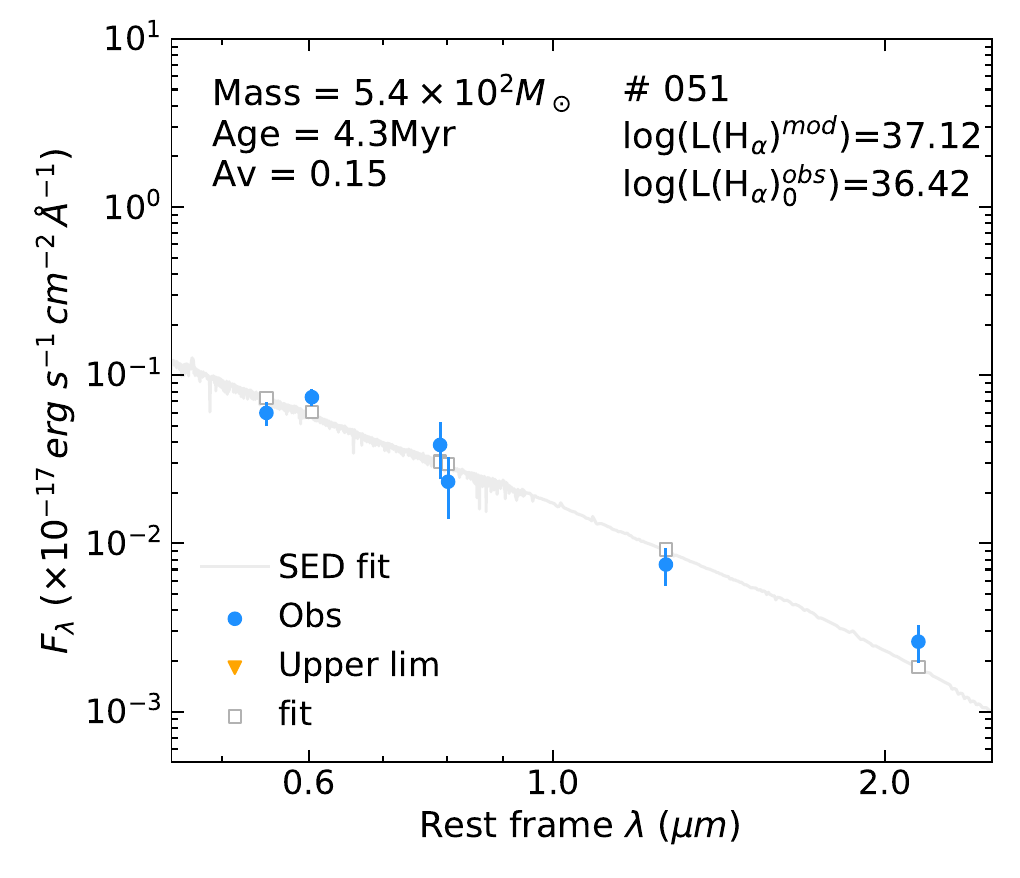}
			\includegraphics[width=0.35\textwidth]{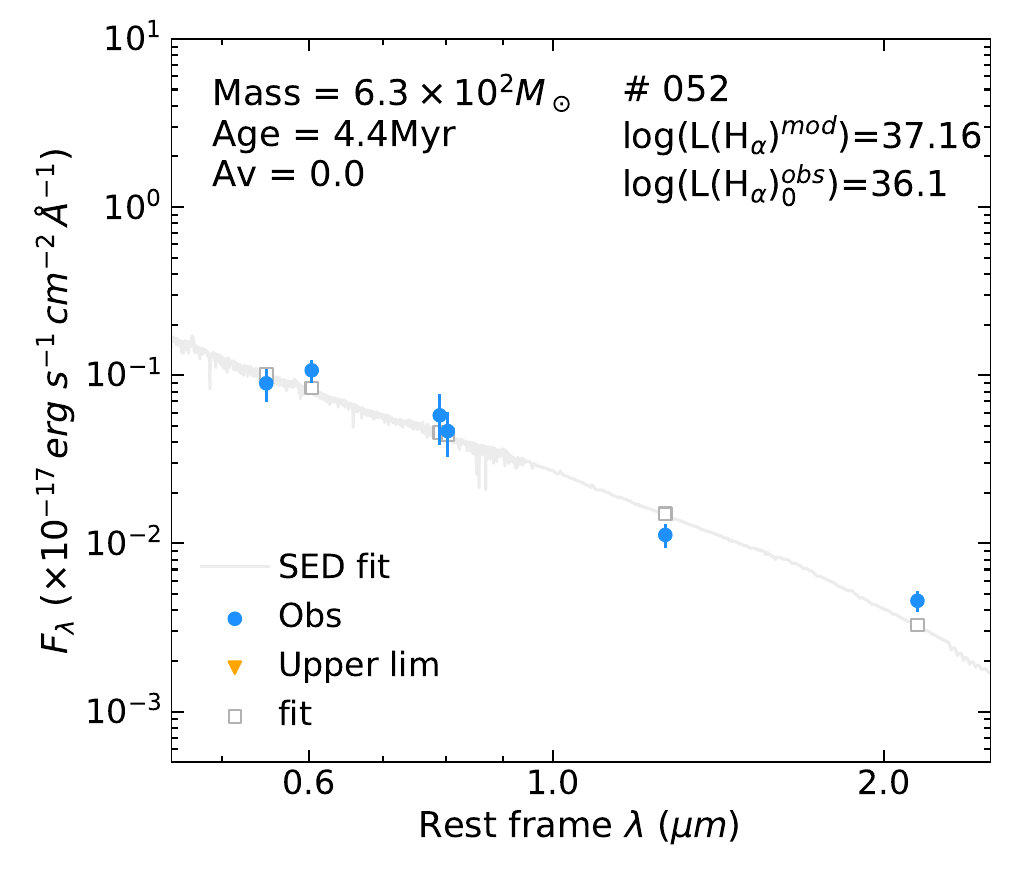}}
\contcaption{Results of the SED fit for each cluster.}
\end{figure*}     

\begin{figure*}
     \centering
     \subfigure{	\includegraphics[width=0.35\textwidth]{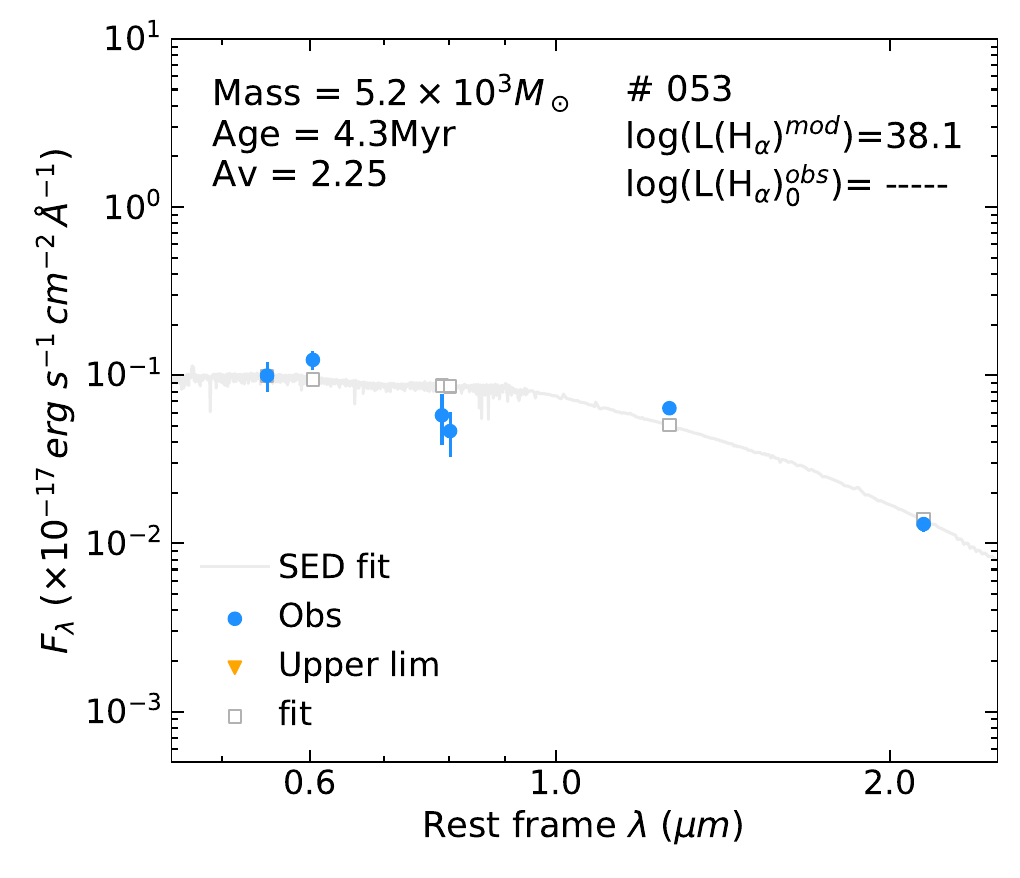}
      			\includegraphics[width=0.35\textwidth]{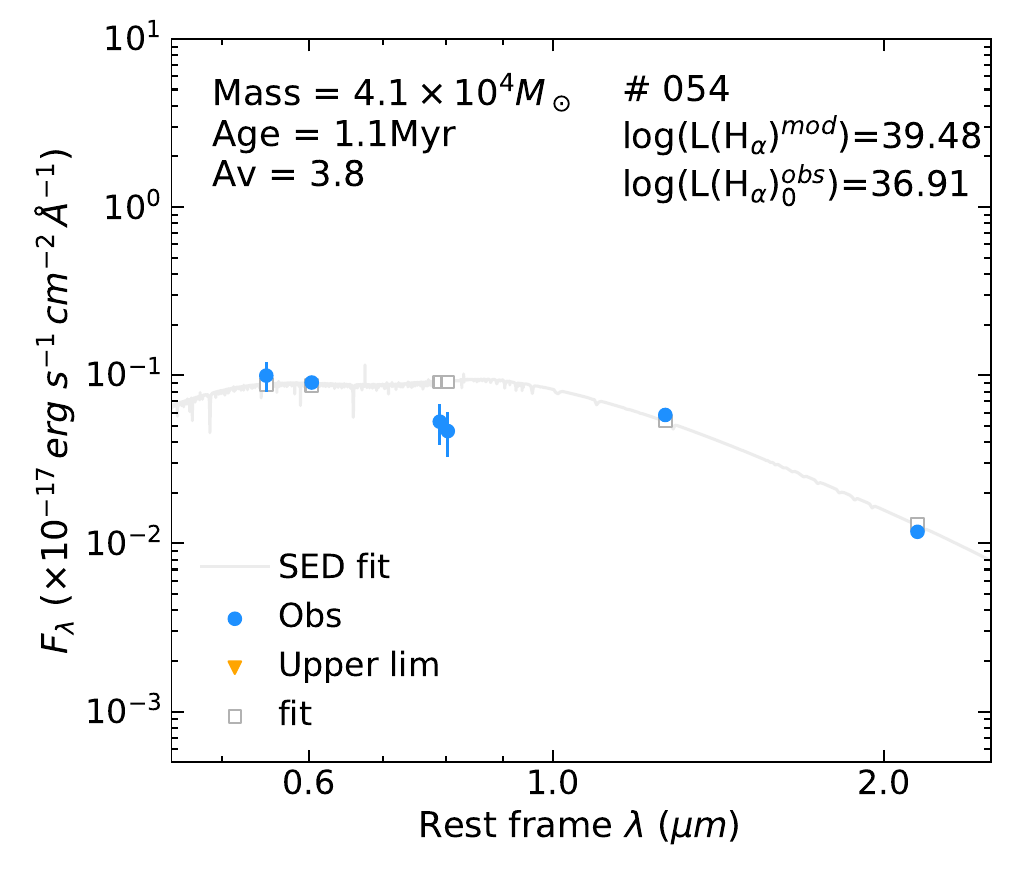}
			\includegraphics[width=0.35\textwidth]{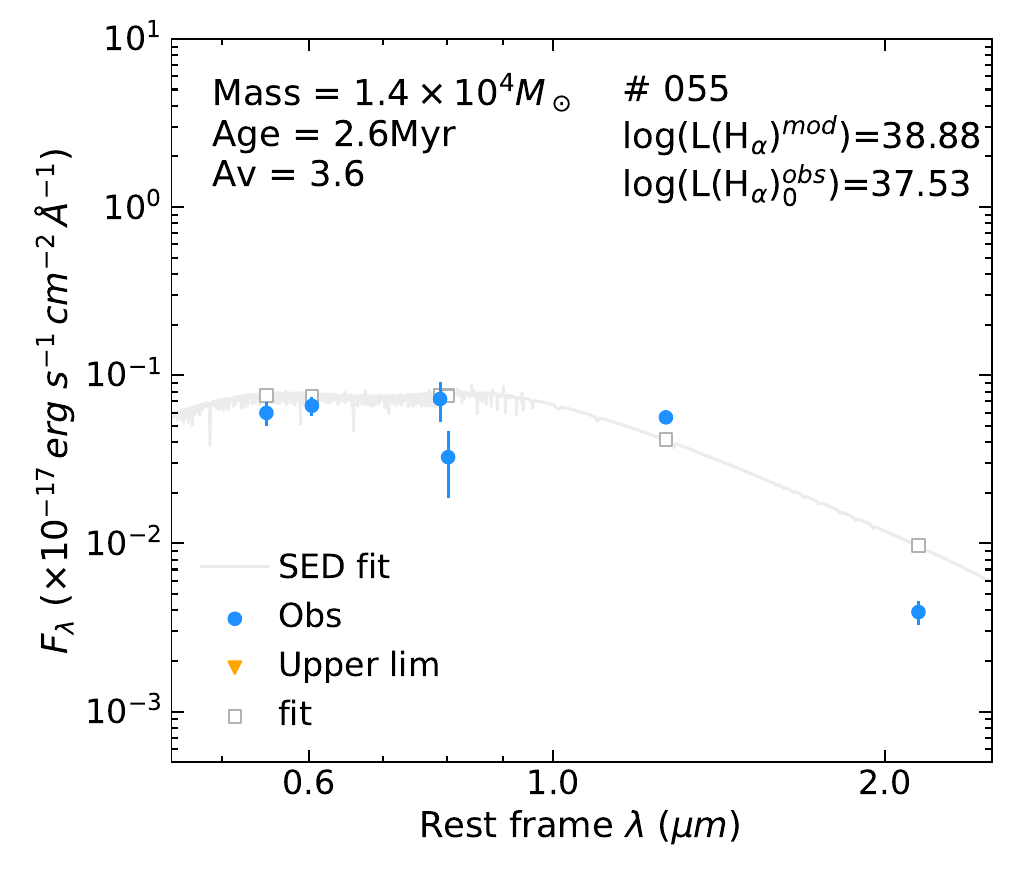}}
     \subfigure{	\includegraphics[width=0.35\textwidth]{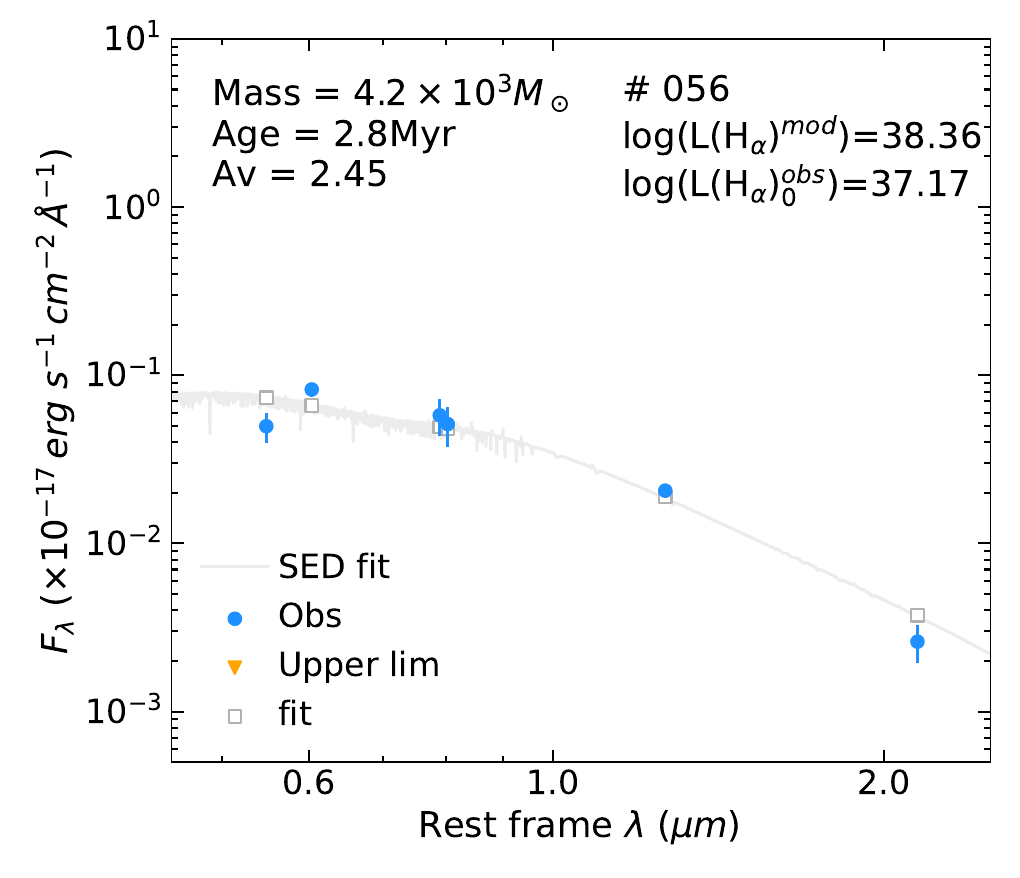}
      			\includegraphics[width=0.35\textwidth]{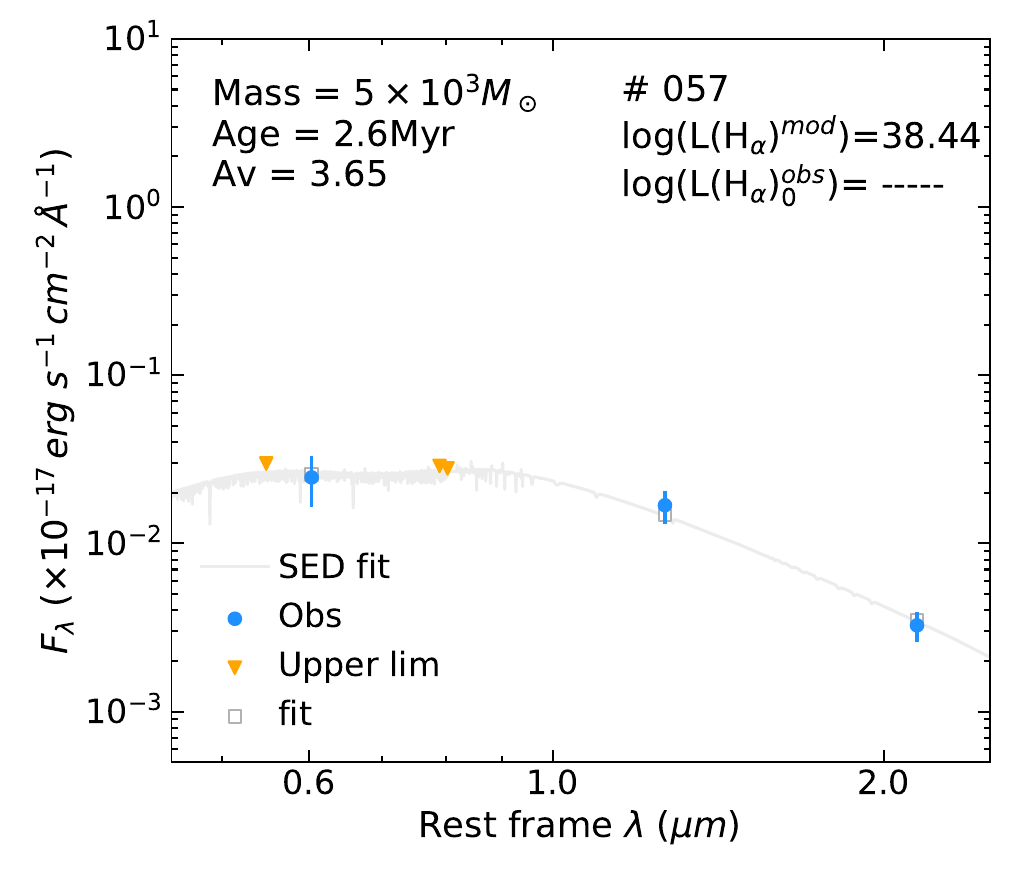}
			\includegraphics[width=0.35\textwidth]{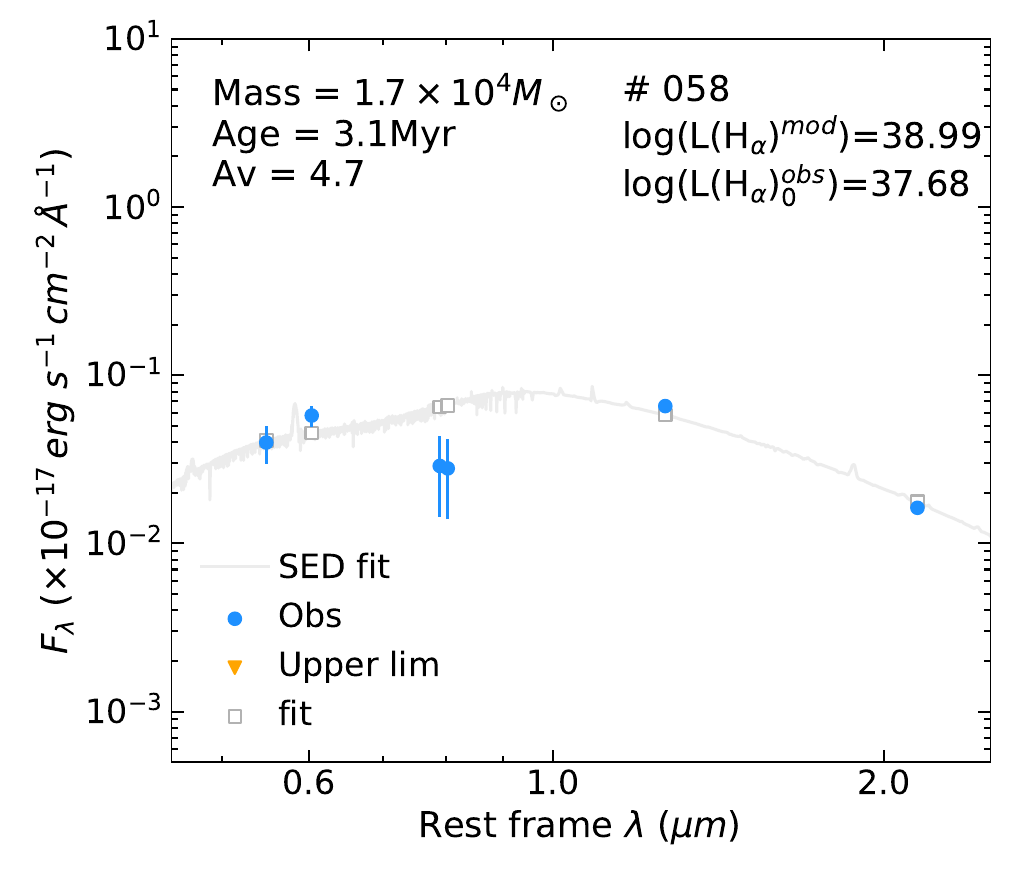}}
     \subfigure{	\includegraphics[width=0.35\textwidth]{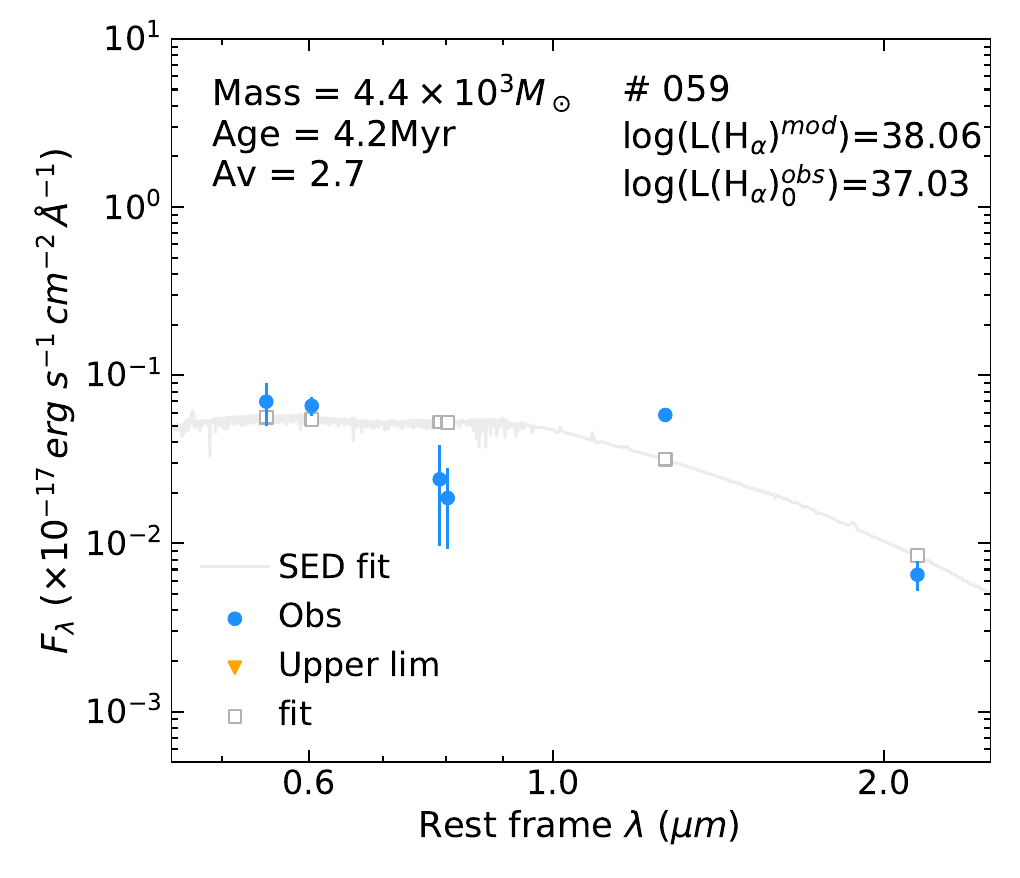}}
\contcaption{Results of the SED fit for each cluster.}
\end{figure*}


\bsp	
\label{lastpage}
\end{document}